\documentclass[11pt]{article}
\usepackage[margin=1in]{geometry}
\usepackage{amssymb,amsthm}
\usepackage{amsmath,amsfonts,cancel}
\usepackage{thmtools,enumitem}
\usepackage{pifont}

\usepackage{mathtools}
\usepackage{dsfont}
\usepackage{tcolorbox}
\usepackage{booktabs}
\usepackage{bbm}
\usepackage{xspace}
\usepackage{subfigure}
\usepackage{tensor}
\usepackage{bm}
\usepackage{tikz}
\usepackage{placeins}
\usepackage{setspace}
\usepackage{hyperref}
\usepackage[capitalize]{cleveref}

\newcommand{\Deff}{\Delta_{\text{\it efficiency}}}
\newcommand{\Denv}{\Delta_{\text{\it EF}}}
\newcommand{\Dprop}{\Delta_{\text{\it prop}}}
\newcommand{\Denvhind}{\textsc{Envy}}
\newcommand{\Denvplus}{\Delta_{\text{\it EF}}^+}

\newcommand{\RR}{\mathbb{R}}

\newcommand{\A}{\mathcal{A}}

\newcommand{\Ind}[1]{\mathds{1}_{\left[ #1 \right]}}

\newcommand{\Exp}[1]{\mathbb E \left[ #1 \right]} 
\newcommand{\Var}[1]{\mathrm{Var} \left[ #1 \right]}
\renewcommand{\Pr}{\mathbb{P}}

\newcommand{\B}{\mathcal{B}}
\newcommand{\C}{\mathcal{C}}
\newcommand{\E}{\mathcal{E}}
\newcommand{\F}{\mathcal{F}}

\newcommand{\HopeGuardrail}{\textsc{Guarded-Hope}\xspace}

\newcommand{\conf}{\textsc{Conf}}
\newcommand{\abs}[1]{\left\lvert #1 \right\rvert}
\newcommand{\Bav}{\beta_{avg}}

\newcommand{\multconf}{\gamma}

\usepackage{multirow}
\newcommand{\CE}{\textsc{CE}\xspace}
\newcommand{\ResolveCE}{\textsc{Resolve CE}\xspace}
\newcommand{\StaticAllocation}{\textsc{Static-Allocation}\xspace}
\newcommand{\xmark}{\ding{55}}%

\DeclareMathOperator*{\tsum}{\sum}

\mathchardef\mhyphen="2D 
\ifdefined \argmin
\else \DeclareMathOperator*{\argmax}{arg\,min}
\fi
\ifdefined \argmax
\else \DeclareMathOperator*{\argmax}{arg\,max}
\fi
\DeclarePairedDelimiter{\norm}{\lVert}{\rVert}

\let\originalleft\left
\let\originalright\right
\renewcommand{\left}{\mathopen{}\mathclose\bgroup\originalleft}
\renewcommand{\right}{\aftergroup\egroup\originalright}

\usepackage{color-edits}
\addauthor{sb}{violet}
\addauthor{srs}{orange}
\addauthor{cy}{purple}
\addauthor{gj}{blue}

\newenvironment{rproof}[1]{ \ifdefined\informs \proof{Proof of #1.}
\else \begin{proof} \fi }{ \ifdefined\informs 
\endproof \else \end{proof} \fi  }

\newtheorem{informaltheorem}{Informal Theorem}
\usepackage[numbers]{natbib}

\usepackage{booktabs} 
\usepackage[ruled]{algorithm2e} 

\SetAlFnt{\small}
\SetAlCapFnt{\small}
\SetAlCapNameFnt{\small}
\SetAlCapHSkip{0pt}
\IncMargin{-\parindent}

\allowdisplaybreaks
\newcommand{\Halmos}{{}}
\newtheorem{theorem}{Theorem}
\numberwithin{theorem}{section}
\newtheorem{definition}[theorem]{Definition}

\newtheorem{lemma}[theorem]{Lemma}

\newtheorem{proposition}[theorem]{Proposition}

\begin{document}
	\title{Sequential Fair Allocation: Achieving the Optimal Envy-Efficiency Tradeoff Curve}
	\author{Sean R. Sinclair
	        \,\, Gauri Jain
			\,\, Siddhartha Banerjee  
			\,\, Christina Lee Yu \\
			School of Operations Research and Information Engineering \\
			Cornell University}
	\date{}
	\maketitle

	\begin{abstract}
    We consider the problem of dividing limited resources to individuals arriving over $T$ rounds. Each round has a random number of individuals arrive, and individuals can be characterized by their type (i.e. preferences over the different resources).  A standard notion of fairness in this setting is that an allocation simultaneously satisfy envy-freeness and efficiency.  The former is an individual guarantee, requiring that each agent prefers their own allocation over the allocation of any other; in contrast, efficiency is a global property, requiring that the allocations clear the available resources. For divisible resources, when the number of individuals of each type are known upfront, the above desiderata are simultaneously achievable for a large class of utility functions.  However, in an online setting when the number of individuals of each type are only revealed round by round, no policy can guarantee these desiderata simultaneously, and hence the best one can do is to try and allocate so as to approximately satisfy the two properties.

We show that in the online setting, the two desired properties (envy-freeness and efficiency) are in direct contention, in that any algorithm achieving additive {counterfactual} envy-freeness up to a factor of $L_T$ necessarily suffers an efficiency loss of at least $1 / L_T$.  We complement this uncertainty principle with a simple algorithm, \HopeGuardrail, which allocates resources based on an adaptive threshold policy and is able to achieve any fairness-efficiency point on this frontier.  Ours is the first result to provide guarantees for fair online resource allocation with high probability for multiple resource and multiple type settings.  In simulation results, our algorithm provides allocations close to the optimal fair solution in hindsight, motivating its use in practical applications as the algorithm is able to adapt to any desired fairness efficiency trade-off.
	\end{abstract}
	\newpage
	\setcounter{tocdepth}{2}
	\tableofcontents
	\newpage
	
	\section{Introduction}
\label{sec:introduction}

Our work here is motivated by a problem faced by a collaborating food-bank (Food Bank for the Southern Tier of New York (FBST)~\citep{fbst}) in operating their mobile food pantry program.  Recent demands for food assistance have climbed at an enormous rate, and an estimated fourteen million children are not getting enough food due to the COVID-19 epidemic in the United States~\citep{kulish_2020,brookings}.  With sanctions on operating in-person stores, many foodbanks have increased their mobile food pantry services.  In these systems, the mobile food pantry must decide on how much food to allocate to a distribution center on arrival without knowledge of demands in future locations.  This model also extends as a representation of broader stockpile allocation problems (such as vaccine and medical supply allocation) and reservation mechanisms.

As a simplified example (see \cref{sec:preliminary} for the full model, including multiple resources and individual types), every day, the mobile food pantry uses a truck to deliver $B$ units of food supplies to individuals over $T$ rounds (where each round can be thought of as a distribution location: soup kitchens, pantries, nursing homes, etc).  When the truck arrives at a site $t$ (or round $t$), the operator observes $N_t$ individuals and chooses how much to allocate to each individual $(X_t \in \mathbb{R}^{N_t})$ before moving to the next round.  The number of people assembling at each site changes from day to day, and the operator typically does not know the number of individuals at later sites (but has a sense of the distribution based on previous visits).

In \emph{offline} problems, where the number of individuals at each round $(N_t)_{t \in [T]}$ are known to the principal in advance, there are many well-studied notions of fair allocation of resources.  One guarantee, envy-freeness, requires that each individual prefers their own allocation over the allocation of any other.  In contrast, efficiency is a global property, requiring that the allocations clear the available resources.  For divisible resources, the above desiderata are simultaneously achievable for a large class of utility functions, with multiple resources, and is easily computed (via a convex program) by maximizing the Nash Social Welfare (NSW) objective subject to allocation constraints~\citep{varian1973equity,eisenberg1961aggregation}.  As an example, in this (simplified) setting, the fair allocation is easily computed by allocating $X^{opt} = \frac{B}{N}$ to each individual, where $N = \sum_{t \in [T]} N_t$ is the total number of individuals across all rounds.  This allocation is clearly envy-free (as each individual receives an equal allocation), and is efficient (as all of the resources are exhausted); its also easy to see that this is the \emph{only} allocation that satisfies these two properties simultaneously.

Many practical settings, however, operate more akin to the FBST mobile food pantry, where the principal makes allocation decisions \emph{online} with incomplete knowledge of the demand for future locations.  However, these principals do have access to historical data allowing them to generate histograms over the number of individuals for each round (or potentially just first moment information).  Designing good allocation algorithms in such settings necessitates harnessing the Bayesian information of future demands to ensure equitable access to the resource, while also adapting to the online realization of demands as they unfold to ensure efficiency.

Satisfying any one of these properties is trivially achievable in online settings.  The solution that allocates $X_t = 0$ to each individual satisfies hindsight envy-freeness as each individual is given an equal allocation.  The solution that allocates $X_1 = B / N_1$ to individuals at the first location and $X_t = 0$ for $t \geq 2$ satisfies efficiency as the entire budget is exhausted at the first location.  Another difficult challenge in this setting is achieving low counterfactual envy, ensuring that the allocations made by the algorithm ($X_t$) are close to what each individual \emph{should} have received with the fair solution in hindsight ($B/N$).  More meaningful is understanding how these different criteria interact.  Here we tackle these important challenges by defining meaningful notions of \emph{approximately-fair online allocations} and develop algorithms which are able to utilize distributional knowledge to achieve allocations that strike a balance between the competing objectives of envy and efficiency.


\subsection{Overview of our Contributions}
\label{section:contributions}

\ifdefined\fixcolor 
\definecolor{red}{rgb}{1, 0, 0}
\else
\fi

\ifdefined\informs

\begin{figure}
\centering     
\subfigure[\textbf{$\Denv$\,-\,$\Deff$}]{\label{fig:diagram_aa}\scalebox{.85}{
\tikzset{every picture/.style={line width=0.75pt}} 
\ifdefined\informs 

\begin{tikzpicture}[x=0.75pt,y=0.75pt,yscale=-1,xscale=1]

\draw  [draw opacity=0][fill={rgb, 255:red, 184; green, 233; blue, 134 }  ,fill opacity=1 ] (220,2.68) -- (489.4,2.68) -- (489.4,100.4) -- (220,100.4) -- cycle ;
\draw  [draw opacity=0][fill={rgb, 255:red, 240; green, 180; blue, 190 }  ,fill opacity=1 ] (220,2.68) -- (251.68,2.68) -- (251.68,260.06) -- (220,260.06) -- cycle ;
\draw  [draw opacity=0][fill={rgb, 255:red, 240; green, 180; blue, 190 }  ,fill opacity=1 ] (220,2.68) -- (488.5,2.68) -- (488.5,260.06) -- (220,260.06) -- cycle ;
\draw  [dash pattern={on 4.5pt off 4.5pt}]  (252.63,100.58) -- (252.63,260.06) ;
\draw  [draw opacity=0][fill={rgb, 255:red, 184; green, 233; blue, 134 }  ,fill opacity=1 ] (490.57,246.11) .. controls (470.61,244.32) and (449.99,240.68) .. (429.16,235.04) .. controls (339.59,210.81) and (271.88,156.67) .. (252.63,100.58) -- (464.42,104.72) -- cycle ; \draw   (490.57,246.11) .. controls (470.61,244.32) and (449.99,240.68) .. (429.16,235.04) .. controls (339.59,210.81) and (271.88,156.67) .. (252.63,100.58) ;
\draw  [draw opacity=0][fill={rgb, 255:red, 184; green, 233; blue, 134 }  ,fill opacity=1 ] (252.63,2.45) -- (489.43,2.45) -- (489.43,100.58) -- (252.63,100.58) -- cycle ;
\draw  [draw opacity=0][fill={rgb, 255:red, 184; green, 233; blue, 134 }  ,fill opacity=1 ] (461.4,37.4) -- (489.4,37.4) -- (489.4,235.6) -- (461.4,235.6) -- cycle ;
\draw  [draw opacity=0][fill={rgb, 255:red, 184; green, 233; blue, 134 }  ,fill opacity=1 ] (258.6,87.6) -- (489.6,87.6) -- (489.6,106.6) -- (258.6,106.6) -- cycle ;
\draw    (252.63,2.68) -- (252.63,100.58) ;

\draw (194.1,81) node [anchor=north west][inner sep=0.75pt]  [font=\large,rotate=-269.93] [align=left] {$\displaystyle \Deff$};
\draw (447.55,262.06) node [anchor=north west][inner sep=0.75pt]  [font=\large] [align=left] {$\displaystyle \Denv$};
\draw (256.05,230.73) node [anchor=north west][inner sep=0.75pt]  [font=\footnotesize] [align=left] {Impossible via Envy-Efficiency\\[-3ex]Uncertainty Principle (2a)};
\draw (246.65,263.46) node [anchor=north west][inner sep=0.75pt]    {$T^{-1/2}$};
\draw (327.22,77.89) node [anchor=north west][inner sep=0.75pt]   [align=left] {Achievable via \\[-3ex]\HopeGuardrail (3)};
\draw (226,252.06) node [anchor=north west][inner sep=0.75pt]  [font=\footnotesize,rotate=-270] [align=left] {Impossible via\\[-3ex]Statistical Uncertainty (1)};

\end{tikzpicture}

\else

\begin{tikzpicture}[x=0.75pt,y=0.75pt,yscale=-1,xscale=1]

\draw  [draw opacity=0][fill={rgb, 255:red, 184; green, 233; blue, 134 }  ,fill opacity=1 ] (220,2.68) -- (489.4,2.68) -- (489.4,100.4) -- (220,100.4) -- cycle ;
\draw  [draw opacity=0][fill={rgb, 255:red, 240; green, 180; blue, 190 }  ,fill opacity=1 ] (220,2.68) -- (251.68,2.68) -- (251.68,260.06) -- (220,260.06) -- cycle ;
\draw  [draw opacity=0][fill={rgb, 255:red, 240; green, 180; blue, 190 }  ,fill opacity=1 ] (220,2.68) -- (488.5,2.68) -- (488.5,260.06) -- (220,260.06) -- cycle ;
\draw  [dash pattern={on 4.5pt off 4.5pt}]  (252.63,100.58) -- (252.63,260.06) ;
\draw  [draw opacity=0][fill={rgb, 255:red, 184; green, 233; blue, 134 }  ,fill opacity=1 ] (490.57,246.11) .. controls (470.61,244.32) and (449.99,240.68) .. (429.16,235.04) .. controls (339.59,210.81) and (271.88,156.67) .. (252.63,100.58) -- (464.42,104.72) -- cycle ; \draw   (490.57,246.11) .. controls (470.61,244.32) and (449.99,240.68) .. (429.16,235.04) .. controls (339.59,210.81) and (271.88,156.67) .. (252.63,100.58) ;
\draw  [draw opacity=0][fill={rgb, 255:red, 184; green, 233; blue, 134 }  ,fill opacity=1 ] (252.63,2.45) -- (489.43,2.45) -- (489.43,100.58) -- (252.63,100.58) -- cycle ;
\draw  [draw opacity=0][fill={rgb, 255:red, 184; green, 233; blue, 134 }  ,fill opacity=1 ] (461.4,37.4) -- (489.4,37.4) -- (489.4,235.6) -- (461.4,235.6) -- cycle ;
\draw  [draw opacity=0][fill={rgb, 255:red, 184; green, 233; blue, 134 }  ,fill opacity=1 ] (258.6,87.6) -- (489.6,87.6) -- (489.6,106.6) -- (258.6,106.6) -- cycle ;
\draw    (252.63,2.68) -- (252.63,100.58) ;

\draw (194.1,81) node [anchor=north west][inner sep=0.75pt]  [font=\large,rotate=-269.93] [align=left] {$\displaystyle \Deff$};
\draw (447.55,262.06) node [anchor=north west][inner sep=0.75pt]  [font=\large] [align=left] {$\displaystyle \Denv$};
\draw (256.05,230.73) node [anchor=north west][inner sep=0.75pt]  [font=\footnotesize] [align=left] {Impossible via Envy-Efficiency\\Uncertainty Principle (2a)};
\draw (246.65,263.46) node [anchor=north west][inner sep=0.75pt]    {$T^{-1/2}$};
\draw (327.22,77.89) node [anchor=north west][inner sep=0.75pt]   [align=left] {Achievable via \\\HopeGuardrail (3)};
\draw (226,252.06) node [anchor=north west][inner sep=0.75pt]  [font=\footnotesize,rotate=-270] [align=left] {Impossible via\\Statistical Uncertainty (1)};

\end{tikzpicture}

\fi
}}
\subfigure[\textbf{$\Denvhind$\,-\,$\Deff$}]{\label{fig:diagram_abb}\scalebox{.85}{
\tikzset{every picture/.style={line width=0.75pt}} 
\ifdefined\informs 

\tikzset{every picture/.style={line width=0.75pt}} 

\begin{tikzpicture}[x=0.75pt,y=0.75pt,yscale=-1,xscale=1]

\draw  [draw opacity=0][fill={rgb, 255:red, 240; green, 180; blue, 190 }  ,fill opacity=1 ] (220,2.68) -- (488.5,2.68) -- (488.5,260.06) -- (220,260.06) -- cycle ;
\draw  [draw opacity=0][fill={rgb, 255:red, 184; green, 233; blue, 134 }  ,fill opacity=1 ] (251.1,76.69) -- (302.4,120.73) -- (279.29,147.65) -- (227.99,103.61) -- cycle ;
\draw  [draw opacity=0][fill={rgb, 255:red, 184; green, 233; blue, 134 }  ,fill opacity=1 ] (244.22,71.43) -- (318.33,131.92) -- (295.35,160.07) -- (221.24,99.58) -- cycle ;
\draw  [draw opacity=0][fill={rgb, 255:red, 184; green, 233; blue, 134 }  ,fill opacity=1 ] (220,2.68) -- (489.4,2.68) -- (489.4,100.4) -- (220,100.4) -- cycle ;
\draw  [dash pattern={on 4.5pt off 4.5pt}]  (252.63,100.58) -- (252.63,260.25) ;
\draw    (221.24,100.58) -- (304.13,168) ;
\draw  [draw opacity=0][fill={rgb, 255:red, 184; green, 233; blue, 134 }  ,fill opacity=1 ] (461.95,37.4) -- (488.4,37.4) -- (488.4,226) -- (461.95,226) -- cycle ;
\draw  [draw opacity=0][fill={rgb, 255:red, 184; green, 233; blue, 134 }  ,fill opacity=1 ] (254.67,90.25) -- (489.6,90.25) -- (489.6,106.6) -- (254.67,106.6) -- cycle ;
\draw  [draw opacity=0][fill={rgb, 255:red, 184; green, 233; blue, 134 }  ,fill opacity=1 ] (490.13,246.07) .. controls (470.31,244.27) and (449.83,240.64) .. (429.16,235.04) .. controls (339.59,210.81) and (271.88,156.67) .. (252.63,100.58) -- (464.42,104.72) -- cycle ; \draw   (490.13,246.07) .. controls (470.31,244.27) and (449.83,240.64) .. (429.16,235.04) .. controls (339.59,210.81) and (271.88,156.67) .. (252.63,100.58) ;
\draw    (221.24,100.58) -- (252.63,100.58) ;
\draw  [draw opacity=0][fill={rgb, 255:red, 184; green, 233; blue, 134 }  ,fill opacity=1 ] (482.95,220) -- (488.95,220) -- (488.95,245) -- (482.95,245) -- cycle ;

\draw (194.1,81) node [anchor=north west][inner sep=0.75pt]  [font=\large,rotate=-269.93] [align=left] {$\displaystyle \Deff $};
\draw (221.05,232.73) node [anchor=north west][inner sep=0.75pt]  [font=\footnotesize] [align=left] {Impossible via Envy-Efficiency \\[-3ex] Uncertainty Principle (2b)};
\draw (246.65,263.46) node [anchor=north west][inner sep=0.75pt]    {$T^{-1/2}$};
\draw (179,92.4) node [anchor=north west][inner sep=0.75pt]  [font=\tiny]  {$\left( 0,\ \sqrt{T}\right)$};
\draw (288,121.4) node [anchor=north west][inner sep=0.75pt]  [font=\tiny]  {$\left(\frac{2}{\sqrt{T}} ,\ \frac{\sqrt{T}}{2}\right)$};
\draw (329.22,84.89) node [anchor=north west][inner sep=0.75pt]   [align=left] {Achievable via \\[-3ex] \HopeGuardrail (3)};
\draw (443.55,263) node [anchor=north west][inner sep=0.75pt]  [font=\large] [align=left] {$\Denvhind$};

\end{tikzpicture}

\else

\tikzset{every picture/.style={line width=0.75pt}} 

\begin{tikzpicture}[x=0.75pt,y=0.75pt,yscale=-1,xscale=1]

\draw  [draw opacity=0][fill={rgb, 255:red, 240; green, 180; blue, 190 }  ,fill opacity=1 ] (220,2.68) -- (488.5,2.68) -- (488.5,260.06) -- (220,260.06) -- cycle ;
\draw  [draw opacity=0][fill={rgb, 255:red, 184; green, 233; blue, 134 }  ,fill opacity=1 ] (251.1,76.69) -- (302.4,120.73) -- (279.29,147.65) -- (227.99,103.61) -- cycle ;
\draw  [draw opacity=0][fill={rgb, 255:red, 184; green, 233; blue, 134 }  ,fill opacity=1 ] (244.22,71.43) -- (318.33,131.92) -- (295.35,160.07) -- (221.24,99.58) -- cycle ;
\draw  [draw opacity=0][fill={rgb, 255:red, 184; green, 233; blue, 134 }  ,fill opacity=1 ] (220,2.68) -- (489.4,2.68) -- (489.4,100.4) -- (220,100.4) -- cycle ;
\draw  [dash pattern={on 4.5pt off 4.5pt}]  (252.63,100.58) -- (252.63,260.25) ;
\draw    (221.24,100.58) -- (304.13,168) ;
\draw  [draw opacity=0][fill={rgb, 255:red, 184; green, 233; blue, 134 }  ,fill opacity=1 ] (461.95,37.4) -- (488.4,37.4) -- (488.4,226) -- (461.95,226) -- cycle ;
\draw  [draw opacity=0][fill={rgb, 255:red, 184; green, 233; blue, 134 }  ,fill opacity=1 ] (254.67,90.25) -- (489.6,90.25) -- (489.6,106.6) -- (254.67,106.6) -- cycle ;
\draw  [draw opacity=0][fill={rgb, 255:red, 184; green, 233; blue, 134 }  ,fill opacity=1 ] (490.13,246.07) .. controls (470.31,244.27) and (449.83,240.64) .. (429.16,235.04) .. controls (339.59,210.81) and (271.88,156.67) .. (252.63,100.58) -- (464.42,104.72) -- cycle ; \draw   (490.13,246.07) .. controls (470.31,244.27) and (449.83,240.64) .. (429.16,235.04) .. controls (339.59,210.81) and (271.88,156.67) .. (252.63,100.58) ;
\draw    (221.24,100.58) -- (252.63,100.58) ;
\draw  [draw opacity=0][fill={rgb, 255:red, 184; green, 233; blue, 134 }  ,fill opacity=1 ] (482.95,220) -- (488.95,220) -- (488.95,245) -- (482.95,245) -- cycle ;

\draw (194.1,81) node [anchor=north west][inner sep=0.75pt]  [font=\large,rotate=-269.93] [align=left] {$\displaystyle \Deff $};
\draw (221.05,232.73) node [anchor=north west][inner sep=0.75pt]  [font=\footnotesize] [align=left] {Impossible via Envy-Efficiency \\ Uncertainty Principle (2b)};
\draw (246.65,263.46) node [anchor=north west][inner sep=0.75pt]    {$T^{-1/2}$};
\draw (179,92.4) node [anchor=north west][inner sep=0.75pt]  [font=\tiny]  {$\left( 0,\ \sqrt{T}\right)$};
\draw (288,121.4) node [anchor=north west][inner sep=0.75pt]  [font=\tiny]  {$\left(\frac{2}{\sqrt{T}} ,\ \frac{\sqrt{T}}{2}\right)$};
\draw (329.22,84.89) node [anchor=north west][inner sep=0.75pt]   [align=left] {Achievable via \\ \HopeGuardrail (3)};
\draw (443.55,263) node [anchor=north west][inner sep=0.75pt]  [font=\large] [align=left] {$\Denvhind$};

\end{tikzpicture}

\fi
}}
\caption{Graphical representation of the major contributions (\cref{thm:informal}).  Here, the $x$-axis denotes $\Denv$ or $\Denvhind$, and the $y$-axis denotes $\Deff$, the remaining resources.  The dotted line represents the impossibility due to statistical uncertainty in the optimal allocation, and the region below the solid line represents the impossibility due to the envy-efficiency uncertainty principle.}
\label{fig:uncertainty_principle}
\end{figure}
\else 

\begin{figure}
\centering     
\subfigure[\textbf{$\Denv$\,-\,$\Deff$}]{\label{fig:diagram_aa}\scalebox{.85}{
\tikzset{every picture/.style={line width=0.75pt}} 
\ifdefined\informs 

\begin{tikzpicture}[x=0.75pt,y=0.75pt,yscale=-1,xscale=1]

\draw  [draw opacity=0][fill={rgb, 255:red, 184; green, 233; blue, 134 }  ,fill opacity=1 ] (220,2.68) -- (489.4,2.68) -- (489.4,100.4) -- (220,100.4) -- cycle ;
\draw  [draw opacity=0][fill={rgb, 255:red, 240; green, 180; blue, 190 }  ,fill opacity=1 ] (220,2.68) -- (251.68,2.68) -- (251.68,260.06) -- (220,260.06) -- cycle ;
\draw  [draw opacity=0][fill={rgb, 255:red, 240; green, 180; blue, 190 }  ,fill opacity=1 ] (220,2.68) -- (488.5,2.68) -- (488.5,260.06) -- (220,260.06) -- cycle ;
\draw  [dash pattern={on 4.5pt off 4.5pt}]  (252.63,100.58) -- (252.63,260.06) ;
\draw  [draw opacity=0][fill={rgb, 255:red, 184; green, 233; blue, 134 }  ,fill opacity=1 ] (490.57,246.11) .. controls (470.61,244.32) and (449.99,240.68) .. (429.16,235.04) .. controls (339.59,210.81) and (271.88,156.67) .. (252.63,100.58) -- (464.42,104.72) -- cycle ; \draw   (490.57,246.11) .. controls (470.61,244.32) and (449.99,240.68) .. (429.16,235.04) .. controls (339.59,210.81) and (271.88,156.67) .. (252.63,100.58) ;
\draw  [draw opacity=0][fill={rgb, 255:red, 184; green, 233; blue, 134 }  ,fill opacity=1 ] (252.63,2.45) -- (489.43,2.45) -- (489.43,100.58) -- (252.63,100.58) -- cycle ;
\draw  [draw opacity=0][fill={rgb, 255:red, 184; green, 233; blue, 134 }  ,fill opacity=1 ] (461.4,37.4) -- (489.4,37.4) -- (489.4,235.6) -- (461.4,235.6) -- cycle ;
\draw  [draw opacity=0][fill={rgb, 255:red, 184; green, 233; blue, 134 }  ,fill opacity=1 ] (258.6,87.6) -- (489.6,87.6) -- (489.6,106.6) -- (258.6,106.6) -- cycle ;
\draw    (252.63,2.68) -- (252.63,100.58) ;

\draw (194.1,81) node [anchor=north west][inner sep=0.75pt]  [font=\large,rotate=-269.93] [align=left] {$\displaystyle \Deff$};
\draw (447.55,262.06) node [anchor=north west][inner sep=0.75pt]  [font=\large] [align=left] {$\displaystyle \Denv$};
\draw (256.05,230.73) node [anchor=north west][inner sep=0.75pt]  [font=\footnotesize] [align=left] {Impossible via Envy-Efficiency\\[-3ex]Uncertainty Principle (2a)};
\draw (246.65,263.46) node [anchor=north west][inner sep=0.75pt]    {$T^{-1/2}$};
\draw (327.22,77.89) node [anchor=north west][inner sep=0.75pt]   [align=left] {Achievable via \\[-3ex]\HopeGuardrail (3)};
\draw (226,252.06) node [anchor=north west][inner sep=0.75pt]  [font=\footnotesize,rotate=-270] [align=left] {Impossible via\\[-3ex]Statistical Uncertainty (1)};

\end{tikzpicture}

\else

\begin{tikzpicture}[x=0.75pt,y=0.75pt,yscale=-1,xscale=1]

\draw  [draw opacity=0][fill={rgb, 255:red, 184; green, 233; blue, 134 }  ,fill opacity=1 ] (220,2.68) -- (489.4,2.68) -- (489.4,100.4) -- (220,100.4) -- cycle ;
\draw  [draw opacity=0][fill={rgb, 255:red, 240; green, 180; blue, 190 }  ,fill opacity=1 ] (220,2.68) -- (251.68,2.68) -- (251.68,260.06) -- (220,260.06) -- cycle ;
\draw  [draw opacity=0][fill={rgb, 255:red, 240; green, 180; blue, 190 }  ,fill opacity=1 ] (220,2.68) -- (488.5,2.68) -- (488.5,260.06) -- (220,260.06) -- cycle ;
\draw  [dash pattern={on 4.5pt off 4.5pt}]  (252.63,100.58) -- (252.63,260.06) ;
\draw  [draw opacity=0][fill={rgb, 255:red, 184; green, 233; blue, 134 }  ,fill opacity=1 ] (490.57,246.11) .. controls (470.61,244.32) and (449.99,240.68) .. (429.16,235.04) .. controls (339.59,210.81) and (271.88,156.67) .. (252.63,100.58) -- (464.42,104.72) -- cycle ; \draw   (490.57,246.11) .. controls (470.61,244.32) and (449.99,240.68) .. (429.16,235.04) .. controls (339.59,210.81) and (271.88,156.67) .. (252.63,100.58) ;
\draw  [draw opacity=0][fill={rgb, 255:red, 184; green, 233; blue, 134 }  ,fill opacity=1 ] (252.63,2.45) -- (489.43,2.45) -- (489.43,100.58) -- (252.63,100.58) -- cycle ;
\draw  [draw opacity=0][fill={rgb, 255:red, 184; green, 233; blue, 134 }  ,fill opacity=1 ] (461.4,37.4) -- (489.4,37.4) -- (489.4,235.6) -- (461.4,235.6) -- cycle ;
\draw  [draw opacity=0][fill={rgb, 255:red, 184; green, 233; blue, 134 }  ,fill opacity=1 ] (258.6,87.6) -- (489.6,87.6) -- (489.6,106.6) -- (258.6,106.6) -- cycle ;
\draw    (252.63,2.68) -- (252.63,100.58) ;

\draw (194.1,81) node [anchor=north west][inner sep=0.75pt]  [font=\large,rotate=-269.93] [align=left] {$\displaystyle \Deff$};
\draw (447.55,262.06) node [anchor=north west][inner sep=0.75pt]  [font=\large] [align=left] {$\displaystyle \Denv$};
\draw (256.05,230.73) node [anchor=north west][inner sep=0.75pt]  [font=\footnotesize] [align=left] {Impossible via Envy-Efficiency\\Uncertainty Principle (2a)};
\draw (246.65,263.46) node [anchor=north west][inner sep=0.75pt]    {$T^{-1/2}$};
\draw (327.22,77.89) node [anchor=north west][inner sep=0.75pt]   [align=left] {Achievable via \\\HopeGuardrail (3)};
\draw (226,252.06) node [anchor=north west][inner sep=0.75pt]  [font=\footnotesize,rotate=-270] [align=left] {Impossible via\\Statistical Uncertainty (1)};

\end{tikzpicture}

\fi
}}
\subfigure[\textbf{$\Denvhind$\,-\,$\Deff$}]{\label{fig:diagram_abb}\scalebox{.85}{
\tikzset{every picture/.style={line width=0.75pt}} 
\ifdefined\informs 

\tikzset{every picture/.style={line width=0.75pt}} 

\begin{tikzpicture}[x=0.75pt,y=0.75pt,yscale=-1,xscale=1]

\draw  [draw opacity=0][fill={rgb, 255:red, 240; green, 180; blue, 190 }  ,fill opacity=1 ] (220,2.68) -- (488.5,2.68) -- (488.5,260.06) -- (220,260.06) -- cycle ;
\draw  [draw opacity=0][fill={rgb, 255:red, 184; green, 233; blue, 134 }  ,fill opacity=1 ] (251.1,76.69) -- (302.4,120.73) -- (279.29,147.65) -- (227.99,103.61) -- cycle ;
\draw  [draw opacity=0][fill={rgb, 255:red, 184; green, 233; blue, 134 }  ,fill opacity=1 ] (244.22,71.43) -- (318.33,131.92) -- (295.35,160.07) -- (221.24,99.58) -- cycle ;
\draw  [draw opacity=0][fill={rgb, 255:red, 184; green, 233; blue, 134 }  ,fill opacity=1 ] (220,2.68) -- (489.4,2.68) -- (489.4,100.4) -- (220,100.4) -- cycle ;
\draw  [dash pattern={on 4.5pt off 4.5pt}]  (252.63,100.58) -- (252.63,260.25) ;
\draw    (221.24,100.58) -- (304.13,168) ;
\draw  [draw opacity=0][fill={rgb, 255:red, 184; green, 233; blue, 134 }  ,fill opacity=1 ] (461.95,37.4) -- (488.4,37.4) -- (488.4,226) -- (461.95,226) -- cycle ;
\draw  [draw opacity=0][fill={rgb, 255:red, 184; green, 233; blue, 134 }  ,fill opacity=1 ] (254.67,90.25) -- (489.6,90.25) -- (489.6,106.6) -- (254.67,106.6) -- cycle ;
\draw  [draw opacity=0][fill={rgb, 255:red, 184; green, 233; blue, 134 }  ,fill opacity=1 ] (490.13,246.07) .. controls (470.31,244.27) and (449.83,240.64) .. (429.16,235.04) .. controls (339.59,210.81) and (271.88,156.67) .. (252.63,100.58) -- (464.42,104.72) -- cycle ; \draw   (490.13,246.07) .. controls (470.31,244.27) and (449.83,240.64) .. (429.16,235.04) .. controls (339.59,210.81) and (271.88,156.67) .. (252.63,100.58) ;
\draw    (221.24,100.58) -- (252.63,100.58) ;
\draw  [draw opacity=0][fill={rgb, 255:red, 184; green, 233; blue, 134 }  ,fill opacity=1 ] (482.95,220) -- (488.95,220) -- (488.95,245) -- (482.95,245) -- cycle ;

\draw (194.1,81) node [anchor=north west][inner sep=0.75pt]  [font=\large,rotate=-269.93] [align=left] {$\displaystyle \Deff $};
\draw (221.05,232.73) node [anchor=north west][inner sep=0.75pt]  [font=\footnotesize] [align=left] {Impossible via Envy-Efficiency \\[-3ex] Uncertainty Principle (2b)};
\draw (246.65,263.46) node [anchor=north west][inner sep=0.75pt]    {$T^{-1/2}$};
\draw (179,92.4) node [anchor=north west][inner sep=0.75pt]  [font=\tiny]  {$\left( 0,\ \sqrt{T}\right)$};
\draw (288,121.4) node [anchor=north west][inner sep=0.75pt]  [font=\tiny]  {$\left(\frac{2}{\sqrt{T}} ,\ \frac{\sqrt{T}}{2}\right)$};
\draw (329.22,84.89) node [anchor=north west][inner sep=0.75pt]   [align=left] {Achievable via \\[-3ex] \HopeGuardrail (3)};
\draw (443.55,263) node [anchor=north west][inner sep=0.75pt]  [font=\large] [align=left] {$\Denvhind$};

\end{tikzpicture}

\else

\tikzset{every picture/.style={line width=0.75pt}} 

\begin{tikzpicture}[x=0.75pt,y=0.75pt,yscale=-1,xscale=1]

\draw  [draw opacity=0][fill={rgb, 255:red, 240; green, 180; blue, 190 }  ,fill opacity=1 ] (220,2.68) -- (488.5,2.68) -- (488.5,260.06) -- (220,260.06) -- cycle ;
\draw  [draw opacity=0][fill={rgb, 255:red, 184; green, 233; blue, 134 }  ,fill opacity=1 ] (251.1,76.69) -- (302.4,120.73) -- (279.29,147.65) -- (227.99,103.61) -- cycle ;
\draw  [draw opacity=0][fill={rgb, 255:red, 184; green, 233; blue, 134 }  ,fill opacity=1 ] (244.22,71.43) -- (318.33,131.92) -- (295.35,160.07) -- (221.24,99.58) -- cycle ;
\draw  [draw opacity=0][fill={rgb, 255:red, 184; green, 233; blue, 134 }  ,fill opacity=1 ] (220,2.68) -- (489.4,2.68) -- (489.4,100.4) -- (220,100.4) -- cycle ;
\draw  [dash pattern={on 4.5pt off 4.5pt}]  (252.63,100.58) -- (252.63,260.25) ;
\draw    (221.24,100.58) -- (304.13,168) ;
\draw  [draw opacity=0][fill={rgb, 255:red, 184; green, 233; blue, 134 }  ,fill opacity=1 ] (461.95,37.4) -- (488.4,37.4) -- (488.4,226) -- (461.95,226) -- cycle ;
\draw  [draw opacity=0][fill={rgb, 255:red, 184; green, 233; blue, 134 }  ,fill opacity=1 ] (254.67,90.25) -- (489.6,90.25) -- (489.6,106.6) -- (254.67,106.6) -- cycle ;
\draw  [draw opacity=0][fill={rgb, 255:red, 184; green, 233; blue, 134 }  ,fill opacity=1 ] (490.13,246.07) .. controls (470.31,244.27) and (449.83,240.64) .. (429.16,235.04) .. controls (339.59,210.81) and (271.88,156.67) .. (252.63,100.58) -- (464.42,104.72) -- cycle ; \draw   (490.13,246.07) .. controls (470.31,244.27) and (449.83,240.64) .. (429.16,235.04) .. controls (339.59,210.81) and (271.88,156.67) .. (252.63,100.58) ;
\draw    (221.24,100.58) -- (252.63,100.58) ;
\draw  [draw opacity=0][fill={rgb, 255:red, 184; green, 233; blue, 134 }  ,fill opacity=1 ] (482.95,220) -- (488.95,220) -- (488.95,245) -- (482.95,245) -- cycle ;

\draw (194.1,81) node [anchor=north west][inner sep=0.75pt]  [font=\large,rotate=-269.93] [align=left] {$\displaystyle \Deff $};
\draw (221.05,232.73) node [anchor=north west][inner sep=0.75pt]  [font=\footnotesize] [align=left] {Impossible via Envy-Efficiency \\ Uncertainty Principle (2b)};
\draw (246.65,263.46) node [anchor=north west][inner sep=0.75pt]    {$T^{-1/2}$};
\draw (179,92.4) node [anchor=north west][inner sep=0.75pt]  [font=\tiny]  {$\left( 0,\ \sqrt{T}\right)$};
\draw (288,121.4) node [anchor=north west][inner sep=0.75pt]  [font=\tiny]  {$\left(\frac{2}{\sqrt{T}} ,\ \frac{\sqrt{T}}{2}\right)$};
\draw (329.22,84.89) node [anchor=north west][inner sep=0.75pt]   [align=left] {Achievable via \\ \HopeGuardrail (3)};
\draw (443.55,263) node [anchor=north west][inner sep=0.75pt]  [font=\large] [align=left] {$\Denvhind$};

\end{tikzpicture}

\fi
}}
\caption{Graphical representation of the major contribution (\cref{thm:informal}).  Here, the $x$-axis denotes $\Denv$ (the maximum difference between utility individuals receive from the algorithm and the fair allocation in hindsight) or $\Denvhind$ (the maximum difference between utility individuals receive from the algorithm), and the $y$-axis denotes $\Deff$, the remaining resources.  The dotted line represents the impossibility due to statistical uncertainty in the optimal allocation, and the region below the solid line represents the impossibility due to the envy-efficiency uncertainty principle.  In the trade-off between $\Denvhind$ and $\Deff$ we have an additional discrepancy between the lower and upper bounds (represented in yellow) which reduces via $O(\frac{1}{T})$.}
\label{fig:uncertainty_principle}
\end{figure}
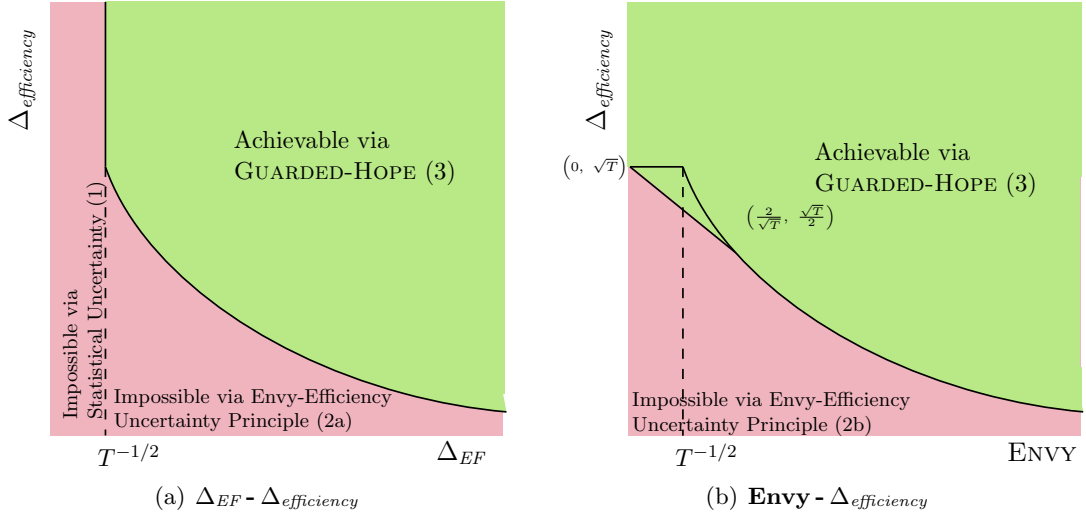

\fi

\ifdefined\fixcolor 
\definecolor{red}{rgb}{0, 0, 0}
\else
\fi


In sequential settings, one way to measure the \emph{(un) fairness} of any \emph{online} allocation ($X^{alg}$) is in terms of its counterfactual distance (for both envy and efficiency) when compared to the \emph{optimal fair allocation in hindsight} (i.e., \emph{offline} allocation $X^{opt}$).  Another measure is hindsight envy (when compared only to allocations made by the algorithm).  In particular, we define the \emph{counterfactual envy} as $\Denv = \norm{u(X^{opt}_\theta, \theta) - u(X^{alg}_\theta, \theta)}_{\infty}$ to be the maximum difference in utility between the algorithm's allocation and the offline allocation where agents are characterized by their type $\theta$, define the \emph{hindsight envy} as $\Denvhind = \max_{t,t',\theta,\theta'} u(X_{t', \theta'}^{alg}, \theta) - u(X_{t, \theta}^{alg}, \theta)$ to be the maximum difference between the utility individuals would have received if given someone else's allocations, and let $\Deff = B - \sum_{t, \theta} N_{t,\theta} X_{t,\theta}^{alg}$ be the algorithm's total leftover resources. These are all very stringent metrics, akin to the notion of regret in online decision-making settings.  

In these settings with competing objectives, practitioners often resort to ad-hoc rules of thumb, heuristics, and trial-and-error adjustments of the system to attempt to manage the balance between objectives.  How these criteria interact and trade-off amongst one another is often not well understood or characterized, and furthermore there typically does not exist a single best “ranking”, or a clear single objective function that determines which tradeoffs are better than others.  In fact, minimizing some combination of $(\Denv, \Denvhind, \Deff)$ can be formulated as a Markov decision process (MDP).  However, as these metrics depend on the entire allocation, the complexity of finding the optimal policy is exponential in the number of rounds and may be difficult to interpret~\citep{manshadi2021fair}. Moreover, it is much harder to use MDP formulations to explore the tradeoff between the objectives.

Our main technical contribution is to provide a complete characterization of the achievable pairs of $(\Denv,\Denvhind,\Deff)$.  Our results hold in expectation and with high probability, under multiple divisible resources, and with a finite set of individual types with linear utilities.  In particular, we show the following informal theorem (see \cref{fig:uncertainty_principle} for a graphical representation).

\begin{informaltheorem}[See~\cref{thm:lower_bound,thm:envy-efficiency,thm:hindsight_envy-efficiency,thm:upper_bound} for full versions]
\label{thm:informal}
Under mild regularity conditions on the distribution of $N_t$, we have the following (where $\gtrsim$ ignores problem dependent constants, logarithmic factors of $T$, and $o(1)$ factors):
\begin{enumerate}
\item[1.] \emph{(Statistical Uncertainty Principle)}: Any online allocation algorithm must suffer counterfactual envy of at least $\Denv \gtrsim \frac{1}{\sqrt{T}}$.
\item[2a.] \emph{(Counterfactual Envy-Efficiency Uncertainty Principle)}: Any {online allocation} algorithm necessarily suffers $\Deff \gtrsim \min\{\sqrt{T}, 1 / \Denv\}.$
\item[2b.] \emph{(Hindsight Envy-Efficiency Uncertainty Principle)}: Any online allocation algorithm necessarily suffers $\Deff \gtrsim \min\{\sqrt{T}, 1 / \Denvhind \}$.
\item[3] \emph{(Upper Bound via \HopeGuardrail)}: For any choice of $L_T$, with probability at least $1-\delta$, \HopeGuardrail with parameter $L_T$ achieves:
    \begin{align*}
    \Denvhind \leq L_T \quad\quad \Denv \lesssim \max\{1 / \sqrt{T}, L_T\} \quad\quad \Deff \lesssim \min\{\sqrt{T}, 1 / L_T\}.
    \end{align*}
\end{enumerate}
\end{informaltheorem}

In short, our results show that envy and waste must be inversely proportional to one another such that decreasing envy requires increasing waste and vice versa.  The lower bounds $(1$ and $2)$ are established using anti-concentration arguments alongside understanding the fundamental gap in ensuring enough resources to allocate close to the estimated optimal solution while simultaneously trying to eliminate waste. 

Furthermore, we provide a simple algorithm, \HopeGuardrail, which achieves the correct trade-off between envy and waste, matching the lower bound in terms of $T$ up to logarithmic factors. Given an input of $L_T$, our algorithm satisfies a hindsight envy bound of $\Denvhind \lesssim L_T$, counterfactual envy bound of $\Denv \lesssim \max\{1 / \sqrt{T}, L_T\}$, with waste bounded by ${\Deff} \lesssim \max\{\sqrt{T}, 1 / L_T\}$.   Our algorithm achieves this using novel concentration arguments on the optimal Nash Social Welfare solution, utilizing a sensitivity argument on the \emph{solution} to the optimization problem instead of the objective (as commonly used for competitive ratio guarantees) to learn a lower guardrail on the optimal solution in hindsight.  Given this, we construct an upper guardrail to satisfy the desired $\Denv$ and $\Denvhind$ bound.  We then achieve the proper trade-off by carefully balancing allocating the established lower guardrail with the upper guardrail while simultaneously ensuring the algorithm never runs out of budget.

To get some intuition into the envy-efficiency uncertainty principle, consider the simple food bank example described above for a single resource (with arrivals $N_t$ in each location, and $X^{opt} = B/N$ where $N = \sum_{t \in [T]} N_t$). For convenience we temporarily assume that each agents utility is directly proportional to their allocation (i.e. $u(X, \theta) = X$).  Consider allocation $X_1$ at the first location: via standard concentration arguments, one can find a high probability lower confidence bound for $B/N$ with a half-width on the order of $1/\sqrt{T}$. Now it's not hard to argue that allocating according to the lower confidence bound at \emph{all} locations achieves counterfactual envy of $\Denv \approx 1/\sqrt{T}$, $\Denvhind = 0$, and $\Deff \approx {\sqrt{T}}$. This corresponds to the cusp of the efficiency-envy trade-off curves in~\cref{fig:uncertainty_principle}. 

Now if we relax the $\Denv$ or $\Denvhind$ constraint to $ \approx 1/T^{1/3}$ and use the naive static policy of always allocating via the now looser lower confidence bound, we get a waste of $T\cdot T^{-1/3} = T^{2/3}$. Our algorithm instead takes a different approach, using the lower confidence bound of order $1 / \sqrt{T}$ as the lower guardrail allocation, and sets the upper guardrail allocation to be the lower one plus the desired bound on $\Denv$ or $\Denvhind$.  If we were to establish that the algorithm always allocates within the guardrails, we automatically have the desired bound on $\Denv$ and $\Denvhind$.  The main additional factor in achieving the tradeoff for $\Deff$ is ensuring we properly allocate according to the upper threshold while ensuring we do not run out of budget to ensure the lower threshold allocation.  With this \HopeGuardrail achieves $\Deff \approx T^{1/3}$, which furthermore is the best possible.  Moreover, we complement our theoretical results with experiments highlighting the empirical performance of different algorithms (both on synthetic settings, as well as a dataset based on mobile food pantry operations), which shows that \HopeGuardrail has much lower waste and envy compared to static underallocation, as well as other certainty equivalence based heuristics~\cite{bertsekas2012dynamic}.

While fairness in resource allocation is well-studied in offline and adversarial settings, fairness metrics for the sequential stochastic setting are poorly understood (especially when individuals are arriving online).  {Our proposed metrics and results give a novel way of extending Varian's definitions of fairness to the sequential setting.  Moreover, ours is the first result to provide guarantees for fair online resource allocation with high probability for multiple resource and multiple type settings.} 
Most existing work aims to show competitive ratio or additive guarantees on the Nash social welfare objective~\citep{gorokh2020fair} or focus on the max-min objective~\citep{lien2014sequential,manshadi2021fair}. Such guarantees are dangerously misleading in that the resultant allocations may exhibit clear unfairness in hindsight.  Similarly, an \emph{ex ante} or probabilistic guarantee may also be perceived as unfair -- both allocating $1$ unit with certainty, and allocating $10$ units with probability $1/10$, give the same ex ante guarantee.
In contrast, our chosen metrics and theoretical results provides a firm basis for \emph{counterfactual and ex-post individual fairness guarantees}. While we do not believe our work gives a final answer in the theoretical and practical understanding of fairness in online allocation, we hope it will add to the conversation of incorporating ethics into sequential AI algorithms.  More discussion on the advantages and disadvantages of our proposed fairness metrics is in \cref{app:varian}.

\subsection{Other Motivating Examples}
\label{section: examples}

In addition to the mobile food pantry allocation problem which forms the focus of our work, we believe our ideas can prove useful in several other settings:\\
\noindent \textbf{Stockpile Allocation.}  In many healthcare systems or resource allocation problems, government mechanisms decide how to allocate critical resources to states, individuals, or hospitals.  For example, the US federal government was tasked with distributing Remdesivir, an antiviral drug used early in the panedemic for COVID-19 treatment~\citep{lupkin_2020}.  More recently relevant, states and government organizations have been deciding how to allocate COVID-19 (or influenza) vaccines to various population demographics across several rounds~\citep{yi2015fairness,jaberi2014optimal}.  In these scenarios on a monthly basis, each state is given a fixed amount of the resource (say COVID-19 vaccinations) and is tasked with distributing these to individuals across various distribution locations.  While the primary goal is to develop efficient allocations, an alternative objective may be to ensure equitable access to the resource~\citep{shadmi2020health, donahue2020fairness, manshadi2021fair}.    


\noindent \textbf{Reservation Mechanisms.}
These are key for operating shared high-performance computing (HPC) systems~\citep{ghodsi2011dominant}. Cluster centers for HPC receive numerous requests online with varying demands for CPUs and GPUs. Algorithms must allocate resources to incoming jobs with only distributional knowledge of future resource demands. Important to these settings is the large number of resources (number of GPUs, RAM, etc available at the center), requiring algorithms that scale to higher-dimensional problems.


    \section{Related Work}
\label{sec:related_work}

Fairness in resource allocation and the use of Nash Social Welfare was pioneered by Varian in his seminal work~\citep{varian1973equity,varian1976two}. Since then researchers have investigated fairness properties for both offline and online allocation, in settings with divisible or indivisible resources, and when either the individuals or resources arrive online. We now briefly discuss some related works; see~\cite{aleksandrov2019online} for a comprehensive survey. What distinguishes our setting from many of the previous works is that we consider the online Bayesian setting with a known distribution. Many previous works are either limited to offline or non-adaptive algorithms, or consider adversarial online arrivals.  {Trade-offs between various fairness metrics has alsos been considered previously in the literature, but for classification-based fairness metrics on protected attributes instead of allocation based ones~\citep{kleinberg2016inherent}.}



\noindent \textbf{Food Bank and Health Care Operations}: There is a growing body of work in the operations research literature addressing logistics and supply chain issues in the area of humanitarian relief, health-care, and food distribution~\citep{sengul2017modeling,orgut2016achieving,alkaabneh2020unified,yi2015fairness,jaberi2014optimal}.  The research focuses on designing systems which balance efficiency, effectiveness, and equity.  In \cite{eisenhandler2019humanitarian} they study the logistical challenges of managing vehicles with limited capacity to distribute food and provide routing and scheduling protocols.  In \cite{lien2014sequential,manshadi2021fair} they consider sequential allocation with an alternative objective of maximizing the minimum utility (also called the leximin in the literature \citep{moulin2004fair}).  We instead consider sequential allocation of resources under the objectives of achieving approximate fairness notions with regards to envy and efficiency.  


\noindent \textbf{Cake Cutting}: Cake cutting serves as a model for dividing a continuous object (whether that be a cake, advertisement space, land, etc)~\citep{brams1995envy,procaccia2013cake}.  Under this model, prior work considers situations where individuals arrive and depart during the process of dividing a resource, where the utility of an agent is a set-function on the interval of the resource received.  Researchers analyze the offline setting to develop algorithms to allocate the resource with a minimal number of cuts \citep{brams1996fair}, or online under adversarial arrivals~\citep{walsh2011online}.  Our model instead imposes stochastic assumptions on the number of arriving individuals and characterizes probabilistic instead of sample-path fairness criteria.


\noindent \textbf{Online Resources}: One line of work considers the resource (here to be thought of as the units of food, processing power, etc) are online and the agents are fixed~\citep{benade2018make,aleksandrov2015online,mattei2018fairness,mattei2017mechanisms,aleksandrov2019monotone,gorokh2020fair,bansal2020online,bogomolnaia2021fair}.  In \cite{zeng2019fairness} they study the tradeoffs between fairness and efficiency when items arrive under several adversarial models.  Another common criteria is designing algorithms which are \textit{envy-free up to one item}, where researchers design algorithms that can reallocate previously allocated items, but try to minimize these adjustments~\citep{ijcai2019-49,aziz2016control}.  These problems are in contrast to our model where instead the resources are fixed and depleting over time, and individuals arrive online.


\noindent \textbf{Online Individuals}: The other setting more similar to our work considers agents as arriving online and the resources as fixed.  In \cite{kalinowski2013social} they consider this setting where the resources are indivisible with the goal of maximizing utilitarian welfare (or the sum of utilities) which provides no guarantees on individual fairness.  Another approach in \cite{gerding2019fair} considers a scheduling setting where agents arrive and depart online.  Each agent has a fixed and known arrival time, departure time, and demand.  The goal then is to determine a schedule and allocation which is Pareto-efficient and envy-free.  {Another line of work~\citep{friedman_2017,cole_2013,10.1145/2764468.2764495} considers fair division with minimal disruptions on previous allocations.  Their \emph{fairness ratio} can be viewed as a competitive ratio form of our \emph{counterfactual envy} definition (\cref{def:distance}).}


\noindent \textbf{Non-adaptive Allocations}:  A separate line of research considers fairness questions for resource allocations in a similar setting where the utilities across groups are drawn from known probability distributions \citep{donahue2020fairness,elzayn2019fair}.  They investigate probabilistic versions of fairness, where the goal is to quantify the discrepancy between the objectives of ensuring the expected utilization of the resources is large (ex-ante Pareto-optimal), while the probability of receiving the resource is proportional across groups (ex-ante proportional).  However, they consider algorithms which decide on the entire allocation for each agent upfront before observing the demand rather than adaptive policies. 

\noindent \textbf{Adaptive Allocations}: In contrast, we consider a model where the principal makes decisions on how much of the resources after witnessing the number of individuals in a round.  Most similar to our work is recent work analyzing a setting where individuals arrive over time and do not depart, so that the algorithm can allocate additional resources to individuals who arrived in the past~\citep{kash2014no}.  We instead consider a stochastic setting where individuals arrive and depart in the same step with the goal of characterizing allocations that cannot reallocate to previous agents.  Other papers either seek competitive ratios in terms of the Nash Social Welfare objective~\citep{azar2010allocate,bateni2018fair,gorokh2020fair}, or derive allocation algorithms which perform well in terms of max-min~\citep{lien2014sequential,manshadi2021fair}.  {Our work differs from these in that we impose additional distribution assumptions (notably that the variance of the demand is smaller order than its mean, more common in real-world scenarios).  The results in \cite{manshadi2021fair} can be viewed as highlighting a trade-off between efficiency and the max-min objective, although achieving efficiency of zero is trivial in that setting as the algorithm designer is not penalized for giving all leftover resources at the last location.  In contrast, under our setting eliminating the resources at the final round penalizes the algorithm in terms of both $\Denv$ and $\Denvhind$, requiring a more nuanced discussion on the trade-off between efficiency and envy.}
    \section{Preliminaries}
\label{sec:preliminary}

We use $\mathbb{R}_+$ to denote the set of non-negative reals, and $\norm{X}_{\infty} = \max_{i,j} |X_{i,j}|$ to denote the matrix maximum norm, and $cX$ to denote entry-wise multiplication for a constant $c$. When comparing vectors, we use $X \leq Y$ to denote that each component $X_i \leq Y_i$.

\subsection{Model and Assumptions}
A principal is tasked with dividing $K$ divisible resources among a population of individuals who are divided between $T$ distinct rounds -- these could represent $T$ locations visited sequentially by the principal (for example, food distribution sites visited by a mobile pantry), or $T$ consecutive time periods (for example, days over which a hospital must stretch some limited medical supply before it is restocked).

Each resource $k \in [K]$ has a fixed initial budget $B_k$ that the principal can allocate across these rounds.  
Each round has a (possibly random) set of {distinct} individuals arriving to request a share of the resources.  Individuals are characterized by their type $\theta \in \Theta$, corresponding to their preferences over the $K$ resources, where individuals of type $\theta$ receive utility $u(x, \theta) : \mathbb{R}^{K} \times \Theta \rightarrow \mathbb{R}$ for an allocation $x$. We henceforth assume that the set of possible types has finite cardinality $|\Theta|$, and denote $(N_{t, \theta})_{\theta \in \Theta}$ to be the vector containing the number of arrivals of each type in round $t$, where $N_{t, \theta}$ denotes the number of type-$\theta$ arrivals. $(N_{t, \theta})_{\theta \in \Theta, t \leq T}$ is drawn from some \emph{known} distribution $\F$; note that these distributions across rounds need not be identical.

In the ex-post or \emph{offline} setting, the number of individuals per round $(N_{t, \theta})_{t \in [T], \theta \in \Theta}$ are known in advance and can be used by the principal to choose allocations $X \in \mathbb{R}^{T \times |\Theta|  \times K}$ for individuals in each round $t$ of type $\theta$.  In the \emph{online} setting the principal considers each round sequentially in a fixed order $t = 1, \ldots, T$, is informed of the number of individuals $(N_{t, \theta})_{\theta \in \Theta}$ in that round, chooses allocation $X_t^{alg} \in \mathbb{R}^{|\Theta| \times K}$ before continuing on to the next round, where $X_{t, \theta, k}^{alg}$ denotes the allocation of resource $k$ earmarked for each of the $N_{t, \theta}$ individuals of type $\theta$ in that round.  This assumption includes not only i.i.d demands, but can also be extended to distributions that arise from Markov chains or latent variable models (see \cref{sec:proof} for more details).  We impose the additional assumption that the algorithm allocates the same allocation to each of the $N_{t, \theta}$ individuals of type $\theta$.  This is without loss of generality, as one of the primary goals of the paper is to investigate envy, whereby one out of any two individuals of type $\theta$ in round $t$ will envy the other unless their allocations are the same. Allocation decisions are irreversible, and must obey the overall budget constraints.

\smallskip

\noindent \textbf{Assumptions}: {We assume that for every $t \in [T]$ and $\theta \in \Theta$, $N_{t, \theta} \geq 1$ almost surely.  We also assume that $N_{t, \theta}$ are independent, with variance $\Var{N_{t, \theta}} = \sigma_{t, \theta} > 0$, and mean absolute deviation $|N_{t, \theta} - \Exp{N_{t, \theta}}| = \rho_{t, \theta} < \infty$ almost surely. We additionally denote $\sigma^2_{min} = \min_{t, \theta} \sigma_{t, \theta}^2, \sigma^2_{max} = \max_{t, \theta} \sigma_{t, \theta}^2$ and $\mu_{max} = \max_{t, \theta} \Exp{N_{t, \theta}}$, and assume that $\sigma^2_{min},\sigma^2_{max},\mu_{max}$ are given constants. These assumptions are for ease of notation and clarity of presentation; in particular, our results only depend on mild conditions on the expectation and tails of the sums of future arrivals $\sum_{t'>t}N_{t', \theta}$ of each type.  Extensions will be discussed in \cref{sec:proof}.  We define $\Bav = \frac{B}{\sum_{\theta \in \Theta} \sum_{t \in [T]} \Exp{N_{t, \theta}}} \in \mathbb{R}^k$ as the average resource per individual; for ease of understanding, $\Bav$ can be viewed as being a constant, but our results hold for any $\Bav$.} 

We also focus on utility functions that are linear, i.e., where $u(x, \theta) = \langle w_\theta, x \rangle$, where the latent individual type $\theta$ is characterized by $w_\theta \in \mathbb{R}_{\geq 0}^{K}$ as a vector of \emph{preferences} over each of the different resources. {For example, the type $\theta$ could refer to a ``vegetarian'' type with preferences $[2,0,1]$ over the set of resources $[$produce, meat, canned soup$]$ indicating a marginal utility of zero for any allocated meat and increased preference for produce.  The relative \emph{scale} of the weights help indicate preference for one food resource over another.}

{
The assumption that agents' preferences over resources are linear is limiting, in that it does not account for settings in which resources exhibit {\it complementarities} (modeled via e.g., Leontief, or filling, utilities), in addition to omitting popular utility functions in the extant literature (e.g., Cobb-Douglas utilities).  Our algorithmic techniques naturally extend to more general utility functions (so long as the Eisenberg-Gale program can be solved efficiently).  However, we leave understanding both the upper and lower bounds on the achievable envy and efficiency pairs to future work.  More details on modeling individual utilities for the experiments are in~\cref{sec:experiments}.
}

Finally, we assume that our resources are \emph{divisible}, in that allocations can take values in $\mathbb{R}_+^K$. In our particular regimes of interest where we scale the number of rounds and budgets, this is easy to relax to integer allocations with vanishing loss in performance.

\smallskip

\noindent \textbf{Additional Notation}: We use $B = (B_1, \ldots, B_K)$ to be the budget vector.   For any location $t$ and type $\theta$, we use $N_{\geq t, \theta}$ to denote $\sum_{t' \geq t} N_{t', \theta}$.  If the subscript $t$ is omitted we use $N_\theta = \sum_{t=1}^T N_{t, \theta}$ to denote the total number of individuals of type $\theta$.  We additionally let $\bar{\rho}_{\geq t, \theta} = \frac{1}{T - t} \sum_{t' \geq t} \rho_{t, \theta}$ and similarly for $\bar{\sigma}_{\geq t, \theta}^2$ and $\bar{\mu}_{\geq t, \theta}$. A table with all our notation is provided in the Appendix.

\smallskip

\noindent \textbf{Limitations and Extensions}: The assumption that latent types $\Theta$ are finite is common in decision-making settings, as in practice, the set of possible types is approximated from historical data. 
One limiting assumption is that in the online setting, the principal only knows the number of individuals from one location at a time.  In reality the principal could have some additional information about future locations, e.g. via calling ahead, that could be incorporated in deciding an allocation.  Our algorithmic approach naturally incorporates such additional information.  {Additionally we assume a distinct set of individuals across each round, and consider the rounds $t$ as fixed and distinct locations.}

\subsection{Fairness and Efficiency in Offline Allocations}

To define an ex-post fair allocation, i.e., with known number of individuals $(N_{t, \theta})_{t \in [T], \theta \in \Theta}$ across rounds in $[T]$, we adopt an approach proposed by~\cite{varian1973equity} (commonly referred to as ``Varian Fairness''), which is widely used in the operations research and economics literature.  We will refer to this as fairness for brevity; for a more detailed discussion on the advantages and limitations of this model, see~\citep{sugden1984fairness} or \cref{app:varian}.
\begin{definition}[Fair Allocation]
\label{def:fairness}
Given types $\Theta$, number of individuals of each type $(N_{t, \theta})_{t \in [T], \theta \in \Theta}$ and utility functions $(u(\cdot,\theta))_{\theta \in \Theta}$, an allocation $X = \{X_{t, \theta}\in\mathbb{R}^K_+ \mid \sum_{t=1}^T \sum_{\theta \in \Theta} N_{t, \theta} X_{t, \theta} \leq B\}$ is said to be fair if it simultaneously satisfies the following:
\begin{enumerate}
\item \textit{Envy-Freeness} (EF): For every pair of rounds $t,t'$ and types $\theta, \theta'$, we have $u(X_{t, \theta}, \theta) \geq u(X_{t', \theta'}, \theta)$.
\item \textit{Pareto-Efficiency} (PE): For any allocation ${Y}\neq X$ such that $u(Y_{t, \theta},\theta)> u(X_{t, \theta},\theta)$ for some round $t$ and type $\theta$, there exists some other round $t'$ and type $\theta'$ such that $u(Y_{t', \theta'},\theta')< u(X_{t', \theta'}, \theta')$. 
\item \textit{Proportional} (Prop): For any round $t$, type $\theta$ we have $u(X_{t, \theta}, \theta) \geq u(B/N, \theta)$ where $N = \sum_{t=1}^T \sum_{\theta \in \Theta} N_{t, \theta}$.
\end{enumerate}
\end{definition}
\noindent 

While the three properties form natural desiderata for a fair allocation, the power of this definition lies in that asking for them to hold simultaneously rules out many natural (but unfair) allocation policies. 
In particular, allocation rules based on maximizing a global function such as utilitarian welfare (sum of individual utilities) or egalitarian welfare (the maximin allocation, or more generally, the leximin allocation~\citep{bogomolnaia2001new,lien2014sequential,manshadi2021fair}) are Pareto-efficient, but tend to violate individual envy-freeness, as they focus on global optimality rather than per-individual guarantees. 
A remarkable exception to this, however, is the Nash Social Welfare, whose maximization leads to an allocation that is Pareto-efficient, envy-free, and proportional, and hence fair.

\begin{proposition}[Theorem 2.3 in \citep{varian1973equity}]
\label{prop:fair}
For allocation $X$, the Nash Social Welfare is: 
\begin{align}
\label{eq:nsw}
NSW(X) = \left( \prod_{t \in [T]} \prod_{\theta \in \Theta} u(X_{t, \theta}, \theta)^{N_{t, \theta}} \right)^{1 / \sum_{t, \theta} N_{t, \theta}}.
\end{align}
Under linear utilities, an allocation $X$ that maximizes $NSW(X)$ is Pareto-efficient, envy-free, and proportional.
\end{proposition}

In addition to simultaneously ensuring PE, EF and Prop properties, the NSW maximizing solution can also be efficiently computed via the following convex program called the \emph{Eisenberg-Gale} (EG) program~\citep{eisenberg1961aggregation}, obtained by taking the logarithm of the Nash Social Welfare:
\begin{align}
\label{eq:offline_nsw}
\max_{X \in \mathbb{R}_+^{T \times \Theta \times K}} \, \sum_{t=1}^T  \sum_{\theta \in \Theta} N_{t, \theta}  \log\left( u(X_{t, \theta}, \theta) \right) 
\quad \quad  \text{ s.t. } \,\sum_{t=1}^T \sum_{\theta \in \Theta} N_{t, \theta} X_{t, \theta} \leq B 
\end{align}

Important to note in our setting is that the optimal fair allocation in hindsight, which solves \cref{eq:offline_nsw} with a given number of individuals of each type across all rounds $(N_{t, \theta})_{t \in [T], \theta \in \Theta}$, does not depend on the round $t$.  Indeed, any envy-free allocation can be formulated so $X_{t, \theta} = X_{t', \theta}$ (by setting $X_\theta = \frac{1}{t} \sum_t X_{t, \theta}$) and so we can instead consider the solution to:
\begin{align}
\label{eq:eg}
\max_{X \in \mathbb{R}_+^{\Theta \times K}} \, \sum_{\theta \in \Theta} N_{\theta}  \log\left( u(X_{\theta}, \theta) \right) 
\quad \quad \text{ s.t. } \, \sum_{\theta \in \Theta} N_{ \theta} X_{\theta} \leq B 
\end{align}
where we use $N_{\theta} = \sum_{t \in [T]} N_{t, \theta}$ to denote the total number of individuals across all rounds of type $\theta$.  The fact that the optimal solution in hindsight does not depend on the round $t$ forms the basis for our algorithm \HopeGuardrail.

\subsection{Approximate Fairness and Efficiency in Online Allocations}

Recall that in our online setting the principal allocates resources across each round in a fixed order $t = 1, \ldots, T$, whereupon at round $t$ the principal sees $(N_{t, \theta})_{\theta \in \Theta}$ and decides on an allocation before continuing to the next round.  A natural (albeit naive) approach in this setting could be to try and obtain allocations which satisfy Pareto-efficiency and envy-freeness on all sample paths. However, such an approach is not feasible even in the simplest online setting, as the optimal solution in hindsight is often a unique function of the realized number of individuals across each rounds.
\begin{proposition}
\label{lemma:lower_bound}
For $T = 2$ rounds, $|\Theta|=1$ type, single resource, and linear utilities, for any non-trivial distribution $\F_2$, no online algorithm can guarantee ex-post envy-freeness and Pareto-efficiency almost surely.
\end{proposition}
\begin{rproof}{\cref{lemma:lower_bound}}
Let $\F_2 \sim 1+\textsc{Bernoulli}(p)$ with $p\in(0,1)$.  For any value of $N_1$ with probability $p$ the optimal solution is $X_{opt} = B / (N_1 + 1)$, else $X^{opt} = B / (N_1 + 2)$. As any algorithm must decide how much to allocate at round $t = 1$ without knowledge of $N_2$, no algorithm can match the ex-post fair solution almost surely. \Halmos
\end{rproof}

\cref{lemma:lower_bound} shows that trying to simultaneously achieve ex-post envy-freeness and Pareto-efficiency is futile, and hence we need to consider approximate fairness notions. To this end, we define \emph{counterfactual envy}, \emph{hindsight envy}, and \emph{efficiency}.
\begin{definition}[Counterfactual Envy, Hindsight Envy, and Efficiency]
\label{def:distance}
Given individuals with types $\Theta$, sizes $(N_{t, \theta})_{t \in [T], \theta \in \Theta}$, and resource budgets $(B_k)_{k \in [K]}$, for any online allocation $(X^{alg}_{t, \theta})_{t \in [T], \theta \in \Theta} \in \mathbb{R}^{k}$, we define:
\begin{itemize}
    \item\emph{Counterfactual Envy}: The counterfactual distance of  $X^{alg}$ to envy-freeness as \begin{align*}
    \Denv \triangleq \max_{t \in [T], \theta \in \Theta} \norm{u(X^{alg}_{t, \theta}, \theta) - u(X^{opt}_{t, \theta}, \theta)}_{\infty}\end{align*}
    where $X^{opt}$ is the optimal fair allocation in hindsight, i.e. the solution to \cref{eq:eg} with true values $(N_{t, \theta})_{t \in [T], \theta \in \Theta}$.
    
    \item \emph{Hindsight Envy}: The hindsight distance of $X^{alg}$ to envy-freeness as \begin{align*}
        \Denvhind \triangleq \max_{t,t' \in [T]^2, \theta,\theta' \in \Theta^2} u(X^{alg}_{t', \theta'},\theta) - u(X^{alg}_{t, \theta}, \theta).
    \end{align*}
    \item\emph{Efficiency}: The distance to efficiency as
    \begin{align*}\Deff \triangleq \sum_{k \in K} \left(B_k - \sum_{t \in [T]} \sum_{\theta \in \Theta} N_{t, \theta} X_{t, \theta, k}^{alg}\right)
\end{align*}
\end{itemize}
\end{definition}
Our algorithm also provides ex-post guarantees on hindsight proportionality defined via 
$\Dprop  \triangleq \max_{t, \theta} u\left(\frac{B}{\sum_{t, \theta} N_{t, \theta}}, \theta\right) - u(X^{alg}_{t, \theta}, \theta).$

These approximate fairness definitions are motivated by the problems faced by the FBST.  Hindsight envy measures the algorithm's ability to ensure individuals are not envious of the allocations given to any other.  While this might serve as a natural first step at a definition, an algorithm achieving low hindsight envy does not necessarily imply that individuals are eager to participate.  In particular, the algorithm which allocates $X_{t, \theta} = 0$ for all $t$ and $\theta$ trivially achieves hindsight envy of zero (while suffering from large efficiency).  Another consideration is ensuring allocations are close to what they \emph{should have been given} based on observed information along the trajectory.  Our measure of \emph{counterfactual envy} addresses that, penalizing allocation algorithms based on how close they were at addressing individual's utility versus the optimal solution in hindsight.  In fact, this metric has been considered in the literature in a competitive ratio instead of additive sense~\citep{friedman_2017,10.1145/2764468.2764495}.  Lastly, efficiency is a natural yardstick for measuring an algorithm in order to ensure all of the resources which can be utilized are used.

We also note that all of these metrics are much stronger than the existing metrics in the literature because we provide hindsight guarantees that hold with high probability with respect to the distribution as opposed to weaker ex-ante guarantees that only hold in expectation.  Moreover, most approaches in the literature focus on defining a single optimization problem with a specified objective embodying ``fairness'' that attempts to capture desired goals.  This is fundamentally flawed as the definition of the objective or choice of metric itself biases the outcomes towards a particular point along the tradeoff curve between different criteria.  The \emph{important challenge} in this setting then is considering meaningful trade-offs between these metrics in the online setting (see \cref{fig:uncertainty_principle}) and designing algorithms which achieve \emph{any} point along the tradeoff curve.

As highlighted earlier, the definition of counterfactual envy and efficiency are related.  
By using the fact that the optimal solution in hindsight $X^{opt}$ is efficient, we can naively bound  $\Deff$ using $\Denv$,
\begin{align*}
    \Deff 
    = \tsum_{t \in [T]} \tsum_{\theta \in \Theta} \tsum_{k \in K} N_{t, \theta} (X_{t, \theta, k}^{opt} - X_{t, \theta, k}^{alg})
    \leq \frac{T K \Denv}{\norm{w}_{min}} \tsum_{\theta \in \Theta} N_{\theta}.
\end{align*}
This naive bound is loose, with unnecessary dependence on the number of locations $T$ (see the \emph{Counterfactual Envy-Efficiency Uncertainty Principle} in \cref{sec:uncertainty_principle}).

    \section{Uncertainty Principles}
\label{sec:uncertainty_principle}

In this section we show (1), (2a), and (2b) from \cref{thm:informal} concerning a lower bound on the achievable $\Denv$, and the relationship between $\Denv$, $\Denvhind$, and $\Deff$ due to the envy-efficiency uncertainty principle.  In all of these proofs we consider the case of a single resource, single type, and assume that $u(X, \theta) = X$ for brevity and clarity in the presentation.  However, the proofs extend directly to multiple resources where one considers the setting with $|\Theta| = K$ and each type $\theta$ desires a unique resource.

We begin with (1), the statistical uncertainty principle on the optimal fair allocation in hindsight, showing that no online algorithm is able to achieve counterfactual envy smaller than order $1 / \sqrt{T}$.  This arises due to the uncertainty in the number of individuals arriving in the future, forcing the algorithm to make a non-trivial decision on the allocation made to to individuals in the first round.

\begin{theorem}[Statistical Uncertainty Principle] \label{thm:lower_bound}

Let $\alpha$ be a constant with $\alpha + \frac{C \rho_{max}}{\sigma_{min}^3 \sqrt{T}} < \frac{1}{2}$ where $C$ is an absolute constant.  Then with probability at least $\alpha$ any online algorithm must incur
\begin{align*}
    \Denv & \geq \Bav \frac{3 \Phi^{-1}(1 - \alpha - \frac{C \rho_{max}}{\sigma_{min}^3 \sqrt{T}}) \sigma_{min}}{4\sqrt{T}}.
\end{align*}
\end{theorem}
\begin{rproof}{\cref{thm:lower_bound}}
We use the generalized Berry-Esseen theorem~\citep{berry1941accuracy}. Recall that for all $t$, $\Var{N_t} =\sigma_t > 0$ and $\Exp{|N_t - \Exp{N_t}|} = \rho_t < \infty$, and moreover, $X^{opt}_t = B/N$ for all $t$ where $N = \sum_{t \in [T]} N_t$. Let us denote $\bar{\sigma}^2 = \frac{1}{T} \sum_{t\in[T]}\sigma_t^2$ and $\bar{\rho} = \frac{1}{T} \sum_{t\in[T]} \rho_t$, and let $\Phi$ be the CDF of a standard normal. Using Berry-Esseen it holds that for an absolute constant $C$,
for all $z \in \RR$, 
$$\Phi(z) - \frac{C\bar{\rho}}{\bar{\sigma}^{3} \sqrt{T}} \leq \Pr\left(X_{opt} \geq \frac{B}{\Exp{N}+z \bar{\sigma} \sqrt{T}} \right) \leq \Phi(z) + \frac{C\bar{\rho}}{\bar{\sigma}^{3} \sqrt{T}}.$$

Taking $z = - y$ and using the lower bound we have that with probability at least $\Phi(-y) - \frac{C \bar{\rho}}{\bar{\sigma}^3 \sqrt{T}}$,
\begin{align*}
    X^{opt} & \geq \frac{B}{\Exp{N} - y \bar{\sigma} \sqrt{T}} 
     \geq \frac{B}{\Exp{N}} \left( 1 + \frac{y \bar{\sigma} \sqrt{T}}{\Exp{N}}\right).
\end{align*}
Taking $z = y$ and using the upper bound we have that with probability at least $\Phi(-y) - \frac{C \bar{\rho}}{\bar{\sigma}^3 \sqrt{T}}$,
\begin{align*}
    X^{opt} & \leq \frac{B}{\Exp{N} + y \bar{\sigma} \sqrt{T}} 
     \leq \frac{B}{\Exp{N}} \left(1 - \frac{y \bar{\sigma}\sqrt{T}}{2 \Exp{N}}\right).
\end{align*}

Note that these intervals are non-overlapping for $y > 0$.  As the algorithm must decide on a value $X^{alg}_1$ to allocate for the first round then with probability at least $\Phi(-y) - \frac{C \bar{\rho}}{\bar{\sigma}^3 \sqrt{T}} \geq \Phi(-y) - \frac{C \rho_{max}}{\sigma_{min}^3\sqrt{T}}$:
\begin{align*}
    \norm{X^{alg} - X^{opt}}_{\infty} & \geq \min_{x \in \mathbb{R}} \max \left(\abs{\frac{B}{\Exp{N}} \left(1 - \frac{y \bar{\sigma}\sqrt{T}}{2\Exp{N}}\right) - x}, \abs{\frac{B}{\Exp{N}} \left( 1 + \frac{y \bar{\sigma} \sqrt{T}}{\Exp{N}}\right) - x} \right)
     = \frac{B}{\Exp{N}} \frac{3y \bar{\sigma} \sqrt{T}}{4\Exp{N}}.
\end{align*}
Taking $y = \Phi^{-1}(1 - \alpha - \frac{C \rho_{max}}{\sigma_{min}^3 \sqrt{T}})$ which is positive then we get with probability at least $\alpha$ that $$\norm{X^{alg} - X^{opt}}_\infty \geq \frac{B}{\Exp{N}} \frac{3\Phi^{-1}(1 - \alpha - \frac{C \rho_{max}}{\sigma_{min}^3 \sqrt{T}}) \bar{\sigma}}{4\Exp{N}} \geq \Bav \frac{ 3\Phi^{-1}(1 - \alpha - \frac{C \rho_{max}}{\sigma_{min}^3 \sqrt{T}}) \sigma_{min}}{4\sqrt{T}}. \Halmos$$
\end{rproof}

We next show the first part of the Envy-Efficiency Uncertainty principle (2a), highlighting that any online algorithm which achieves a factor of $L_T$ on counterfactual envy necessarily suffers efficiency of at least $1 / L_T$.  This result follows from the statistical uncertainty in the number of individuals arriving in the final $L_T^{-2}$ rounds, and the fact that ensuring a bounded envy requires any online algorithm to save enough budget to allocate a minimum allocation to all future arriving individuals.

\begin{theorem}[Counterfactual Envy-Efficiency Uncertainty Principle]
\label{thm:envy-efficiency}
Let $\alpha < \frac{1}{8}$ be a constant such that $3 \alpha + \frac{C \rho_{max}}{\sigma_{min}^3 \sqrt{T}} < \frac{1}{2}$ for an absolute constant $C$. Any online algorithm which achieves $ \Denv \leq L_T = o(1) $
with probability at least $1 - \alpha$ must also incur waste $\Deff \geq \tilde{C} \min \{\sqrt{T},  1 / L_T\}$ where
$
\tilde{C} = \left(\Bav - o(1) \right)^2 \frac{\Phi^{-1}(1 - 3 \alpha - \frac{C \rho_{max}}{\sigma_{min}^3 \sqrt{T}})^2 \sigma_{min}^2}{24 \sqrt{2 \rho_{max}^2 \log(T/\alpha)} \mu_{max}}
$
with probability at least $\frac{1}{12} - o(1)$.
\end{theorem}

\begin{rproof}{\cref{thm:envy-efficiency}}
In order for the algorithm to guarantee that $\Denv = \norm{X^{alg} - X^{opt}}_\infty \leq L_T$ with probability at least $1 - \alpha$ it must limit all allocations made to the interval $[\frac{B}{N} - L_T, \frac{B}{N}+L_t]$ as $X^{opt} = \frac{B}{N}$.  Moreover, a straightforward application of Hoeffding's inequality shows that $|N - \Exp{N}| \leq \tilde{c}\sqrt{T}$ with probability $1 - \alpha$ where $\tilde{c} = \sqrt{2 \rho_{max}^2 \log(T / \alpha)}$.  Using this, algebraic manipulations, and simplifying, one can show that the following event
\[\mathcal{D} = \bigcap_{t\in[T]}\left\{X_t^{alg} \in \left[ \frac{B}{\Exp{N}} - \frac{\tilde{c}\sqrt{T}}{\Exp{N}} - L_T, \frac{B}{\Exp{N}} + \frac{2\tilde{c}\sqrt{T}}{\Exp{N}} + L_T \right]\right\}\]
occurs with probability at least $1 - 2 \alpha$.  We interpret these lower and upper thresholds on allocations made by the algorithm as \emph{guardrails}.

Recall that we use the notation $B_t^{alg}$ to denote the budget remaining for the algorithm at the start of round $t$.  We begin by defining three events for a fixed round $t \leq T$ and constant $z = \Phi^{-1}(1 - 3 \alpha - \frac{C \rho_{max}}{\sigma_{min}^3 \sqrt{T}}) > 0$:
\begin{align*}
    \A & = \{ N_{\geq t} \leq \Exp{N_{\geq t}} \} \\
    \B & = \{ N_{\geq t} \geq \Exp{N_{\geq t}} + z \bar{\sigma}_{\geq t} \sqrt{T - t + 1} \} \\
    \C & = \{ B_t^{alg} \geq \left(\frac{B}{\Exp{N}} - \frac{\tilde{c}\sqrt{T}}{\Exp{N}}- L_T\right)\left(\Exp{N_{\geq t}} + z \bar{\sigma}_{\geq t} \sqrt{T - t + t} \right)\}.
\end{align*}
By Berry-Esseen Theorem \citep{berry1941accuracy} we know that $\Pr(\A) \geq \frac{1}{2} - \frac{C \bar{\rho}_{\geq t}}{\bar{\sigma}_{\geq t}^3 \sqrt{T - t + 1}} \geq \frac{1}{2} - \frac{C \rho_{max}}{\sigma_{min}^3 \sqrt{T}}$ and that $\Pr(\B) \geq \Phi(-z) - \frac{C \bar{\rho}_{\geq t}}{\bar{\sigma}_{\geq t}^3 \sqrt{T-t+1}} \geq 3 \alpha$ by choice of $z$.

We first show that $\neg \C \cap \B$ implies $\neg\mathcal{D}$ (or equivalently by taking the contrapositive that $\mathcal{D}$ implies that $\B$ implies $\C$) which gives us that $\Pr(\neg \C \cap \B) \leq \Pr(\neg\mathcal{D})$. These two conditions ($\mathcal{D}$ and $\B$) dictates that the algorithm must have a lot of budget by allocating within the guardrails based on the number of individuals arriving in the future being small.  Indeed, under events $\mathcal{D}$ and $\B$ we have:
\begin{align*}
    B_t^{alg} & \geq \sum_{t' \geq t} N_{t'} X_{t'}^{alg} \geq N_{\geq t} \left(\frac{B}{\Exp{N}} - \frac{\tilde{c}\sqrt{T}}{\Exp{N}}- L_T\right) \geq (\Exp{N_{\geq t}} + z \bar{\sigma}_{\geq t} \sqrt{T - t + 1})\left(\frac{B}{\Exp{N}} - \frac{\tilde{c}\sqrt{T}}{\Exp{N}}- L_T\right).
\end{align*}
Moreover, due to the fact that the algorithm is non-anticipatory, we know that the events $\C$ and $\B$ are independent.  Thus we have that $\Pr(\neg \C \cup \B) = \Pr(\neg \C) \Pr(\B)$.  Using the bound on $\Pr(\B)$ and the fact that $\Pr(\neg \C \cup \B) \leq \Pr(\neg \mathcal{D}) \leq 2\alpha$ we get that $\Pr(\neg \C) \leq \frac{2}{3}$.

Now we consider the event $\C \cap \A \cap \mathcal{D}$.  Using that the allocations must be bounded by event $\mathcal{D}$ we have that the waste will be at least:
\begin{align*}
\Deff & = B - \tsum_{i=1}^T X_i^{alg} N_i = B_t^{alg} - \tsum_{i \geq t} X_i^{alg} N_i \\
& \geq \left(\Exp{N_{\geq t}} + z \bar{\sigma}_{\geq t}\sqrt{T-t+1}\right) \left(\frac{B}{\Exp{N}} - \frac{\tilde{c}\sqrt{T}}{\Exp{N}}- L_T\right) - \left(\frac{B}{\Exp{N}} + \frac{2\tilde{c}\sqrt{T}}{\Exp{N}} + L_T\right) \Exp{N_{\geq t}} \\
& \geq \left(\frac{B}{\Exp{N}} - \frac{\tilde{c}\sqrt{T}}{\Exp{N}}- L_T\right) z \bar{\sigma}_{\geq t}\sqrt{T-t+1} - 3\left( L_T + \frac{\tilde{c}\sqrt{T}}{\Exp{N}} \right) \Exp{N_{\geq t}}.
\end{align*}
The inequality follows from lower bounding $B_t^{alg}$ with the amount required to be reserved up to location $t$ (i.e. event $\C$), and upper bounding the maximum amount of budget that can be expended for locations $i \geq t$ when $N_{\geq t} \leq \Exp{N_{\geq t}}$ (i.e. event $\A$).

Recall that $\Exp{N_{\geq t}} = (T-t+1) \bar{\mu}_{\geq t}$, so that while the first term increases with $(T-t+1)$, the second term decreases with $(T-t+1)$.  Solving for the maximum value in terms of $t$ yields
\begin{align*}
    \Deff & \geq 
    \frac{1}{12} \left(\frac{B}{\Exp{N}} - \frac{\tilde{c} \sqrt{T}}{\Exp{N}} - L_T\right)^2 \frac{z^2 \bar{\sigma}_{> t}^2}{(L_T+\frac{\tilde{c}\sqrt{T}}{\Exp{N}}) \bar{\mu}_{\geq t}} \geq \frac{1}{12} \left(\frac{B}{\Exp{N}} - \frac{\tilde{c} \sqrt{T}}{\Exp{N}} - L_T\right)^2 \frac{z^2 \sigma_{min}^2}{(L_T + \tilde{c} / \sqrt{T}) \mu_{max}}.
\end{align*}
The probability of this event is lower bounded by $\Pr(\C \cap \A \cap \mathcal{D}) \geq \Pr(\C \cap \A) - \Pr(\neg\mathcal{D}) \geq \Pr(\C) \Pr(\A) - 2\alpha \geq (1 - \frac{2}{3}) (\frac{1}{2} - \frac{C \rho_{max}}{\sigma_{min}^3 \sqrt{T}}) - 2 \alpha \geq \frac{1}{12} - o(1)$.  Plugging in the value of $z$ and simplifying terms yields the final result. \Halmos
\end{rproof}

Lastly we show the second part of the Envy-Efficiency Uncertainty principle (2b), highlighting that any online algorithm which achieves a factor of $L_T$ on hindsight envy necessarily suffers efficiency of at least $\min\{\sqrt{T}, \frac{1}{L_T}\}$.  This result follows from the previous lower bound (2a), combined with an almost sure relationship between $\Denv$ and $\Denvhind$.  We start with this brief lemma relating the two notions of envy.

\begin{lemma}[Relation between $\Denvhind$ and $\Denv$]
\label{lemma:envy_relation}
For any valid online allocation algorithm we have the following almost surely
\begin{align*}
    \Denv - \frac{1}{N} \Deff \leq \Denvhind \leq 2 \Denv.
\end{align*}
\end{lemma}

\begin{rproof}{\cref{lemma:envy_relation}}
The upper bound follows immediately from applying the triangle inequality around $X^{opt} = \frac{B}{N}$.  For the lower bound we instead show that $\frac{1}{N} \Deff \geq \Denv - \Denvhind$.  Here we set $L = \min_i X_i$ and $U = \max_i X_i$ to be the maximum and minimum allocations given out by the algorithm.    Note that $\Denvhind = U - L$ and $\Denv = \max\{|\frac{B}{N} - U|, |\frac{B}{N} - L|\}$.  First notice no algorithm can have $L > \frac{B}{N}$ due to the feasibility of allocations made by the algorithm.  Thus we get that $\Denv = \max\{|\frac{B}{N} - U|, \frac{B}{N} - L\}$.  We show the inequality breaking into cases on the side which achieves the $\max$.

\noindent \textbf{Case 1}: $\Denv = \frac{B}{N} - L$.

In this setting we have that $\Denv = \frac{B}{N} - L$, $\Denvhind = U - L$.  Using this we can show
\begin{align*}
    \Deff & = B - \sum_i N_i X_i = \sum_i N_i(\frac{B}{N} - X_i) \geq \sum_i N_i(\frac{B}{N} - U) = N(\frac{B}{N} - U) = N(\Denv - \Denvhind). 
\end{align*}

\noindent \textbf{Case 2}: $\Denv = |\frac{B}{N} - U|$.

This implies that $L \leq \frac{B}{N} \leq U$ as otherwise the maximum would have been achieved by $\frac{B}{N} - L$.  Thus we get that $\Denv - \Denvhind = U - \frac{B}{N} - (U - L) = L - \frac{B}{N}$ which is negative, so the inequality is trivially true.
\Halmos
\end{rproof}

Using this we are able to show (2b) in the Envy-Efficiency Uncertainty Principle, relating the necessary trade-off between $\Denvhind$ and $\Deff$.
\begin{theorem}[Hindsight Envy-Efficiency Uncertainty Principle]
\label{thm:hindsight_envy-efficiency}
Let $\alpha < \frac{1}{8}$ be a constant such that $3 \alpha + \frac{C \rho_{max}}{\sigma_{min}^3 \sqrt{T}} < \frac{1}{2}$ for an absolute constant $C$. Any online algorithm which achieves $ \Denvhind \leq L_T = o(1) $
with probability at least $1 - \alpha$ must also incur waste $\Deff \geq \tilde{C} \min\{\sqrt{T}, 1 / L_T\} - o(1)$ where $\tilde{C}$ is as in \cref{thm:envy-efficiency} 
with probability at least $\frac{1}{12} - o(1)$.
\end{theorem}
\begin{rproof}{\cref{thm:hindsight_envy-efficiency}}
Suppose the online algorithm achieves $\Denvhind \leq L_T$ with probability at least $1 - \alpha$.  However, using \cref{lemma:envy_relation} we get that $\Denv - \frac{1}{N}\Deff \leq \Denvhind \leq L_T$.  Hence we have that $\Denv \leq L_T  + \frac{1}{N} \Deff$ with probability at least $1 - \alpha$.  Denote by $\tilde{C}$ as the terms on the right hand side of \cref{thm:envy-efficiency} and applying the result there for the case when the $1/L_T$ term attains the minimum we get that
$
\Deff \geq \tilde{C} \frac{1}{L_T + \frac{1}{N}\Deff}.
$
Rearranging the inequality gives us that
$
\Deff \geq \frac{N}{2}\left[ \sqrt{\frac{4 \tilde{C}}{N} + L_T^2} - L_T \right].
$
The final bound comes from taking the first term of the Taylor series about infinity with the additional $o(1)$ factor.
\Halmos
\end{rproof}

    \section{Sensitivity and Concentration on Counterfactual Optimal Fair Allocation}
\label{sec:concentration}
The lower bounds presented in \cref{sec:uncertainty_principle} highlight a key facet of algorithm design in this setting: generating lower and upper guardrail allocations.  Suppose we were able to construct \emph{envy-free} allocations $(\underline{X}_\theta)_{\theta \in \Theta}$ and $(\overline{X}_\theta)_{\theta \in \Theta}$ such that $\norm{\overline{X}_\theta - \underline{X}_\theta}_\infty \lesssim L_T$ for a given parameter $L_T$.  If the algorithm was able to ensure that all of the allocations made to individuals of type $\theta$ are within $[\underline{X}_\theta, \overline{X}_\theta]$, it is not difficult to show that $\Denvhind \lesssim L_T$.  However, if we additionally desire a bound of $\Denv \lesssim L_T$, the same philosophy requires that we are able to establish that with high probability:
\begin{align}
\label{eq:bound_diff}
    & u(\underline{X}_\theta, \theta) \leq u(X^{opt}_\theta, \theta) \leq u(\overline{X}_\theta, \theta) \quad \forall \theta \in \Theta \text{ with } L_T \gtrsim 1 / \sqrt{T}.
\end{align}
Motivated by these two use-cases, we turn our attention to sensitivity and concentration properties on solutions to the Eisenberg-Gale program.  Unfortunately, the true EG program for the counterfactual optimal fair allocation depends on the unknown vector of number of individuals of each type $(N_{t, \theta})_{t \in [T], \theta \in \Theta}$.  As such, our algorithms are motivated by solving \emph{information relaxed} versions of the EG program, appealing to sensitivity and concentration on the \emph{optimizers} of the program, instead of the \emph{objective value} as is typically done in competitive ratio analysis.

For the time being, we assume that we are given concentration inequalities of the form: with probability at least $1 - \delta$ we have that for every $t$ and $\theta$, $|\Exp{N_{> t, \theta}} - N_{> t, \theta}| \leq \conf_{t, \theta}$. 
As this concentration only depends on the assumptions on the variables $N_{t, \theta}$, we include a simple form of $\conf_{t, \theta}$ scaling as $\sqrt{T-t}$ using Hoeffding's inequality in \cref{lem:hoeffding_app}, but see \cref{sec:proof} for extensions.  

Consider the Eisenberg-Gale program from \cref{sec:preliminary} with multiple types $\theta$ and $K$ resources as specified in \cref{eq:eg}.  Recall that the dual variables corresponding to the budget feasibility constraint $p_k$, can be thought of as \emph{prices} for the corresponding resources~\citep{nisan_roughgarden_tardos_vazirani_2007}.  
We start with a lemma showing properties of the optimal solution to the Eisenberg-Gale program with various number of individuals of each type vectors $(N_\theta)_{\theta \in \Theta}$.
\begin{lemma}[Sensitivity of solutions to the Eisenberg-Gale Program]
\label{lem:kkt_eg}
Let $x((N_\theta)_{\theta \in \Theta})$ and $p((N_\theta)_{\theta \in \Theta})$ denote the optimal primal and dual solution to the Eisenberg-Gale program (\cref{eq:eg}) for a given vector of individuals of each type $(N_\theta)_{\theta \in \Theta}$.  Then we have that:
\begin{enumerate}
    \item \textit{Scaling}: If $\tilde{N}_\theta = (1 + \zeta) N_{\theta}$ for every $\theta \in \Theta$ and $\zeta \geq 0$ then we have that:
    \begin{align*}
    x((\tilde{N}_\theta)_{\theta \in \Theta}) & = \frac{x((N_\theta)_{\theta \in \Theta})}{1 + \zeta} \\
    p((\tilde{N}_\theta)_{\theta \in \Theta}) & = (1 + \zeta)p((N_\theta)_{\theta \in \Theta}) \\
    u(x((N_\theta)_{\theta \in \Theta})_\theta, \theta) - u(x((\tilde{N}_\theta)_{\theta \in \Theta})_\theta, \theta) & = \left(1 - \frac{1}{1 + \zeta}\right) \max_k \frac{w_{\theta, k}}{p((N_\theta)_{\theta \in \Theta})_k}.
    \end{align*}
    \item \textit{Monotonicity}: If $N_\theta \leq \tilde{N}_{\theta}$ for every $\theta \in \Theta$ then we have 
    \begin{align*}
         p((\tilde{N}_{\theta})_{\theta \in \Theta}) & \geq p((N_{\theta})_{\theta \in \Theta})\\
        u(x((\tilde{N}_\theta)_{\theta \in \Theta})_\theta, \theta) & \leq u(x((N_\theta)_{\theta \in \Theta})_\theta, \theta) \quad \forall \theta \in \Theta
    \end{align*}
\end{enumerate}
\end{lemma}

These Lipschitz properties follow from the Fisher market interpretation of the Eisenberg-Gale optimum, which corresponds to market-clearing allocations in a setting with $|\Theta|$ agents, each with an endowment or budget of $N_{\theta}$.  {The second property is a generalization of the {\it competitive monotonicity} property~\cite{devanur2002market}.}  See \cref{app:conc_proof} for the full proof.

Recall that our goal is to construct lower and upper threshold allocations about $X^{opt} = x((N_{\theta})_{\theta \in \Theta})$ where $N_\theta = \sum_t N_{t, \theta}$ is the true (random) number of individuals of type $\theta$ arriving over all rounds. First suppose we were able to construct $\overline{n}_\theta \geq N_\theta$ for all $\theta$ and set $\underline{n}_\theta = (1 - \gamma) \overline{n}_\theta$ for some constant $\gamma$ chosen based on $L_T$.  Setting the guardrails as $\underline{X}_\theta = x((\overline{n}_\theta)_{\theta \in \Theta})$ and $\overline{X}_\theta = x((\underline{n}_\theta)_{\theta \in \Theta})$ will be envy free (by \cref{prop:fair}), and for the correct value of $\gamma$ satisfy the bounds needed to ensure $\Denvhind \lesssim L_T$ (by $(1)$ of \cref{lem:kkt_eg}).  However, for a large enough $\gamma$ (or equivalently a large enough $L_T$) we can additionally ensure that $\underline{n}_\theta \leq N_\theta$ and appeal to the monotonicity property $(2)$ of \cref{lem:kkt_eg} to ensure \cref{eq:bound_diff}.  These assumption includes not only i.i.d. demands, but also demand distributions that arise from Markov chains, latent variable models, or known cumulative sums, from different Chernoff style arguments (see \cref{sec:proof} for more details).  The following lemma shows the final construction of our guardrails by appropriately choosing $\overline{n}_\theta$ and $\underline{n}_\theta$ and appealing to \cref{lem:kkt_eg}.

\begin{theorem}[Construction of Guardrail Allocations]
\label{thm:concentration}
Let $X^{opt} = x((N_{\theta})_{\theta \in \Theta})$ denote the optimal solution to the Eisenberg-Gale program for a given vector of individuals of each type $(N_{\theta})_{\theta \in \Theta}$.  Further suppose that with probability at least $1 - \delta$ we have for all $\theta \in \Theta$: $|N_{\theta} - \Exp{N_{\theta}}| \leq \conf_\theta$.  Given any $L_T \geq 0$ and setting:
\begin{align*}
    \overline{n}_\theta & = \Exp{N_{\theta}}\left(1 + \max_\theta \frac{\conf_\theta}{\Exp{N_\theta}}\right)\\
    \underline{n}_\theta & = \Exp{N_{\theta}}\left(1 - c\right) \quad \text{ for } c = \frac{\norm{w}_{min} \norm{\Bav}_{min}}{\norm{w}^2_{\infty}} L_T \left(1 + \max_\theta \frac{\conf_\theta}{\Exp{N_\theta}}\right) - \max_\theta \frac{\conf_\theta}{\Exp{N_\theta}}
\end{align*}
then almost surely we have that: 
\begin{enumerate}
    \item $u(x((\underline{n}_{\theta})_{\theta \in \theta})_\theta, \theta) - u(x((\overline{n}_{\theta})_{\theta \in \theta})_\theta, \theta) \leq L_T$
    \item $\norm{x((\overline{n}_{\theta})_{\theta \in \Theta}) - x((\underline{n}_{\theta})_{\theta \in \theta})}_{\infty} \geq L_T \frac{\norm{\Bav}^2_{min} \norm{w}_{min}}{\norm{w}_\infty}$
    \item $\norm{x((\overline{n}_{\theta})_{\theta \in \Theta}) - x((\underline{n}_{\theta})_{\theta \in \theta})}_{\infty} \leq L_T \frac{\norm{B}_\infty \norm{\Bav}_{min} \norm{w}_{min}}{\norm{w}_\infty}$.
\end{enumerate}
If in addition $L_T \geq 2 \frac{\norm{w}^2_{\infty}}{\norm{w}_{min} \norm{\Bav}_{min}} \max_{\theta} \frac{\conf_\theta}{\Exp{N_\theta}}$ then with probability at least $1 - \delta$ we have:
\begin{enumerate}
    \item[\emph{4.}] $\underline{n}_\theta \leq N_\theta \leq \overline{n}_\theta$
    \item[\emph{5.}] $u(x((\overline{n}_{\theta})_{\theta \in \theta})_\theta, \theta) \leq u(X^{opt}_\theta, \theta) \leq u(x((\overline{n}_{\theta})_{\theta \in \theta})_\theta, \theta)$
\end{enumerate}
\end{theorem}
See \cref{app:conc_proof} for the full proof.  Using a straightforward application of Hoeffding's inequality we notice that this construction ensures that we are able to guarantee a bound of $L_T$ on the difference in utilities for any
$L_T \geq 2 \frac{\norm{w}^2_\infty}{\norm{w}_{min} \norm{\Bav}_{min}} \sqrt{\frac{2 \rho_{max}^2 \log(T |\Theta| / \delta)}{T}}.$
    \section{\HopeGuardrail}
\label{sec:hope}

Here we define our algorithm \HopeGuardrail.  The algorithm takes as input a budget $B$, expected number of each type $(\Exp{N_{\theta}})_{\theta \in \Theta}$, confidence terms $(\conf_{t,\theta})_{\theta \in \Theta}$, and a desired bound $L_T$ on the $\Denv$ and $\Denvhind$.  Assuming the lower and upper threshold allocations are constructed such that we can guarantee the results from \cref{sec:concentration} our algorithm is able to achieve any envy-efficiency tradeoff as developed in \cref{thm:informal}.  The algorithm relies on two main components, both of which we believe to be necessary in developing an algorithm to achieve the envy-efficiency uncertainty principle (as removing any one of them leads to breakdowns - as will be discussed in \cref{sec:proof}).  We start by describing the high level ideas needed in the algorithm before describing the pseudocode (with full algorithm description in Algorithm~\ref{alg:brief}).  The proof that \HopeGuardrail achieves the desired bounds will be deferred to \cref{sec:proof}.

\medskip

\noindent \textbf{Guardrails on Optimal Fair Allocation in Hindsight}
\\*
As a result of \cref{thm:lower_bound} we saw that no online algorithm can guarantee $\Denv \lesssim \frac{1}{\sqrt{T}}$.  Moreover, the proof highlighted that any algorithm which satisfies a bound on $\Denv$ or $\Denvhind \lesssim L_T$ must limit allocations based on guardrails with high probability.  As such, our algorithm uses the construction from \cref{sec:concentration} to obtain estimates $\overline{X} = x((\underline{n}_\theta)_{\theta \in \Theta})$ and $\underline{X} = x((\overline{n}_\theta)_{\theta \in \Theta})$ which are both envy-free and satisfy that $\max_{t, \theta} |u(\overline{X}_{t, \theta}, \theta) - u(\underline{X}_{t, \theta}, \theta)| \leq L_T.$  If in addition $L_T \gtrsim 1 / \sqrt{T}$ we have that $u(\underline{X}_\theta, \theta) \leq u(X^{opt}_\theta, \theta) \leq u(\overline{X}_\theta, \theta)$.

The allocations $\overline{X}_{\theta}$ and $\underline{X}_{\theta}$ are used by the algorithm as \emph{guardrails}, where all allocations made by the algorithm for a type $\theta$ are forced to fall within $\{\underline{X}_\theta, \overline{X}_\theta\}$.  With this requirement, on sample paths where we do not run out of budget, then we trivially have an upper bound on $\Denvhind \lesssim L_T$ and $\Denv \leq \max\{1 / \sqrt{T}, L_T\}$.  Thus `accepting' the first round loss in envy-freeness allows us to limit all future allocations to the \emph{guardrails} generated by that uncertainty.

\medskip
\noindent \textbf{Minimizing Waste via `Online Stochastic Packing'}
\\*
Once the guardrails $\overline{X}_{\theta}$ and $\underline{X}_{\theta}$ are found to verify a bound on the approximate envy up to a factor of $L_T$, we change the focus to instead try and minimize the loss of efficiency.  Thanks to our guardrails, we develop the algorithm to match a clairvoyant benchmark policy which minimizes the resource waste with the knowledge of $(N_{t, \theta})_{t \in [T], \theta \in \Theta}$ while simultaneously limiting the allocations to lie between $\{\underline{X}_{\theta}, \overline{X}_{\theta}\}$.  This can be thought of as solving an online stochastic packing problem (whose objective is to minimize efficiency loss) with the addition of \emph{guardrail constraints} (i.e. our minimum and maximum allocation constraints).  In this setting, after writing down the stochastic packing optimization program, a competitive algorithm arises naturally by ensuring that the budget remaining for the algorithm is enough to satisfy a high probability bound on the resources required to allocate $\underline{X}$ to every individual arriving in the future.  This idea is formalized in \cref{sec:proof}, and takes motivation in recent developments on Bayesian prophet benchmarks for online bin packing problems~\citep{vera2019bayesian}.

\medskip
\noindent \textbf{Algorithm Description}
\\* Let $B_{t,k}^{alg}$ denote the budget remaining to the principal for resource $k$ at iteration $t$, i.e. $B_k - \sum_{t' < t} \sum_{\theta} N_{t',\theta} X_{t',\theta,k}$.
Assume the algorithm is given the expected demands $(\Exp{N_{\theta}})_{\theta \in \Theta}$ and confidence terms $\conf_{t, \theta}$ such that $|N_{> t, \theta} - \Exp{N_{> t, \theta}}| \leq \conf_{t, \theta}$ with high probability. 
Let $\multconf = \max_{\theta} \frac{\conf_{0, \theta}}{\Exp{N_{\geq 1, \theta}}}$.
Given a desired bound on envy $L_T$, the algorithm computes the guardrails by
\begin{align*}
    \underline{X} & = x((\overline{n}_\theta)_{\theta \in \Theta}) \text{ for } \overline{n}_\theta = \left(1 + \max_\theta \frac{\conf_\theta}{\Exp{N_\theta}}\right) \Exp{N_\theta}\\
    \overline{X} & = x((\underline{n}_\theta)_{\theta \in \Theta}) \text{ for } \underline{n}_\theta = (1 - c) \Exp{N_\theta} \\
    & \text{ for } c = \frac{\norm{w}_{min} \norm{\Bav}_{min}}{\norm{w}^2_{\infty}} L_T \left(1 + \max_\theta \frac{\conf_\theta}{\Exp{N_\theta}}\right) - \max_\theta \frac{\conf_\theta}{\Exp{N_\theta}}.
\end{align*}
where $x(\cdot)$ denotes the solution to \cref{eq:eg}. Note that as long as $L_T \gtrsim \sqrt{\log(|\Theta|T/\delta) / T}$, the utility of the optimal allocation is sandwiched by the utilities of $\underline{X}$ and $\overline{X}$ according to \cref{sec:concentration}.

Our algorithm allocates to type $\theta$ according to these thresholds $\underline{X}_{\theta}$ and $\overline{X}_{\theta}$ in order to ensure the guarantee of $\Denv$ of at most $L_T$, while simultaneously trying to eliminate as much waste as possible. At each time $t$,  for each resource $k \in [K]$, 
\begin{enumerate}
    \item \textbf{(Insufficient budget):} If $B_{t, k}^{alg} \leq N_{t, \theta} \underline{X}_{\theta, k}$ then divide the resources equally among all remaining individuals for this round.
    \item \textbf{(Sufficient budget to promise lower threshold):} If $B_{t, k}^{alg} \geq \overline{X}_{\theta, k}N_{t, \theta} + \underline{X}_{\theta, k} \left( \Exp{N_{>t, \theta}} + \conf_{t, \theta} \right)$ then set $X_{t, \theta, k}^{alg} = \overline{X}_{\theta, k}$ for each $\theta \in \Theta$.
    \item \textbf{(Insufficient budget to promise lower threshold):} Otherwise set $X_{t, \theta, k}^{alg} = \underline{X}_{\theta, k}$ for each $\theta \in \Theta$.
\end{enumerate}

Our algorithm is easy to implement in practice - in particular, it requires solving the Eisenberg-Gale program ~\cref{eq:eg} \emph{only twice} to obtain $\underline{X}$ and $\overline{X}$. In contrast, other versions of certainty equivalence algorithms require frequent resolves of the Eisenberg-Gale (see \cref{app:ceq}).  This allows \HopeGuardrail to scale easily to multiple resources and larger number of types (as it only involves solving for the `optimistic' and `pessimistic' allocation rules which are done offline with historical data).  Moreover, it allows practitioners to leverage work on poly-time algorithms for solving the Eisenberg-Gale program \citep{devanur2002market}.  It also extends easily to more complex information structures (see \cref{sec:proof} for a discussion).

\section{Envy and Efficiency Bound for \HopeGuardrail}
\label{sec:proof}

We are now ready to show the bound on $\Denv$, $\Denvhind$, and $\Deff$, for \HopeGuardrail, relying on the construction of $\overline{X}_{\theta}$ and $\underline{X}_{\theta}$ from \cref{sec:concentration}.  We note that these guarantees match the envy-efficiency uncertainty principles from \cref{sec:uncertainty_principle} up to problem-dependent constants and logarithmic terms in $T$.  {Afterwards we comment on extending the distributional assumptions on $N_{t, \theta}$ to more robust settings.}

\begin{theorem}
\label{thm:upper_bound}
Given budget $B$, expected number of types $(\Exp{N_{\theta}})_{\theta \in \Theta}$, and confidence terms $(\conf_{t, \theta})_{\theta \in \Theta}$ such that with probability at least $1 - \delta$, $|N_{> t, \theta} - \Exp{N_{> t, \theta}}| \leq \conf_{t, \theta} \text{ for all } t \in [T], \, \theta \in \Theta$, \HopeGuardrail with parameter $L_T$ is able to achieve with probability at least $1 - \delta$ (where $\lesssim$ drops poly-logarithmic factors of $T$, $o(1)$ terms, and absolute constants):
\begin{align*}
    \Denvhind \leq L_T \quad\quad \Denv \lesssim \max\{1 / \sqrt{T}, L_T\} \quad\quad \Deff \lesssim \min\{\sqrt{T}, 1 / L_T\} \quad\quad \Dprop \lesssim \max\{1 / \sqrt{T}, L_T\}.
\end{align*}
\end{theorem}
\begin{rproof}{\cref{thm:upper_bound}}
Define the event $\mathcal{E} = \{ \forall t \in [T], \forall \theta \in \Theta : |N_{\geq t, \theta} - \Exp{N_{\geq t, \theta}}| \leq \conf_{t-1, \theta}\}$.  By assumption we know that $\Pr(\mathcal{E}) \geq 1 - \delta$.  The following lemma shows the algorithm ensures it has enough budget to allocate according to the lower threshold for everyone arriving in the future.  Recall that $B_{t,k}^{alg}$ denotes the budget remaining at the start of round $t$ for resource $k$.

\begin{lemma}
\label{lem:enough-budget}
Under the event $\mathcal{E}$, for every resource $k$ and time $t \in [T]$ it follows that
$$B_{t, k}^{alg} \geq \sum_{\theta \in \Theta} N_{\geq t, \theta} \underline{X}_{\theta, k}.$$
As a result, the algorithm is able to guarantee that at every iteration $X_{t, \theta, k}^{alg}$ is either $\overline{X}_{\theta, k}$ or $\underline{X}_{\theta, k}$.
\end{lemma}
\begin{rproof}{\cref{lem:enough-budget}}
We show the first statement by induction on $t$.  The second statement follows immediately.

\noindent \textbf{Base Case} $t = 1$.  Here we have that $B_1^{alg} = B$ and by construction of $\underline{X}_\theta$ we have that:
\begin{align*}
    B_{1, k}^{alg} & \geq \tsum_{\theta \in \Theta} \overline{n}_{\theta} \underline{X}_{\theta, k} \text{(by feasibility)} \, \geq \tsum_{\theta \in \Theta} N_\theta \underline{X}_{\theta, k} \text{(by event } \mathcal{E}).
\end{align*}
\noindent \textbf{Step Case} $t - 1 \rightarrow t$.
We split into two cases based on the allocation. If $X_{t-1, \theta, k}^{alg} = \underline{X}_{\theta, k}$, then by the induction hypothesis
$
    B_{t, k}^{alg} = B_{t-1, k}^{alg} - \tsum_{\theta \in \Theta} N_{t-1, \theta} \underline{X}_{\theta, k} \geq \tsum_{\theta \in \Theta} N_{\geq t, \theta} \underline{X}_{\theta, k}.
$
If $X_{t-1, \theta, k}^{alg} = \overline{X}_{\theta, k}$, then 
\begin{align*}
    B_{t, k}^{alg} & = B_{t-1, k}^{alg} - \tsum_{\theta \in \Theta} N_{t-1, \theta} X_{t-1, \theta, k}^{alg} \overset{(a)}{\geq} \tsum_{\theta \in \Theta} \underline{X}_{\theta, k} \left(\Exp{N_{\geq t, \theta}} + \conf_{t-1, \theta}\right) \overset{(b)}{\geq} \tsum_{\theta \in \Theta} N_{\geq t, \theta} \underline{X}_{\theta, k}.
\end{align*}
where (a) holds by the condition for allocating $\overline{X}_{\theta, k}$, and (b) holds under event $\mathcal{E}$. \Halmos
\end{rproof}

The next lemma shows that the algorithm is adaptively cautious, i.e., after some point, \HopeGuardrail switches to allocating according to the lower threshold.
\begin{lemma}
\label{lem:switching_point_last}
For each resource $k$, let $t_{0, k}$ be the last time that $X_{t, \theta, k}^{alg} \neq \overline{X}_{\theta, k}$ (or else $0$ if the algorithm always allocates according to $\overline{X}_{\theta, k}$).  Then under the event $\mathcal{E}$ for some $c = \tilde{\Theta}(1)$ we have that for all $k$, $t_{0, k} \geq \max\{0, T - 2 c L_T^{-2}\}$.
\end{lemma}
\begin{rproof}{\cref{lem:switching_point_last}}
We will show the result by contradiction (for the setting when $T - 2cL_T^{-2} > 0$). For some resource $k$, assume that $0 < t_{0,k} < T- 2 c L_T^{-2}$. By definition of $t_{0,k}$, it must be that the algorithm allocated $\underline{X}_{\theta,k}$ at time $t_{0,k}$ and allocated $\overline{X}_{\theta,k}$ for all subsequent times. Given the assumption, it must be that for any $t > t_{0,k}$
\begin{align*}
&\tsum_{\theta} \underline{X}_{\theta,k} \left(\Exp{N_{>t,\theta}} + \conf_{t, \theta}\right) 
\overset{(a)}{\leq} B_{t,k}^{alg} - \tsum_{\theta} N_{t,\theta} \overline{X}_{\theta,k} \overset{(b)}{=} B_{t_{0,k}}^{alg} - \tsum_{\theta} \underline{X}_{\theta,k} N_{t_0,\theta} - \tsum_{\theta} \overline{X}_{\theta,k} \tsum_{i = t_{0,k} + 1}^{t} N_{i,\theta} \\
&\hspace{1in}\overset{(c)}{<} \tsum_{\theta} \underline{X}_{\theta,k} \left(\Exp{N_{>t_{0,k},\theta}} +\conf_{t_{0,k},\theta}\right) + \tsum_{\theta}(\overline{X}_{\theta,k} - \underline{X}_{\theta,k}) N_{t_{0,k},\theta} - \tsum_{\theta} \overline{X}_{\theta,k} \tsum_{i = t_{0,k} + 1}^{t} N_{i,\theta}
\end{align*}
where (a) follows from the condition in the algorithm for $X_{t,\theta,k}^{alg} = \overline{X}_{\theta,k}$, (b) follows from the definition of $t_{0,k}$ and the choice of $t > t_{0,k}$, and (c) follows from the condition in the algorithm for $X_{t_{0,k},\theta,k}^{alg} = \underline{X}_{\theta,k}$.
By rearranging the inequality we get that 
\begin{align*}
&\tsum_{\theta} \overline{X}_{\theta,k} N_{(t_{0,k},t],\theta} < \tsum_{\theta} (\overline{X}_{\theta,k} - \underline{X}_{\theta,k}) N_{t_{0,k},\theta} + \tsum_{\theta} \underline{X}_{\theta,k} \Exp{N_{(t_{0,k},t],\theta}} + \tsum_{\theta}\underline{X}_{\theta,k} \left( \conf_{t_{0,k},\theta} -  \conf_{t,\theta}\right) \\
&\iff \tsum_{\theta} \overline{X}_{\theta,k} (N_{(t_{0,k},t],\theta} - \Exp{N_{(t_0,t],\theta}}) < \tsum_{\theta} (\overline{X}_{\theta,k} - \underline{X}_{\theta,k}) \left(N_{t_{0,k},\theta} -  \Exp{N_{(t_{0,k},t],\theta}}\right) \\
&\hspace{2.5in} + \tsum_{\theta}\underline{X}_{\theta,k} \left( \conf_{t_{0,k},\theta} -  \conf_{t,\theta}\right).
\end{align*}
Using the fact that $\overline{X}_{\theta,k} - \underline{X}_{\theta,k} \leq L_T \frac{\norm{B}_\infty \norm{\Bav}_{min} \norm{w}_{min}}{\norm{w}_\infty}$, and plugging in an upper bound for the demand at time $t_{0,k}$, a lower bound on the expected demand, and the confidence terms from \cref{lem:hoeffding_app}, the right hand side of the above inequality can be bounded above by
\begin{align*}
& L_T \frac{\norm{B}_\infty \norm{\Bav}_{min} \norm{w}_{min}}{\norm{w}_\infty} |\Theta| (\rho_{max} + \mu_{max}) - L_T \frac{\norm{B}_\infty \norm{\Bav}_{min} \norm{w}_{min}}{\norm{w}_\infty} |\Theta| (t - t_{0,k}) \\
& +  \norm{B}_\infty |\Theta| \sqrt{2 \rho_{max}^2 \log(T |\Theta| / \delta)}\left( \sqrt{T - t_{0,k}} - \sqrt{T - t}\right).
\end{align*}
Moreover, the left hand side can be bounded below under event $\mathcal{E}$ via
$$\tsum_{\theta} \overline{X}_{\theta, k} (N_{(t_{0,k},t],\theta} - \Exp{N_{(t_0,t],\theta}}) \geq -\norm{\Bav}_\infty \sqrt{2 \rho_{max}^2 \log(T \Theta| / \delta) (t - t_{0,k})}.$$
Plugging in the value of $t = T - c L_T^{-2}$ and by assumption that $t_{0,k} < T - 2cL_T^{-2}$, it follows that for $\xi = \sqrt{2 \rho_{max}^2 \log(T |\Theta| / \delta)}$, $\zeta = \frac{\norm{B}_\infty \norm{\Bav}_{min} \norm{w}_{min}}{\norm{w}_\infty}$,
\begin{align*}
    - \norm{\Bav}_{\infty} \xi \sqrt{ c L_T^{-2}} \leq L_T |\Theta| \zeta (\rho_{max} + \mu_{max}) - L_T |\Theta| \zeta c L_T^{-2} + \norm{B}_\infty |\Theta| \xi \left(\sqrt{2cL_T^{-2}} - \sqrt{c L_T^{-2}}\right). 
\end{align*}
Relabeling $x = \sqrt{c}$ to show a contradiction we need to find a value of $x$ such that
$-\frac{a_1}{L_T}x^2 + \frac{a_2}{L_T}x + a_3 L_T \leq 0.$
Noting that the cusp of the quadratic is at $\frac{a_2}{a_1}$ we see that taking the constant $$c = \frac{\norm{B}_\infty |\Theta| \xi + \norm{\Bav}_\infty \xi}{|\Theta| \zeta}$$ (independent of $L_T$) suffices to show the contradiction $(\Rightarrow \Leftarrow)$. \Halmos
\end{rproof}

The main result follows from the above two lemmas. We start by giving the upper bound on $\Deff$.  Following from Lemma \ref{lem:switching_point_last}, which states that if $t_{0,k}$ denotes the last time that $X_{t,\theta,k}^{alg} \neq \overline{X}_{\theta,k}$, then for all $k$, $t_{0,k} \geq \max\{0, T - 2 c L_T^{-2}\}$ with high probability. 
This implies that for all $k$,
\begin{align*}
\Deff &= \tsum_{k} \left(B_k - \tsum_{t\in[T]} \tsum_{\theta} X_{t,\theta,k}^{alg} N_{i,\theta}\right)
= \tsum_{k} \left(B_{t_{0, k}}^{alg} - \tsum_{t \geq t_{0, k}} \tsum_{\theta \in \Theta} N_{t, \theta} X_{t, \theta, k}^{alg} \right)\\
&\overset{(a)}{=} \tsum_{k} B_{t_{0,k},k}^{alg} - \tsum_\theta \left(\underline{X}_{\theta,k} N_{t_{0,k},\theta} + \tsum_{t > t_{0,k}} \overline{X}_{\theta,k} N_{t,\theta} \right) \\
&\overset{(b)}{<} \tsum_k \tsum_{\theta} \left(\underline{X}_{\theta,k} \left(\Exp{N_{>t_{0,k},\theta}}  +\conf_{t_{0,k},\theta} - N_{>t_{0,k},\theta} \right) {- (\overline{X}_{\theta,k} - \underline{X}_{\theta,k}) (N_{>t_{0,k},\theta} -  N_{t_{0,k}, \theta}})\right),
\end{align*}
where (a) follows from the fact that by the definition of $t_{0,k}$ the algorithm allocated the lower allocation at time $t_{0,k}$ and the upper allocation for all $t > t_{0,k}$, and (b) follows from the condition in the algorithm for allocating the lower allocation at time $t_{0,k}$, which upper bounds $B_{t_{0,k},k}^{alg}$.

However, under $\mathcal{E}$ we know that $\Exp{N_{> t_{0, k}, \theta}} - N_{> t_{0, k}, \theta} \leq 2 \conf_{t_{0, k}, \theta}$.  Plugging in the definition of $\conf_{t_{0,k}, \theta}$ and the bound on $(\overline{X}_{\theta,k} - \underline{X}_{\theta,k})$ from \cref{thm:concentration} we have:
\begin{align*}
\Deff &\leq 2 \tsum_k \tsum_{\theta} \underline{X}_{\theta,k} \conf_{t_{0, k}, \theta} + N_{t_{0, k}, \theta}(\overline{X}_{\theta,k} - \underline{X}_{\theta,k}) \\
& \leq 2 \norm{B}_1 \tsum_{k} \tsum_{\theta} \sqrt{2 
\rho_{max}^2 (T - t_0) \log(T |\Theta| / \delta)} \\ & + (\mu_{max} + \rho_{max}) \frac{2 \norm{B}_{\infty} \norm{\Bav}_{min}^2 \norm{w}_{min} \max_\theta \norm{w_\theta}_1}{\norm{w}_\infty} L_T.
\end{align*}
Taking this and plugging in the value of $t_{0,k}$ we get that:
\begin{align*}
    \Deff \leq & 2 \norm{B}_1 K |\Theta| \sqrt{2 \rho_{max}^2 \log(T|\Theta|/\delta)} \min\{\sqrt{T}, \sqrt{2c} / L_T\} \\
    & + \frac{2 (\mu_{max} + \rho_{max}) \norm{B}_{\infty} \norm{\Bav}_{min}^2 \norm{w}_{min} \max_\theta \norm{w_\theta}_1}{\norm{w}_\infty} L_T.
\end{align*}
Note that $L_T^2 = o(1)$ such that the second term is dominated by the first. 

Next we show the desired bound on $\Denvhind$.  Consider an arbitrary $t, \theta, t', \theta'$.  Then we have that:
\begin{align*}
    & u(X_{t', \theta'}^{alg}, \theta) - u(X_{t, \theta}^{alg}, \theta) = u(X_{t', \theta'}^{alg}, \theta) - u(\underline{X}_{\theta'}, \theta) + u(\underline{X}_{\theta'}, \theta) - u(\underline{X}_\theta, \theta)
     + u(\underline{X}_\theta, \theta) - u(X_{t, \theta}^{alg}, \theta) \\
    & \overset{(a)}{\leq} u(X_{t', \theta'}^{alg}, \theta) - u(\underline{X}_{\theta'}, \theta) + u(\underline{X}_\theta, \theta) - u(X_{t, \theta}^{alg}, \theta)  \overset{(b)}{\leq} \norm{w_{\theta'}}_1 \norm{X_{t', \theta'}^{alg} - \underline{X}_{\theta'}} + \norm{w_\theta}_1 \norm{X_{t, \theta}^{alg} - \underline{X}_\theta} \\
    & \overset{(c)}{\leq} 2 \max_\theta \norm{w_\theta}_1 \norm{\overline{X} - \underline{X}}_\infty \overset{(d)}{\leq} \frac{2 \norm{B}_{\infty} \norm{\Bav}_{min}^2 \norm{w}_{min} \max_\theta \norm{w_\theta}_1}{\norm{w}_\infty} L_T
\end{align*}
Where in $(a)$ we used that $\underline{X}$ is envy free we know the second pair is bounded above by zero, $(b)$ we used the definition of the utilities, and $(c)$ the fact that the algorithm allocates according to guardrails, and $(d)$ the bound in \cref{thm:concentration}.  Taking max over $t, t', \theta, \theta'$ gives the result.

Next we show the bound on $\Denv$.  First consider the setting when $L_T \geq 2 \frac{\norm{w}^2_\infty}{\norm{w}_{min} \norm{\Bav}_{min}} \sqrt{\frac{2 \rho_{max}^2 \log(T |\Theta| / \delta)}{T}}$ such that we satisfy properties four and five of \cref{thm:concentration}.  Using that the algorithm always allocates according to $\overline{X}_\theta$ or $\underline{X}_\theta$ with probability at least $1 - \delta$, we get:
\[
|u(X_{t, \theta}^{alg}, \theta) - u(X^{opt}_\theta, \theta)| \leq |u(\overline{X}_\theta, \theta) - u(\underline{X}_\theta, \theta)| \leq L_T.
\]
However, even for the case that $L_T < 2 \frac{\norm{w}^2_\infty}{\norm{w}_{min} \norm{\Bav}_{min}} \sqrt{\frac{2 \rho_{max}^2 \log(T |\Theta| / \delta)}{T}}$ then for any $\theta$ we have either $u(\underline{X}_\theta, \theta) \leq u(X^{opt}_\theta, \theta) \leq u(\overline{X}_\theta, \theta)$ as in property five, or $u(\underline{X}_\theta, \theta) \leq u(\overline{X}_\theta, \theta) \leq u(X^{opt}_\theta, \theta)$.  If the first case holds then \[
|u(X_{t, \theta}^{alg}, \theta) - u(X^{opt}_\theta, \theta)| \leq |u(\overline{X}_\theta, \theta) - u(\underline{X}_\theta, \theta)| \leq L_T.
\]
Otherwise then we can consider $\tilde{X_\theta}$ to be the upper guardrail solution via the construction from \cref{sec:concentration} with $L_T = 2 \frac{\norm{w}^2_\infty}{\norm{w}_{min} \norm{\Bav}_{min}} \sqrt{\frac{2 \rho_{max}^2 \log(T |\Theta| / \delta)}{T}}$.  There we have
\[
|u(X_{t, \theta}^{alg}, \theta) - u(X^{opt}_\theta, \theta)| \leq |u(\tilde{X}_\theta, \theta) - u(\underline{X}_\theta, \theta)| \leq 2 \frac{\norm{w}^2_\infty}{\norm{w}_{min} \norm{\Bav}_{min}} \sqrt{\frac{2 \rho_{max}^2 \log(T |\Theta| / \delta)}{T}}.
\]

Lastly we show the bound on $\Dprop$.  Recall that \HopeGuardrail satisfies that $\Denv \lesssim \max\{1 / \sqrt{T}, L_T\}$.  However, by definition of $\Denv$ this ensures that for any round $t$ and type $\theta$ that $|u(X_{t, \theta}^{alg}, \theta) - u(X_\theta^{opt}, \theta)| \lesssim \max\{1 / \sqrt{T}, L_T\}$.  Using this and the fact that $X^{opt}$ is proportional we see that
\begin{align*}
    u(\Bav, \theta) - u(X_{t, \theta}^{alg}, \theta) & = u(\Bav, \theta) - u(X_\theta^{opt}, \theta) + u(X_\theta^{opt}, \theta) - u(X_{t, \theta}^{alg}, \theta) \lesssim \max\{1 / \sqrt{T}, L_T\}.
\end{align*}
Taking the max over $t$ and $\theta$ gives the desired bound on $\Dprop$.
\Halmos
\end{rproof}

\noindent \textbf{Randomization and the Convex Envelope}: As stated in \cref{thm:envy-efficiency,thm:hindsight_envy-efficiency,thm:upper_bound}, the given upper and lower bounds are precise up to $o(1)$ factors.  Recall that the performance guarantee on \HopeGuardrail with parameter $L_T$ is $\Deff \lesssim \min\{\sqrt{T}, 1 / L_T\}$.  We can improve the performance of the algorithm for values of $L_T \in [0, 1 / \sqrt{T}]$ by randomization.  Indeed, let $\pi_{L_T}$ denote the \HopeGuardrail allocation policy with parameter $L_T$.  Consider $\pi(\alpha)$ to be the allocation policy which picks $\pi_0$ with probability $(1 - \alpha)$ and $\pi_{1 / \sqrt{T}}$ with probability $\alpha$, playing that policy once chosen across all rounds.  It is easy to see that this policy achieves expected metrics:
\begin{align*}
    \Denvhind(\pi(\alpha)) = \frac{\alpha}{\sqrt{T}} \quad \text{ and} \quad 
    \Deff(\pi(\alpha)) = (1 - \alpha) \sqrt{T} + \frac{\alpha}{\sqrt{T}}.
\end{align*}
This improves the performance on $\Deff$ for values of $L_T$ smaller than $1 / \sqrt{T}$.

\noindent \textbf{Generalizing Distributional Assumptions}:  \cref{thm:upper_bound} considers the setting when $N_{t, \theta} \sim \F_t$ is independent across $\theta$, and a time-dependent process with bounded mean absolute deviation and finite variance.  However, it is simple to see that the proofs only require a bound on the following event:
$$\bigcap_{t, \theta} \E_{t, \theta} \quad \text{ where } \quad \E_{t, \theta} = \{ |N_{> t, \theta} - \Exp{N_{> t, \theta} \mid N_{\leq t, \theta}}| \leq \conf_{t, \theta}(N_{\leq t, \theta})\}$$
with the scaling of $\conf_{t, \theta}$ being on the order of $\sqrt{T-t}$.  The main modification to \HopeGuardrail is to condition on the observed sequence of $N_{\leq t, \theta}$ thus far, particular in step two of the algorithm:
\begin{itemize}
    \item[2.] \textbf{(Sufficient budget to promise lower threshold):} If $B_{t, k}^{alg} \geq \overline{X}_{\theta, k}N_{t, \theta} + \underline{X}_{\theta, k} (\Exp{N_{>t, \theta} \mid N_{\leq t, \theta}} + \conf_{t, \theta}(N_{\leq t, \theta}))$ then set $X_{t, \theta, k}^{alg} = \overline{X}_{\theta, k} \, \forall \theta \in \Theta$
\end{itemize}The concentration arguments and scaling of $\conf_{t, \theta}$ are used in three different sections:
\begin{enumerate}
    \item Construction of the lower and upper guardrails $(\overline{X}_\theta$ and $\underline{X}_\theta)$.  This uses concentration of $\Exp{N_{\theta}}$ which is given by $\conf_{0, \theta}$.
    \item Ensuring the algorithm doesn't run out of budget by saving resources to allocate $\underline{X}_\theta$ to every individual (as in \cref{lem:enough-budget}).  This utilizes the confidence intervals on $N_{> t, \theta}$ which would be given by $\conf_{t, \theta}(N_{\leq t, \theta})$ now conditional on the observed $N_{\leq t, \theta}$ values.
    \item Construction of the time point after which \HopeGuardrail switches to allocating to the lower threshold (as in \cref{lem:switching_point_last}) uses the scaling of $\conf_{t, \theta}$ as $\sqrt{T-t}$.
\end{enumerate}
Each of these steps would be given by $\bigcap_{t, \theta} \E_{t, \theta}$ and so our approach works under distributional assumptions which yield Chernoff bounds on each event $\E_{t, \theta}$:
\begin{itemize}
    \item $N_{t, \theta} \sim \F_t$ is a time-dependent process where each $N_{t, \theta}$ has bounded mean absolute deviation and finite variance (as in \cref{lem:hoeffding_app}).
    \item $N_{t, \theta} \sim \F_t$ is a time-dependent process where each $N_{t, \theta}$ is sub-Gaussian.
    \item $N_{t, \theta}$ are conditionally independent and sub-Gaussian given a latent variable $Z$.  This model naturally encompasses dependence on the weather, other local events, etc.
    \item $N_{\theta} = \sum_t N_{t, \theta}$ is known for each $\theta$.  In this setting the algorithm can simply take $\Exp{N_{> t, \theta} \mid N_{\leq t, \theta}} + \conf_{t, \theta}(N_{\leq t, \theta}) = N_\theta - N_{\leq t, \theta}$.
    \item Each $N_{t, \theta}$ evolves independently across $\theta$ according to different ergodic Markov Chains.  Concentration bounds for these processes can be constructed using recent work on Chernoff-Hoeffding bounds for Markov Chains (Theorem 3.1 in \cite{chung2012chernoff}).  
\end{itemize}


    \section{Numerical Results}
\label{sec:experiments}

\begin{figure}[!t]
    \centering
    \includegraphics[width=.8\columnwidth]{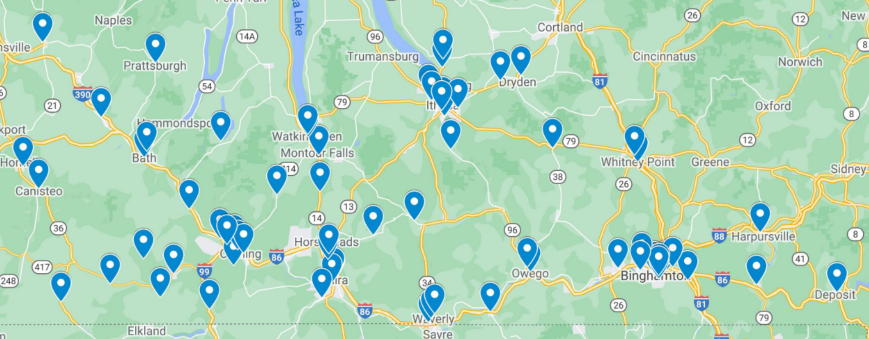}
    \caption{Map of the distribution locations serviced by the mobile-food pantry of the Food Bank of the Southern Tier of New York in 2019.}
    \label{fig:fbst_map}
\end{figure}

Here we complement the theoretical developments on the performance of \HopeGuardrail with an empirical study motivated by the challenges faced by the Food Bank of the Southern Tier when operating their mobile food pantry program.  We first describe the synthetic and data-driven experiments conducted and which aspects of the problem faced by the FBST they model, before later comparing the effectiveness of \HopeGuardrail (with various choices of $L_T$) to other algorithms on our established metrics (\cref{def:distance}).  {These experiments serve as a \emph{proof of concept} of the performance for the algorithms.}  All of the code for the experiments is available at \url{https://github.com/seanrsinclair/Online-Resource-Allocation}.

\subsection{Background}

Our experiments are motivated by the problems faced by the Food Bank of the Southern Tier (FBST) on operating their mobile food pantry program.  Due to recent increase in demands for food assistance from the COVID-19 epidemic in the United States \citep{kulish_2020} and sanctions on operating in-person stores, many food banks have increased their mobile food pantry services to meet the increased demand.  In these systems, the mobile food pantry must decide on how much food to allocate to a distribution center on arrival, without knowledge of demands in future locations.

To be more specific, the FBST operates out of seventy different distribution locations (see \cref{fig:fbst_map} for a map) across the Southern Tier of New York State.  Due to its relatively remote location, throughout the year they receive infrequent shipments of a large amount of resources (hence necessitating modeling a large $T$ for the number of rounds before a new shipment arrives).   While they receive more frequent donations from individuals, the time between larger consistent donations can span up to two or three weeks.  A certain amount of these resources are dedicated to the mobile food pantry, and used to service the seventy different distribution locations until the arrival of the next shipment.  Moreover, the schedule to which the mobile food pantry visits the different distribution locations is fixed and known in advance (and is generated primarily due to time constraints in the locations used for the distribution locations).

We interpret the total number of rounds $T$ as the number of distribution locations the mobile food pantry will visit until the arrival of the next shipment of food resources.  Important to note, is that in our simulation experiments we do not consider the additional challenges faced when trying to design a schedule (i.e. an order to visit the distribution locations) so as to achieve further notions of fairness and equity.  Each round $t$ then corresponds to a visit to a specific distribution location (for example the Tompkins County Community College in Dryden, NY) where the food pantry decides on an allocation of resources to allocate to the (random) number of individuals congregating at that location.

{We conducted four experiments, two with synthetic data referred to as \textbf{Single-Synthetic} and \textbf{Multi-Synthetic}, and two experiments using data provided by the FBST referred to as \textbf{Single-FBST} and \textbf{Multi-FBST}.  In each experiment the distributions $\F_t$ dictating the number of people of each type at a distribution location (i.e. the distribution over $N_{t, \theta}$), and preferences $w_\theta \in \mathbb{R}^{K}$ (with $u(X, \theta) = X^{\top} w_\theta$) are chosen as follows:}

\begin{itemize}
    \item \textbf{Single-Synthetic}: In this experiment we consider a single resource and single type, where the preference $w = 1$ (much like the simulation described in the lower bounds presented in \cref{sec:uncertainty_principle}) results in $u(X, \theta) = X$.  We pick the demand distribution $N_t \sim \F_t = 1+\texttt{Poisson}(1.5)$.
    \item \textbf{Multi-Synthetic}: In this experiment we consider the setting of five types and three resources.  The weights $w_\theta$ for the different types $\theta$ are listed in \cref{tab:weights_synthetic}.    The demand distribution is chosen to be $N_{t, \theta} \sim \F_t = 1+\texttt{Poisson}(1.5, 2.5, 3.5, 4.5, 5.5)$.
\end{itemize}

{In the \textbf{FBST} style simulations we use a dataset provided by the FBST indicating the mean and variance of the number of arrivals at each distribution location in 2019.  As an example, the average demand per visit (i.e. the number of individuals at that location) for the Cayuga Meadows senior living apartment complex in Ithaca was $25.9$ with a standard deviation of $3.3$. The largest distribution location is in the town of Avoca with a demand per visit of $314.6$ with a standard deviation of $57.3$.  As we do not have per-visit data in the simulations we instead sample $N_{t, \theta}$ as a Gaussial distribution with parameters indicated by their historical 2019 demand.  We hope that these experiments help highlight the applicability of the algorithm and to serve as a {\em proof of concept} of the algorithm with realistic parameters.  To be more concrete,}

\begin{itemize}
    \item \textbf{Single-FBST}: Similar to \textbf{Single-Synthetic} we consider a single resource and single type, with preference $w = 1$.  {In each experiment we first pick a random collection of $T$ locations taken from the 2019 historical dataset collected by the FBST and sample $N_{t}$ from the distribution $\F_t = \texttt{Normal}(\mu_t, \sigma^2_t)$ where $\mu_t$ and $\sigma_t^2$ are the 2019 mean and variance of the demand at the $t$'th selected distribution location.} 
    \item \textbf{Multi-FBST}: Here we consider the setting of five resources $K$ (corresponding to cereal, pasta, prepared meals, rice, and meat) and three types $\Theta$ (corresponding to vegetarians, omnivores, and ``prepared-food only'' individuals).  Similar to \textbf{Single-FBST} we first pick a random collection of $T$ locations from the 2019 FBST data, and set $N_{t, \theta}$ to be sampled from $\F_t = D_\theta \texttt{Normal}(\mu_t, \sigma_t^2)$  where $\mu_t$ and $\sigma_t^2$ are the mean and variance of the demand of the $t$'th selected distribution location.  {The distribution over the different types (vegetarians, omnivores, and ``prepared-food only'') is taken as $D_\theta = [.25, .3, .45]$ to be the estimated fraction of the FBST's population with each of those preferences.  To estimate the utility functions for the different types we use the historical prices $p_k$ used in the market mechanism to distribute food resources to food pantries across the United States \citep{prendergast2017food}.  In particular, for each of the types $\theta$ we set the weight $w_{\theta, k} = p_k \Ind{\text{type } \theta \text{ uses resource } k}$.  For example, the indicator is zero for the vegetarian type and the meat resource (see \cref{tab:weights} in the appendix for the full table of weights).}
\end{itemize}

For each of the experiments, we compare the performance of \HopeGuardrail with $L_T = T^{-1/2}$ and $T^{-1/3}$ to a \StaticAllocation algorithm which always allocates according to $\underline{X}$ until running out of resources (also can be interpreted as \HopeGuardrail with $L_T = 0$).  In \HopeGuardrail we use the natural confidence bounds garnered via Hoeffding or Bernstein inequalities.  We also compare against \CE and \ResolveCE as discussed in \cref{app:ceq}, which solve the Eisenberg Gale program (\cref{eq:eg}) at every round $t$ using either the current remaining budget, and future sizes replaced with their expectation (for \ResolveCE) or the initial budget with past sizes as their realizations and future sizes replaced with their expectations (for \CE).

We believe that these experiments both help highlight the theoretical performance of \HopeGuardrail, while simultaneously serving as an example of the practicality of our algorithm for broader stockpile allocation problems.

\subsection{Simulation Results}

\begin{figure}[!t]
\centering     
\subfigure[\textbf{Single-Synthetic}]{\label{fig:aa}\includegraphics[width=.8\linewidth]{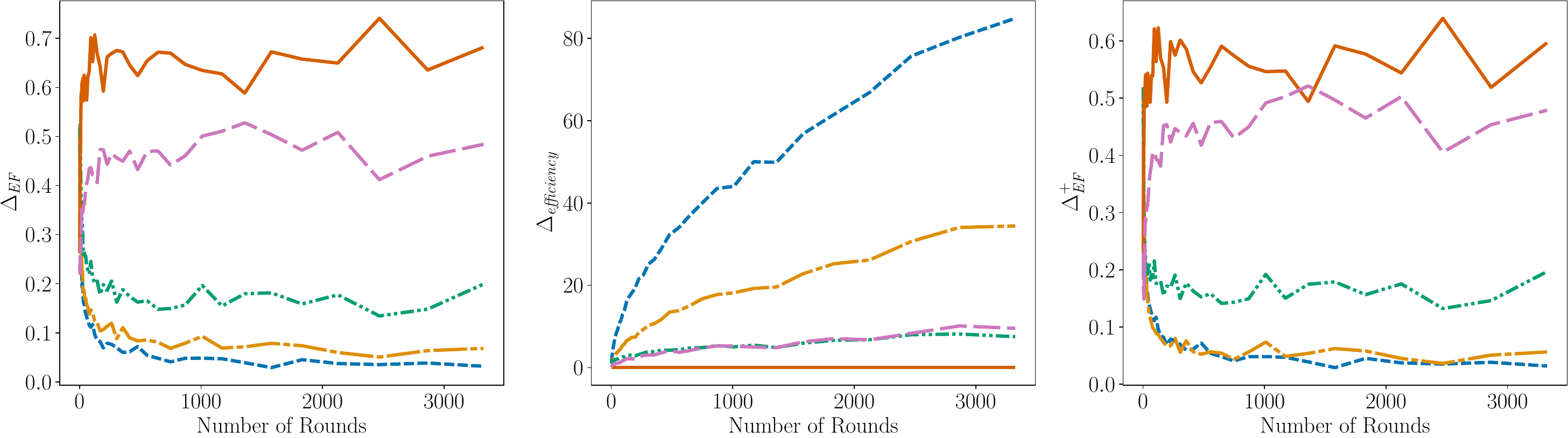}}
\subfigure[\textbf{Single-FBST}]{\label{fig:bb}\includegraphics[width=.8\linewidth]{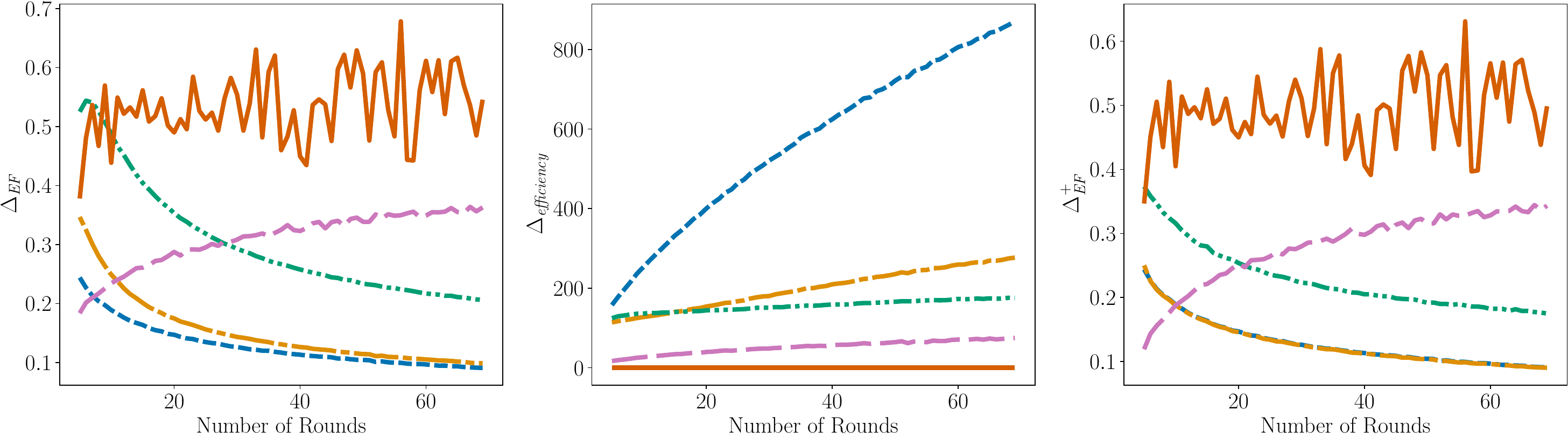}}
\subfigure[\textbf{Multi-Synthetic}]{\label{fig:cc}\includegraphics[width=.8\linewidth]{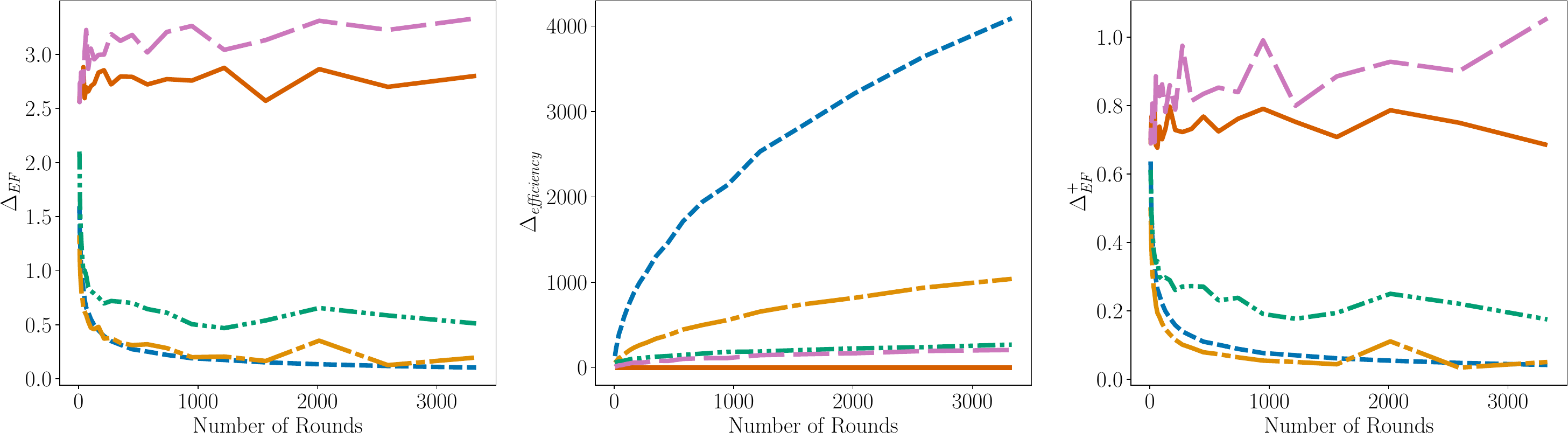}}
\subfigure[\textbf{Multi-FBST}]{\label{fig:dd}\includegraphics[width=.8\linewidth]{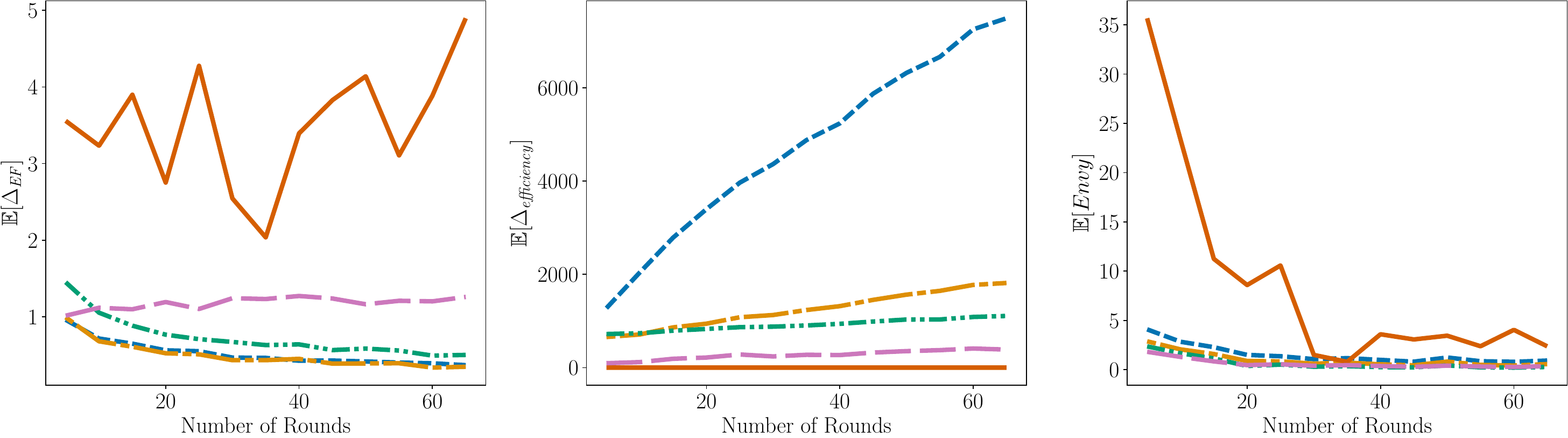}}
\centering \subfigure{\label{fig:ee}\includegraphics[width=\linewidth]{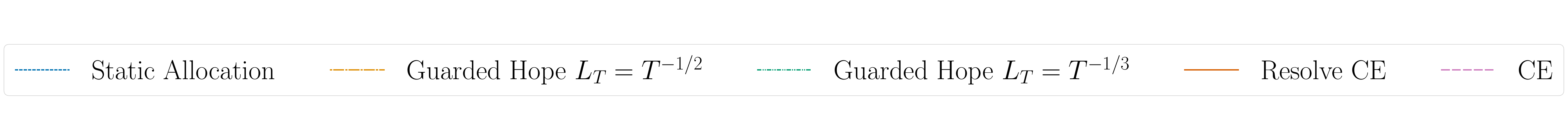}}
\caption{Comparison of \HopeGuardrail for $L_T = T^{1/2}$ and $L_T = T^{1/3}$, \StaticAllocation, \CE, and \ResolveCE on the four simulation settings as we vary the number of rounds $T$.}
\label{fig:line}
\end{figure}

\begin{figure}[!t]
\centering     
\subfigure[\textbf{Single-Synthetic}]{\label{fig:a} \centering \includegraphics[width=.24\linewidth]{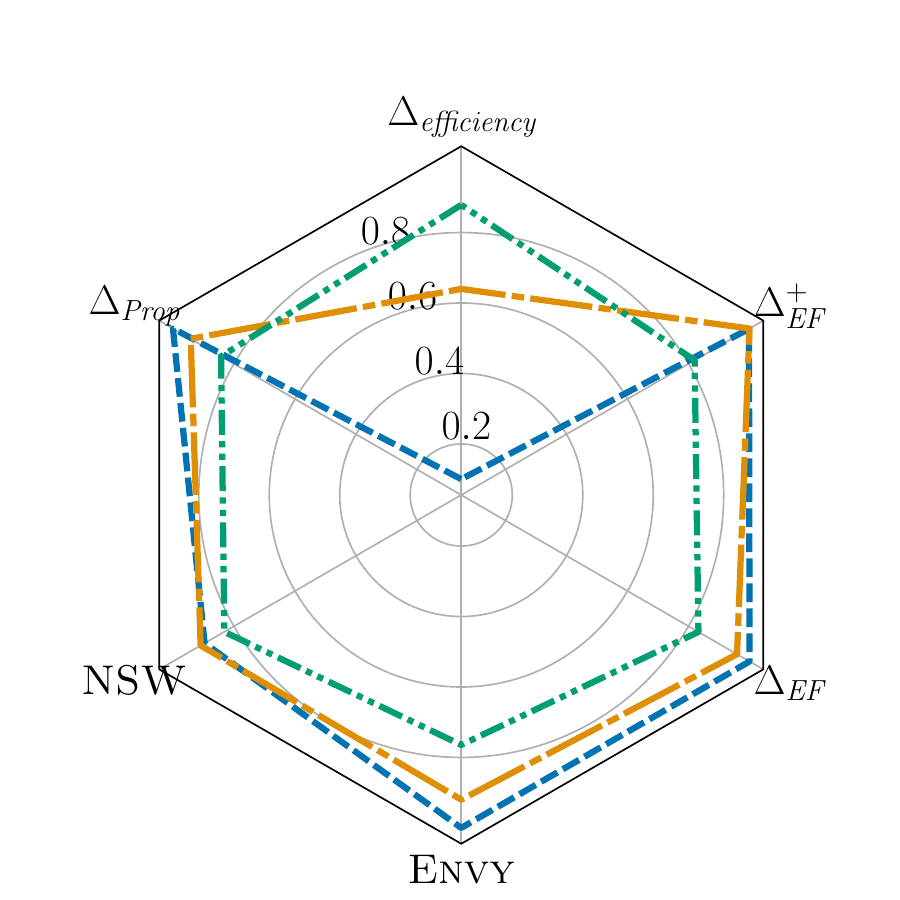} \hfill \includegraphics[width=.24\linewidth]{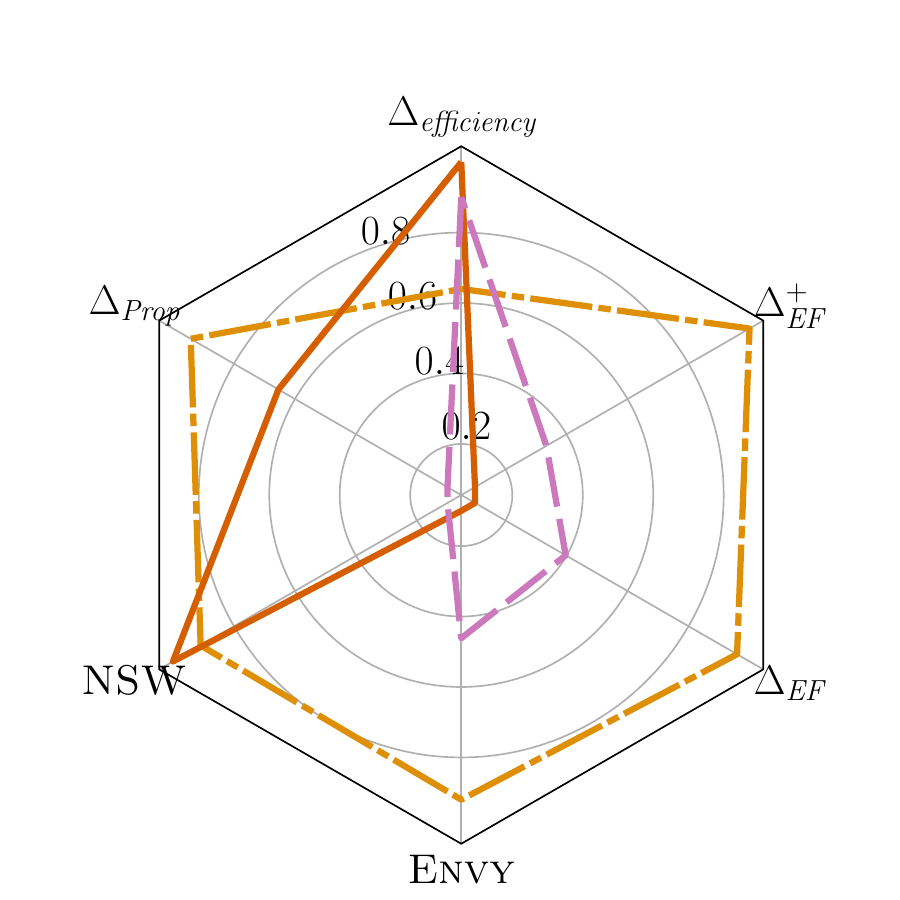}}
\subfigure[\textbf{Single-FBST}]{\label{fig:b}\centering \includegraphics[width=.24\linewidth]{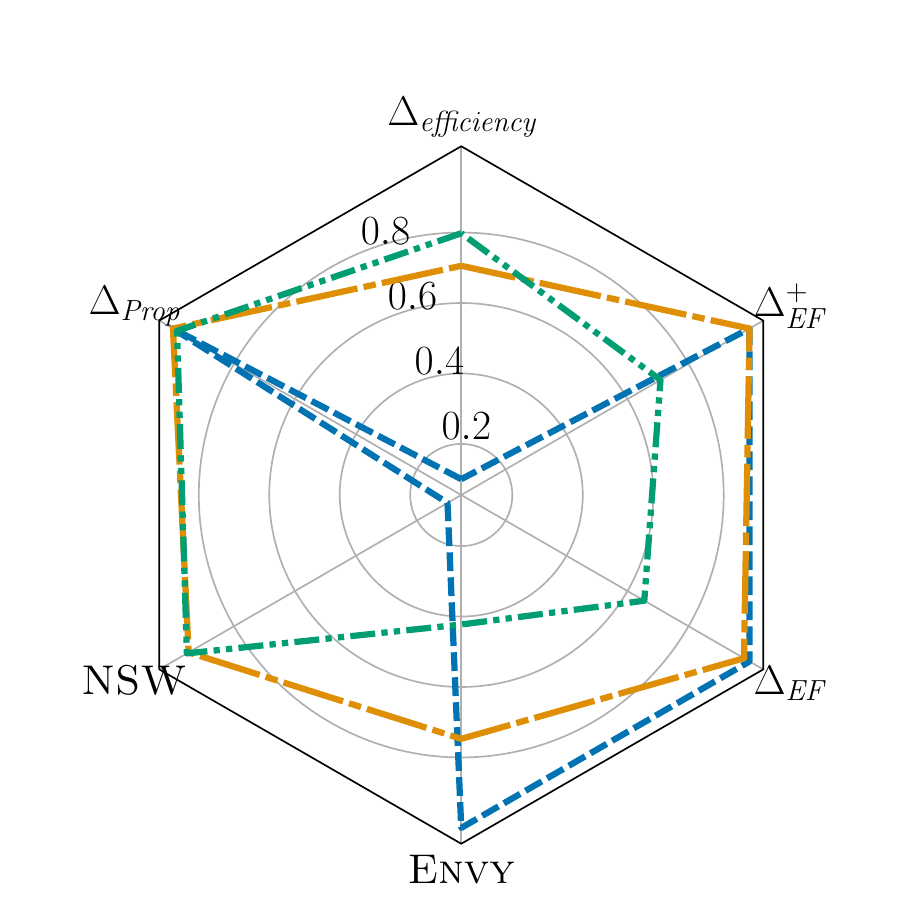} \hfill \includegraphics[width=.24\linewidth]{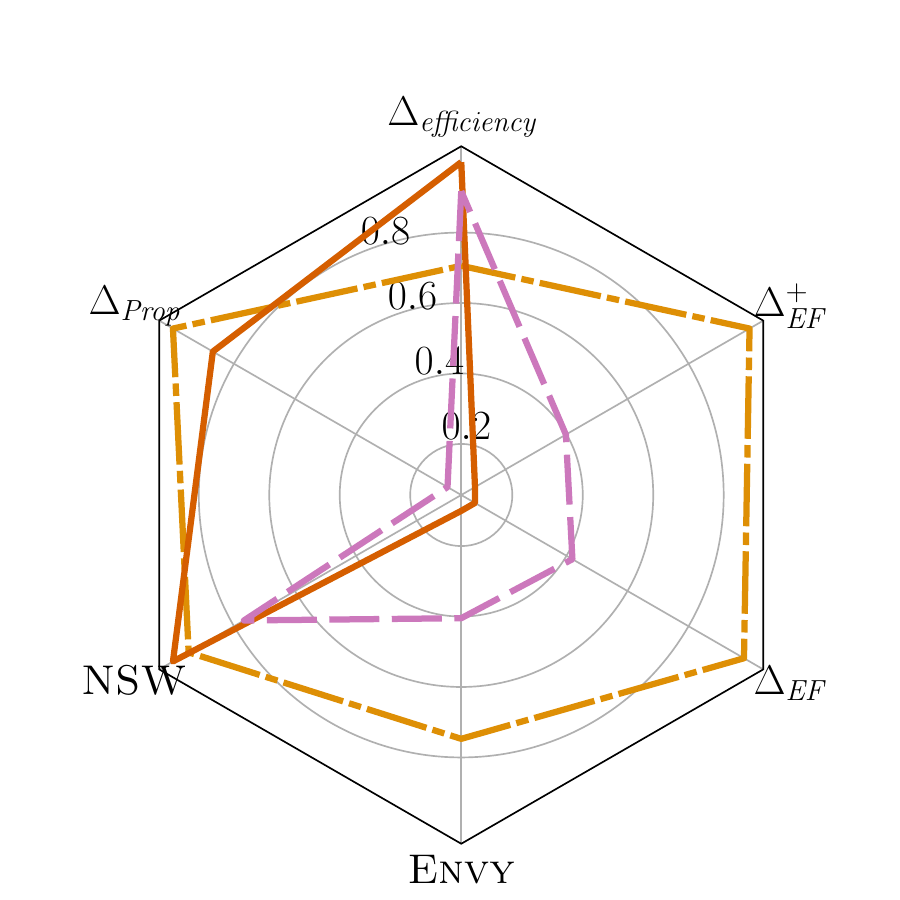}}
\subfigure[\textbf{Multi-Synthetic}]{\label{fig:c}\centering \includegraphics[width=.24\linewidth]{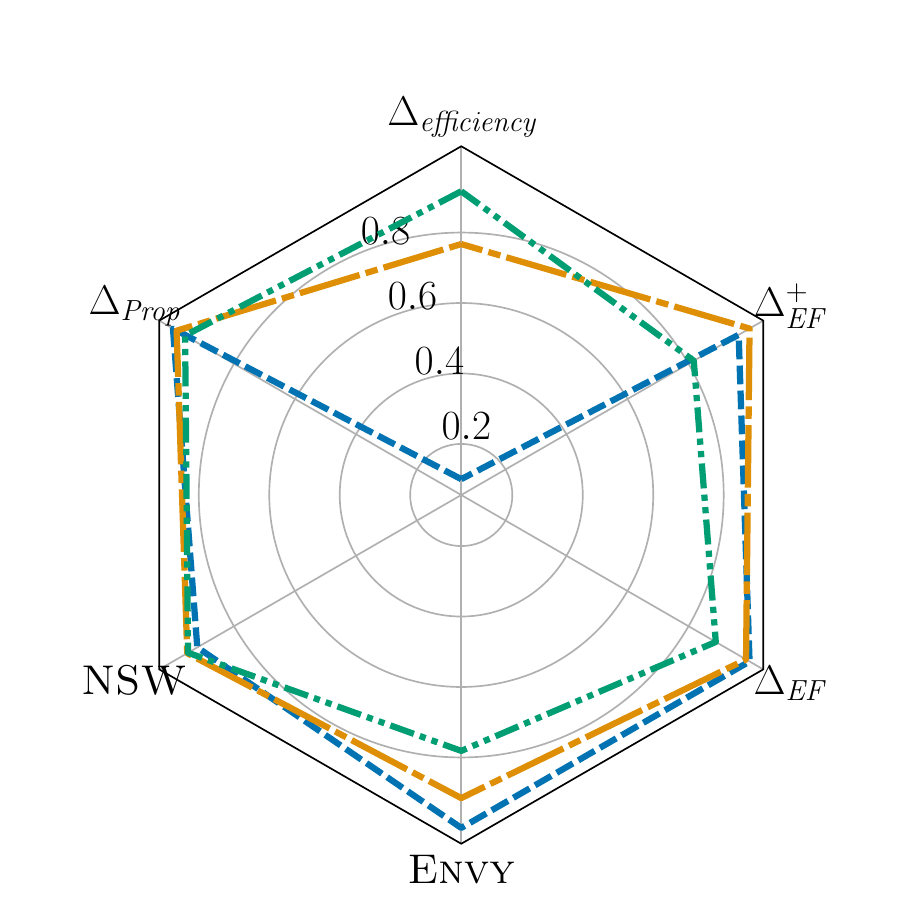} \hfill \includegraphics[width=.24\linewidth]{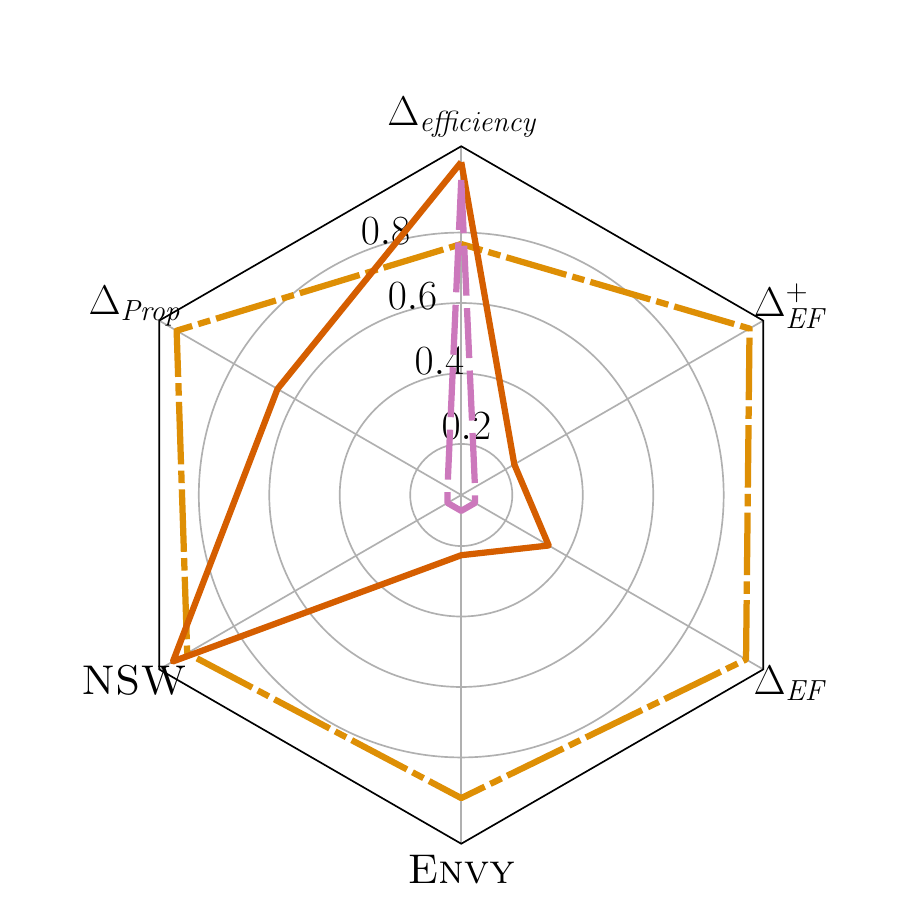}}
\subfigure[\textbf{Multi-FBST}]{\label{fig:d}\centering \includegraphics[width=.24\linewidth]{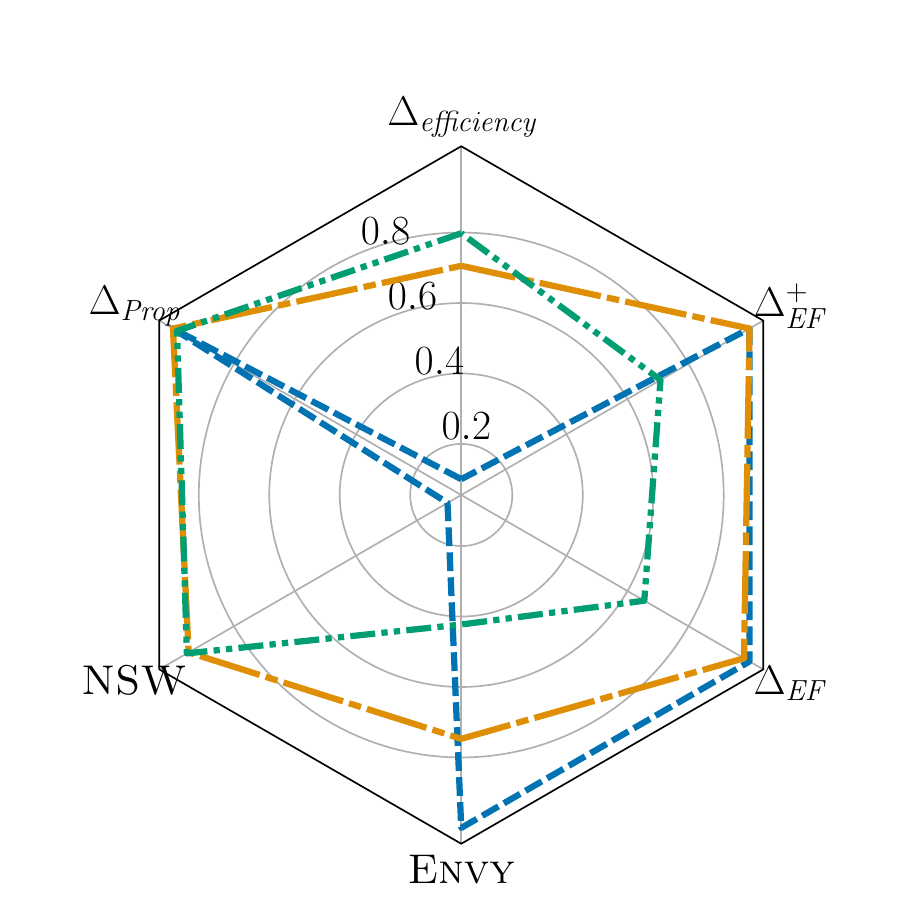} \hfill \includegraphics[width=.24\linewidth]{figures/radar_plots/multi_poisson_heuristic_radar.pdf}}
\centering \subfigure{\label{fig:e}\includegraphics[width=\linewidth]{figures/radar_plots/LABEL_ONLY.pdf}}
\caption{Comparison of \HopeGuardrail for $L_T = T^{1/2}$ and $L_T = T^{1/3}$, \StaticAllocation, \CE, and \ResolveCE, on the four simulation settings where $T = 100$. Values are normalized between $[0,1]$ to better highlight the performance of the algorithms, where larger values correspond to better performance.}
\label{fig:radar}
\end{figure}

For the setups described above, we numerically evaluate the performance of various policies.  In each simulation we set the total budget $B$ to be $\sum_{t, \theta} \Exp{N_{t, \theta}}$ so that the scaling of $\Bav$ remains as a constant as we vary the number of rounds $T$.  We compare the algorithms in terms of

\begin{itemize}
    \item $\Denv = \norm{u(X^{opt}_{t, \theta}, \theta) - u(X^{alg}_{t, \theta}, \theta)}_{\infty}$, the maximum difference between utility individuals receive for the two allocations as we scale the number of rounds $T$ (\cref{def:distance})
    \item $\Deff = \sum_k B_k - \sum_{\theta, t} N_{t, \theta} X_{t, \theta, k}^{alg}$, the leftover resources as we scale the number of rounds $T$ (\cref{def:distance})
    \item $\Denvplus = \max_{t, \theta} \Exp{|u(X^{opt}_{t, \theta}, \theta) - u(X^{alg}_{t, \theta}, \theta)|}$, the ex-ante maximum difference between utility individuals receive for the two allocations as we scale the number of rounds $T$
    \item $\Denvhind = \max_{t, \theta, t', \theta'} u(X_{t', \theta'}^{alg}, \theta) - u(X_{t, \theta}^{alg}, \theta)$, the maximum hindsight envy between any two agents (\cref{def:distance})
    \item $\Dprop = \max_{t, \theta} u(B / N, \theta) - u(X_{t, \theta}^{alg}, \theta)$, the maximum hindsight envy between an agent and equal allocation (\cref{def:distance})
    \item {$NSW = \left(\prod_{t, \theta} u(X_{t, \theta}^{alg}, \theta)^{N_{t, \theta}}\right)^{1 / \sum_{t, \theta} N_{t, \theta}}$, the Nash Social Welfare for the resulting allocation from the algorithm (\cref{eq:nsw}).}
\end{itemize}

For each of the simulations conducted we include three plots (as seen in \cref{fig:radar,fig:line}).  The first of which highlights the measure of $\Denv, \Deff$ and $\Denvplus$ as we vary the number of rounds $T$.  The second and third fixes a value of $T$ and compares the algorithms on each of the six metrics, where the values are normalized and larger scores corresponds to better performance.  We make four key observations:
\begin{enumerate}
    \item \textbf{\StaticAllocation vs \HopeGuardrail}: In \cref{fig:line} we see that our \HopeGuardrail algorithms (for varying values of $L_T$) are able to outperform \StaticAllocation in terms of efficiency, as our algorithms greedily allocate the upper threshold while ensuring budget compliance.
    \item \textbf{\HopeGuardrail for varying $L_T$}: In \cref{fig:line} we see that \HopeGuardrail for larger values of $L_T$ achieves better efficiency performance, but worse performance in terms of $\Denv$ (and illustrating \cref{thm:upper_bound}).
    \item \textbf{\HopeGuardrail vs Certainty Equivalence algorithms}: While \CE and \ResolveCE improve upon \HopeGuardrail in terms of $\Deff$ (as both algorithms use all available resources at the last round), we see that \HopeGuardrail is able to achieve a more nuanced tradeoff between the other metrics.
    \item \textbf{Multi-Objective Radar Plots}:  In \cref{fig:radar} we compare each of the algorithms on several different metrics.  Here we see that \HopeGuardrail performs competitively with respect to \StaticAllocation on the metrics of interest (in particular, $\Denv$ and $\Deff$), with the ability to tune the performance by varying the parameter $L_T$.
\end{enumerate}

These simulations help illustrate the theoretical performance of \HopeGuardrail (as outlined in \cref{thm:upper_bound}) and the Envy-Efficiency Uncertainty Principle (as in \cref{thm:envy-efficiency}).  Moreover, they help illustrate the truly multi-objective landscape of determining a fair allocation algorithm, and the benefits of varying \HopeGuardrail to tune $L_T$ and achieve a desired performance across all of the benchmarks.

    \section{Conclusion}
\label{sec:conclusion}
In this paper we considered the problem of dividing limited resources to individuals arriving over $T$ rounds, where each round can be thought of a distribution location.  In the offline setting (where the number of individuals arriving to each location is known), achieving a fair allocation scheme is found by maximizing the Nash Social Welfare objective subject to budget constraints.  However, in online settings, no online algorithm can achieve fairness properties ex-post.  
We instead consider the objective of minimizing $\Denv$ (the maximum difference between the utility individuals receive from the allocation made by the algorithm and the counterfactual optimal fair allocation in hindsight), $\Denvhind$ (the maximum difference between the utility individuals receive from the allocation made by the algorithm and the allocation given to a different individual), and $\Deff$ (the additive excess of resources).

We show that this objective leads to the envy-efficiency uncertainty principle, an exact characterization between the achievable $(\Denv, \Denvhind, \Deff)$ pairs.  In particular, our result shows that envy and efficiency must be inverseley proportional to one another.  With this analysis, we show that it leads to a simple algorithm, \HopeGuardrail, which is obtained by solving the Eisenberg-Gale program with unknown quantities replaced with their expectation to generate \emph{guardrails} used in the allocation, combined with an adaptive algorithm aimed at minimizing waste.  Through experiments we showed that \HopeGuardrail is able to obtain allocations which achieve any fairness - efficiency tradeoff, with desirable fairness properties compared to several benchmarks.

Several open questions remain, including extending the analysis to more general utility functions (including homothetic, another common model of preferences over resources).  We also believe much of the theoretical results apply to settings where the budget $B$ is instead a stochastic process, accounting for external donations and depletions of the resources independent of the allocations made by the algorithm.  Moreover, we leave the question of matching the upper and lower bounds in terms of problem dependent constants, and the issue of determining the \emph{schedule} to visit locations as future work.

\section*{Acknowledgments}

Part of this work was done while Sean Sinclair and Christina Yu were visiting the Simons Institute for the Theory of Computing for the semester on the Theory of Reinforcement Learning.  We also gratefully acknowledge funding from the NSF under grants ECCS-1847393, DMS-1839346, CCF-1948256, and CNS-1955997, and the ARL under grant W911NF-17-1-0094.  We would also like to thank the Food Bank of the Southern Tier and the Cornell Mathematical Contest in Modeling for their collaboration.
	
    \bibliographystyle{plain}
    {\bibliography{references}}

\begin{thebibliography}{10}

\bibitem{abebe2020roles}
Rediet Abebe, Solon Barocas, Jon Kleinberg, Karen Levy, Manish Raghavan, and
  David~G Robinson.
\newblock Roles for computing in social change.
\newblock In {\em Proceedings of the 2020 Conference on Fairness,
  Accountability, and Transparency}, pages 252--260, 2020.

\bibitem{aleksandrov2019monotone}
Martin Aleksandrov and Toby Walsh.
\newblock Monotone and online fair division.
\newblock In {\em Joint German/Austrian Conference on Artificial Intelligence
  (K{\"u}nstliche Intelligenz)}, pages 60--75. Springer, 2019.

\bibitem{aleksandrov2019online}
Martin Aleksandrov and Toby Walsh.
\newblock Online fair division: {A} survey.
\newblock In {\em The Thirty-Fourth {AAAI} Conference on Artificial
  Intelligence}, pages 13557--13562. {AAAI} Press, 2020.

\bibitem{aleksandrov2015online}
Martin~Damyanov Aleksandrov, Haris Aziz, Serge Gaspers, and Toby Walsh.
\newblock Online fair division: Analysing a food bank problem.
\newblock In {\em Twenty-Fourth International Joint Conference on Artificial
  Intelligence}, 2015.

\bibitem{alkaabneh2020unified}
Faisal Alkaabneh, Ali Diabat, and Huaizhu~Oliver Gao.
\newblock A unified framework for efficient, effective, and fair resource
  allocation by food banks using an approximate dynamic programming approach.
\newblock {\em Omega}, page 102300, 2020.

\bibitem{arrow2012social}
Kenneth~J Arrow.
\newblock {\em Social choice and individual values}, volume~12.
\newblock Yale university press, 2012.

\bibitem{azar2010allocate}
Yossi Azar, Niv Buchbinder, and Kamal Jain.
\newblock How to allocate goods in an online market?
\newblock In {\em European Symposium on Algorithms}, pages 51--62. Springer,
  2010.

\bibitem{aziz2016control}
Haris Aziz, Ildik{\'{o}} Schlotter, and Toby Walsh.
\newblock Control of fair division.
\newblock In Subbarao Kambhampati, editor, {\em Proceedings of the Twenty-Fifth
  International Joint Conference on Artificial Intelligence, {IJCAI} 2016, New
  York, NY, USA, 9-15 July 2016}, pages 67--73. {IJCAI/AAAI} Press, 2016.

\bibitem{gorokh2020fair}
Siddhartha Banerjee, Vasilis Gkatzelis, Artur Gorokh, and Billy Jin.
\newblock Online nash social welfare maximization via promised utilities.
\newblock {\em arXiv preprint arXiv:2008.03564}, 2020.

\bibitem{bansal2020online}
Nikhil Bansal, Haotian Jiang, Sahil Singla, and Makrand Sinha.
\newblock Online vector balancing and geometric discrepancy.
\newblock In {\em Proceedings of the 52nd Annual ACM SIGACT Symposium on Theory
  of Computing}, pages 1139--1152, 2020.

\bibitem{bateni2018fair}
MohammadHossein Bateni, Yiwei Chen, Dragos~Florin Ciocan, and Vahab Mirrokni.
\newblock Fair resource allocation in a volatile marketplace.
\newblock {\em Operations Research}, 70(1):288--308, 2022.

\bibitem{brookings}
Lauren Bauer.
\newblock About 14 million children in the us are not getting enough to eat.
\newblock {\em Brookings}, Jul 2020.

\bibitem{benade2018make}
Gerdus Benade, Aleksandr~M Kazachkov, Ariel~D Procaccia, and
  Christos-Alexandros Psomas.
\newblock How to make envy vanish over time.
\newblock In {\em Proceedings of the 2018 ACM Conference on Economics and
  Computation}, pages 593--610, 2018.

\bibitem{berry1941accuracy}
Andrew~C Berry.
\newblock The accuracy of the gaussian approximation to the sum of independent
  variates.
\newblock {\em Transactions of the american mathematical society},
  49(1):122--136, 1941.

\bibitem{bertsekas2012dynamic}
Dimitri Bertsekas.
\newblock {\em Dynamic programming and optimal control: Volume I}, volume~1.
\newblock Athena scientific, 2012.

\bibitem{bogomolnaia2001new}
Anna Bogomolnaia and Herv{\'e} Moulin.
\newblock A new solution to the random assignment problem.
\newblock {\em Journal of Economic Theory}, 100(2):295--328, 2001.

\bibitem{bogomolnaia2021fair}
Anna Bogomolnaia, Herv{\'e} Moulin, and Fedor Sandomirskiy.
\newblock On the fair division of a random object.
\newblock {\em Management Science}, 68(2):1174--1194, 2022.

\bibitem{brams1995envy}
Steven~J Brams and Alan~D Taylor.
\newblock An envy-free cake division protocol.
\newblock {\em The American Mathematical Monthly}, 102(1):9--18, 1995.

\bibitem{brams1996fair}
Steven~J Brams and Alan~D Taylor.
\newblock {\em Fair Division: From cake-cutting to dispute resolution}.
\newblock Cambridge University Press, 1996.

\bibitem{chung2012chernoff}
Kai-Min Chung, Henry Lam, Zhenming Liu, and Michael Mitzenmacher.
\newblock Chernoff-hoeffding bounds for markov chains: Generalized and
  simplified.
\newblock page 124, 2012.

\bibitem{cole_2013}
Richard Cole, Vasilis Gkatzelis, and Gagan Goel.
\newblock Mechanism design for fair division: Allocating divisible items
  without payments.
\newblock In {\em EC 2013}, 2013.

\bibitem{devanur2002market}
Nikhil~R Devanur, Christos~H Papadimitriou, Amin Saberi, and Vijay~V Vazirani.
\newblock Market equilibrium via a primal-dual-type algorithm.
\newblock In {\em The 43rd Annual IEEE Symposium on Foundations of Computer
  Science, 2002. Proceedings.}, pages 389--395. IEEE, 2002.

\bibitem{donahue2020fairness}
Kate Donahue and Jon Kleinberg.
\newblock Fairness and utilization in allocating resources with uncertain
  demand.
\newblock In {\em Proceedings of the 2020 Conference on Fairness,
  Accountability, and Transparency}, pages 658--668, 2020.

\bibitem{eisenberg1961aggregation}
Edmund Eisenberg.
\newblock Aggregation of utility functions.
\newblock {\em Management Science}, 7(4):337--350, 1961.

\bibitem{eisenhandler2019humanitarian}
Ohad Eisenhandler and Michal Tzur.
\newblock The humanitarian pickup and distribution problem.
\newblock {\em Operations Research}, 67(1):10--32, 2019.

\bibitem{elzayn2019fair}
Hadi Elzayn, Shahin Jabbari, Christopher Jung, Michael Kearns, Seth Neel, Aaron
  Roth, and Zachary Schutzman.
\newblock Fair algorithms for learning in allocation problems.
\newblock In {\em Proceedings of the Conference on Fairness, Accountability,
  and Transparency}, pages 170--179, 2019.

\bibitem{fbst}
{Food Bank of the Southern Tier of New York}.
\newblock \url{https://www.foodbankst.org/}, 2020.

\bibitem{10.1145/2764468.2764495}
Eric Friedman, Christos-Alexandros Psomas, and Shai Vardi.
\newblock Dynamic fair division with minimal disruptions.
\newblock In {\em Proceedings of the Sixteenth ACM Conference on Economics and
  Computation}, EC ’15, page 697–713, New York, NY, USA, 2015. Association
  for Computing Machinery.

\bibitem{friedman_2017}
Eric Friedman, Christos-Alexandros Psomas, and Shai Vardi.
\newblock Controlled dynamic fair division.
\newblock In {\em Proceedings of the 2017 ACM Conference on Economics and
  Computation}, EC '17, page 461–478, New York, NY, USA, 2017. Association
  for Computing Machinery.

\bibitem{gerding2019fair}
Enrico~H. Gerding, Alvaro Perez-Diaz, Haris Aziz, Serge Gaspers, Antonia Marcu,
  Nicholas Mattei, and Toby Walsh.
\newblock Fair online allocation of perishable goods and its application to
  electric vehicle charging.
\newblock In {\em Proceedings of the Twenty-Eighth International Joint
  Conference on Artificial Intelligence, {IJCAI-19}}, pages 5569--5575.
  International Joint Conferences on Artificial Intelligence Organization, 7
  2019.

\bibitem{ghodsi2011dominant}
Ali Ghodsi, Matei Zaharia, Benjamin Hindman, Andy Konwinski, Scott Shenker, and
  Ion Stoica.
\newblock Dominant resource fairness: Fair allocation of multiple resource
  types.
\newblock In {\em Nsdi}, volume~11, pages 24--24, 2011.

\bibitem{ijcai2019-49}
Jiafan He, Ariel~D. Procaccia, Alexandros Psomas, and David Zeng.
\newblock Achieving a fairer future by changing the past.
\newblock In {\em Proceedings of the Twenty-Eighth International Joint
  Conference on Artificial Intelligence, {IJCAI-19}}, pages 343--349.
  International Joint Conferences on Artificial Intelligence Organization, 7
  2019.

\bibitem{jaberi2014optimal}
Majid Jaberi-Douraki and Seyed~M Moghadas.
\newblock Optimal control of vaccination dynamics during an influenza epidemic.
\newblock {\em Mathematical Biosciences \& Engineering}, 11(5):1045, 2014.

\bibitem{kalinowski2013social}
Thomas Kalinowski, Nina Narodytska, and Toby Walsh.
\newblock A social welfare optimal sequential allocation procedure.
\newblock In {\em Twenty-Third International Joint Conference on Artificial
  Intelligence}, 2013.

\bibitem{kash2014no}
Ian Kash, Ariel~D Procaccia, and Nisarg Shah.
\newblock No agent left behind: Dynamic fair division of multiple resources.
\newblock {\em Journal of Artificial Intelligence Research}, 51:579--603, 2014.

\bibitem{kleinberg2016inherent}
Jon Kleinberg, Sendhil Mullainathan, and Manish Raghavan.
\newblock Inherent trade-offs in the fair determination of risk scores.
\newblock {\em arXiv preprint arXiv:1609.05807}, 2016.

\bibitem{kulish_2020}
Nicholas Kulish.
\newblock 'never seen anything like it': Cars line up for miles at food banks,
  Apr 2020.

\bibitem{lien2014sequential}
Robert~W Lien, Seyed~MR Iravani, and Karen~R Smilowitz.
\newblock Sequential resource allocation for nonprofit operations.
\newblock {\em Operations Research}, 62(2):301--317, 2014.

\bibitem{lupkin_2020}
Sydney Lupkin.
\newblock How feds decide on remdesivir shipments to states remains mysterious,
  Aug 2020.

\bibitem{manshadi2021fair}
Vahideh Manshadi, Rad Niazadeh, and Scott Rodilitz.
\newblock Fair dynamic rationing.
\newblock {\em Available at SSRN 3775895}, 2021.

\bibitem{mattei2017mechanisms}
Nicholas Mattei, Abdallah Saffidine, and Toby Walsh.
\newblock Mechanisms for online organ matching.
\newblock In {\em IJCAI}, pages 345--351, 2017.

\bibitem{mattei2018fairness}
Nicholas Mattei, Abdallah Saffidine, and Toby Walsh.
\newblock Fairness in deceased organ matching.
\newblock In {\em Proceedings of the 2018 AAAI/ACM Conference on AI, Ethics,
  and Society}, pages 236--242, 2018.

\bibitem{moulin2004fair}
Herv{\'e} Moulin.
\newblock {\em Fair division and collective welfare}.
\newblock MIT press, 2004.

\bibitem{nisan_roughgarden_tardos_vazirani_2007}
Noam Nisan, Tim Roughgarden, {\'{E}}va Tardos, and Vijay~V. Vazirani, editors.
\newblock {\em Algorithmic Game Theory}.
\newblock Cambridge University Press, 2007.

\bibitem{orgut2016achieving}
Irem~Sengul Orgut, Luther~G Brock~III, Lauren~Berrings Davis, Julie~Simmons
  Ivy, Steven Jiang, Shona~D Morgan, Reha Uzsoy, Charlie Hale, and Earline
  Middleton.
\newblock Achieving equity, effectiveness, and efficiency in food bank
  operations: Strategies for feeding america with implications for global
  hunger relief.
\newblock In {\em Advances in managing humanitarian operations}, pages
  229--256. Springer, 2016.

\bibitem{prendergast2017food}
Canice Prendergast.
\newblock How food banks use markets to feed the poor.
\newblock {\em Journal of Economic Perspectives}, 31(4):145--62, 2017.

\bibitem{procaccia2013cake}
Ariel~D Procaccia.
\newblock Cake cutting: not just child's play.
\newblock {\em Communications of the ACM}, 56(7):78--87, 2013.

\bibitem{sengul2017modeling}
Irem Sengul~Orgut, Julie Ivy, and Reha Uzsoy.
\newblock Modeling for the equitable and effective distribution of food
  donations under stochastic receiving capacities.
\newblock {\em IISE Transactions}, 49(6):567--578, 2017.

\bibitem{shadmi2020health}
Efrat Shadmi, Yingyao Chen, In{\^e}s Dourado, Inbal Faran-Perach, John Furler,
  Peter Hangoma, Piya Hanvoravongchai, Claudia Obando, Varduhi Petrosyan,
  Krishna~D Rao, et~al.
\newblock Health equity and {COVID}-19: global perspectives.
\newblock {\em International journal for equity in health}, 19(1):1--16, 2020.

\bibitem{sugden1984fairness}
Robert Sugden.
\newblock Is fairness good? {A} critique of {V}arian's theory of fairness.
\newblock {\em Nous}, pages 505--511, 1984.

\bibitem{varian1973equity}
Hal~R. Varian.
\newblock {Equity, envy, and efficiency}.
\newblock {\em Journal of Economic Theory}, 9(1):63--91, September 1974.

\bibitem{varian1976two}
Hal~R Varian.
\newblock Two problems in the theory of fairness.
\newblock {\em Journal of Public Economics}, 5(3-4):249--260, 1976.

\bibitem{vera2019bayesian}
Alberto Vera and Siddhartha Banerjee.
\newblock The bayesian prophet: A low-regret framework for online decision
  making.
\newblock {\em ACM SIGMETRICS Performance Evaluation Review}, 47(1):81--82,
  2019.

\bibitem{walsh2011online}
Toby Walsh.
\newblock Online cake cutting.
\newblock In {\em International Conference on Algorithmic Decision Theory},
  pages 292--305. Springer, 2011.

\bibitem{yi2015fairness}
Ming Yi and Achla Marathe.
\newblock Fairness versus efficiency of vaccine allocation strategies.
\newblock {\em Value in Health}, 18(2):278--283, 2015.

\bibitem{zeng2019fairness}
David Zeng and Alexandros Psomas.
\newblock Fairness-efficiency tradeoffs in dynamic fair division.
\newblock In {\em Proceedings of the 21st ACM Conference on Economics and
  Computation}, EC '20, page 911–912, New York, NY, USA, 2020. Association
  for Computing Machinery.

\end{thebibliography}
    
    \newpage
    \onecolumn
    \appendix

    \ifdefined\informs 
\else
\section{Table of Notation}
\label{app:notation}
\fi

\ifdefined\informs \renewcommand{\arraystretch}{.75}
\else \renewcommand{\arraystretch}{1.2} \fi
\begin{table*}[h!]
\begin{tabular}{l|l}
\textbf{Symbol} & \textbf{Definition} \\ \hline
\multicolumn{2}{c}{Problem setting specifications}\\
\hline
$K$ & Number of resources \\
$\Theta$ & Set of types for individuals \\
$T$ & Total number of rounds \\
$B_k$ & Budget for resource $k \in [K]$ \\
$\theta$ & Specification of an individual's type in $\Theta$ \\
$u(x, \theta) : \mathbb{R}^{k} \times \Theta \rightarrow \mathbb{R}_+$ & Utility function for individuals of type $\theta$, here assumed to be $\langle x, w_\theta \rangle$ \\
$N_{t, \theta}$ & Number of individuals of type $\theta$ in round $t$ \\
$\F_t$ & Known distribution over $(N_{t, \theta})_{\theta \in \Theta}$ \\
$N_{\geq t, \theta}$ & $\sum_{t' \geq t} N_{t', \theta}$ \\
$N_{\theta}$ & $\sum_{t' \geq 1} N_{t', \theta}$ \\
$\sigma_{t, \theta}$ & $\Var{N_{t, \theta}}$ \\
$\rho_{t, \theta}$ & Bound on $|N_{t, \theta} - \Exp{N_{t, \theta}}| \leq \rho_{t, \theta}$ \\
$\mu_{t, \theta}$ & $\Exp{N_{t, \theta}}$ \\
$\sigma_{min}^2, \sigma_{max}^2, \rho_{min},$  & The respective maximum and minimum value of each quantity \\
$\rho_{max}, \mu_{min}, \mu_{max}$ & \\
$\Bav$ & $\sum_k B_k / \sum_{\theta \in \Theta} \Exp{N_{\theta}}$ \\
$X^{opt}, X^{alg}$ & Optimal fair allocation in hindsight and allocation by algorithm \\
$\Denv$ & $\max_{t \in [T], \theta \in \Theta} \norm{u(X^{alg}_{t, \theta}, \theta) - u(X^{opt}_{t, \theta}, \theta)}_{\infty}$ \\
$\Denvhind$ & $\max_{t, t' \in [T]^2, \theta, \theta' \in \Theta^2} u(X_{t', \theta'}^{alg}, \theta) - u(X_{t, \theta}^{alg}, \theta)$ \\
$\Deff$ & $\sum_{k} B_k - \sum_{\theta \in \Theta} \sum_{t} N_{t, \theta} X_{t, \theta, k}^{alg}$ \\
\hline
\multicolumn{2}{c}{Algorithm specification}\\
\hline
$B_t^{alg}$ & Budget available to the algorithm at start of round $t$ \\
$\conf_{t, \theta}$ & Confidence bound on $N_{> t, \theta}$ \\
$\multconf$ & Multiplicative confidence bound, i.e. $\max_{\theta} \frac{\conf_{0, \theta}}{\Exp{N_{\geq 1, \theta}}}$ \\
$\overline{n}_{\theta}, \underline{n}_{\theta}$ & $\Exp{N_{\theta}}(1 + \multconf), \Exp{N_{\theta}}(1 - c)$ \\
$\overline{X}_{\theta}, \underline{X}_{\theta}$ & Optimistic and pesimistic solutions to \cref{eq:eg} for $\underline{n}_{\theta}$ and $\overline{n}_\theta$ respectively \\
$\mathcal{E}$ & `Good event' set, where the concentration inequalities hold \\
$L_T$ & Desired bound on $\Denv, \Denvhind$ \\
\hline
\multicolumn{2}{c}{Additional notation used for theory}\\
\hline
$\Phi(\cdot)$ & Standard normal CDF \\
$\bar{\rho}_{\theta}, \bar{\sigma}^2_{\theta}, \bar{\mu}_{\theta}$ & Averages of these quantities, i.e. $\frac{1}{T} \sum_{t} \rho_{t, \theta}$, $\frac{1}{T} \sum_{t} \sigma_{t, \theta}^2$, $\frac{1}{T} \sum_{t} \Exp{N_{t, \theta}}$ \\
\hline
\end{tabular}
\caption{List of common notation}
\label{table:notation}
\end{table*}

\ifdefined\informs 
\else
\section{Algorithm Pseudocode}
\fi

\begin{algorithm}[!h]
\DontPrintSemicolon 
\KwIn{Budget $B_1^{alg} = B$, $(\Exp{N_{\theta}})_{\theta \in \Theta}$, confidence terms $(\conf_{t, \theta})_{\theta \in \Theta}$, and a desired bound on envy $L_T$}
\KwOut{An allocation $X^{alg} \in \mathbb{R}^{T \times |\Theta| \times K}$}
Solve for $\overline{X} = x(\underline{n}_{\theta})$ as the solution to \cref{eq:eg} where $\underline{n}_{\theta} = \Exp{N_{\theta}}(1 - c)$ where $c$ is defined via \cref{thm:concentration} \;
Solve for $\underline{X} = x(\overline{n}_{\theta})$ as the solution to \cref{eq:eg} where $\overline{n}_{\theta} = \Exp{N_{\theta}}(1 + \multconf_1)$ and $\multconf_1 = \max_\theta \frac{\conf_\theta}{\Exp{N_\theta}}$. \tcp*[h]{solve for guardrails}\;
\For{rounds $t = 1, \ldots, T$}{ 
    \For{each resource $k \in [K]$}{
    \uIf(\tcp*[h]{insufficient budget to allocate lower guardrail}){$B_{t,k}^{alg} < \sum_{\theta \in \Theta} N_{t, \theta} \underline{X}_{\theta, k}$}{
        Set $X_{t, \theta, k}^{alg} = \frac{B_{t,k}^{alg}}{\sum_{\theta \in \Theta} N_{t, \theta}}$ for each $\theta \in \Theta$\;}
    \uElseIf(\tcp*[h]{use upper guardrail}){$B_{t,k}^{alg} - \sum_{\theta} N_{t, \theta} \overline{X}_{\theta, k} \geq \sum_{\theta \in \Theta} \underline{X}_{\theta, k} (\Exp{N_{> t, \theta}} + \conf_{t, \theta})$}{
        Set $X_{t, \theta, k}^{alg} = \overline{X}_{\theta, k}$ for each $\theta \in \Theta$\;}
    \uElse(\tcp*[h]{use lower guardrail}){
        Set $X_{t, \theta, k}^{alg} = \underline{X}_{\theta, k}$ for each $\theta \in \Theta$\;}
    }
        Update $B_{t+1}^{alg} = B_t^{alg} - \sum_{\theta \in \Theta} N_{t, \theta} X_{t, \theta}^{alg}$
}
\Return{$X^{alg}$}
	\caption{Histogram of Preference Estimates with Guardrails (\HopeGuardrail)}
	\label{alg:brief}
\end{algorithm}
    \section{Discussion on Varian's Fairness}
\label{app:varian}

In this section we discuss some of the limitations of our definitions as well as compare our individualized envy freeness guarantees from \cref{thm:informal} to competitive ratio guarantees from the literature.

\subsection{Limitation of Fair Allocations}

Economists, computer scientists, and people in the operations research literature have become increasingly interested in questions of fairness~\citep{sugden1984fairness,varian1973equity,varian1976two}.  One particular concept of fairness which has gained wide circulation due to its ease in computability is the so-called notion of `Varian fairness' pioneered by Hal Varian in the 1970s taken in \cref{def:fairness}.  This theory of fairness provides three criteria for judging a given allocation of resources: \textit{envy-freeness}, \textit{Pareto-efficiency}, and \textit{proportionality}, all defined with respect to utilities an individual has for different allocations.  These criteria serve as more of a classification than an optimization perspective, as each of them merely provides a true/false criteria for an allocation to \textit{satisfy} fairness rather than a way an allocation can \textit{approach} fairness.  Numerous other researchers have proposed other definitions of fairness, including $\alpha$-fairness obtained by maximizing~\citep{moulin2004fair,arrow2012social}:
$\sum_{i=1}^n \text{sign}(\alpha)u(X_i, \theta_i)^{\alpha}.$
In this definition, taking $\alpha = 1$ recovers utilitarian welfare, or maximizing the sum of utilities.  Taking $\alpha \rightarrow - \infty$ also recovers the leximin objective, and $\alpha \rightarrow \infty$ the leximax.

One primary critique of `Varian fairness' is that a Varian fair allocation may not exist at all.  Moreover, the implication is that \textit{if} a Varian-fair allocation exists, then it has special merit.  While we specifically consider settings where a `Varian fair' allocation always exists (and is remarkably found as a result of optimizing the Nash Social Welfare objective), it is important to consider some of the several downsides of this model.

\medskip \noindent \textbf{Comparison of Individuals}: Paramount to Varian's definition of proportionality and envy-freeness is that each individual is treated symmetrically.  This ignores systemic factors that inhibit particular individual's access to the resource.  

\medskip \noindent \textbf{Scale Invariance}: The concept of fairness is strictly operational, in the sense that it requires no more information than what is contained in an individual's utility function.  Settings like matching students to local schools via school choice require definitions which measure the `utility of replacement'~\citep{abebe2020roles}.  As an example, a student with preferences (School A, School B, School C) in descending order gets matched to School B.  How can we measure the overall gain to society when the student is instead matched to School A, or School C in comparison to another students list of preferences?  In Varian's definition of fairness, utility functions are only used to exhibit an ordering on preferences, rather than a relative value on different outcomes.  

\medskip

We believe the settings considered in \cref{section: examples} are well suited to Varian's model on fairness.  In the example motivated with the Food Bank of the Southern Tier of New York, rounds correspond to distribution sites, whether that be a soup kitchen, a drop-off location for the mobile food bank, etc.  In these settings, individuals have use for all resources with strictly increasing utility with respect to the resource allocated.  This motivates using scale invariant measures, as every individual will be able to use all available food allocated to them.  Considering processor assignment in cloud computing platforms, each individual request coming in should be treated independently and symmetrically.

\subsection{Competitive Ratio or Individual Guarantees}
\label{app:ce}

One approach on obtaining fairness guarantees for an online algorithm could be in the form of a competitive ratio.  These results find allocation algorithms $X^{alg}$ to which you can construct a bound on the competitive ratio for the Nash Social Welfare (or its logarithm):
\begin{align*}
\quad\quad \frac{\prod_{t,\theta} u(X_{t,\theta}^{alg}, \theta)^{N_{t,\theta}}}{\prod_{t,\theta} u(X_{t,\theta}^{opt}, \theta)^{N_{t,\theta}}} \quad\quad \text{ or } \quad\quad \frac{\sum_{t,\theta} N_{t,\theta} \log(u(X_{t,\theta}^{alg}, \theta))}{\sum_{t,\theta}N_{t,\theta} \log(u(X_{t,\theta}^{opt}, \theta))} .
\end{align*}

While theoretically interesting, these results provide no immediate individual fairness guarantees.  The motivation for the Eisenberg-Gale program arises from the fact that \textit{fairness is a byproduct}.  To some extent, the actual objective value of an allocation is meaningless, and the objective is only taken as it serves as a proxy to obtain fair allocations.  In many applications of resource allocation, stakeholders are more interested in obtaining individualized guarantees than global guarantees on social welfare.  This motivated our alternative approach of designing algorithms with individualized guarantees in mind.  In fact, our guarantees are related to the \emph{fairness ratios} developed in \citep{friedman_2017,cole_2013,10.1145/2764468.2764495}, presented additively instead of multiplicatively.

\begin{table}[!h]
\caption{Table comparing the Eisenberg-Gale formulations used in \CE, \ResolveCE, and \HopeGuardrail.  The first column corresponds to whether the algorithms require resolving the EG program across each round $t$.  The second column indicates the budget value which is fed into the EG program, and the third column the number of arrivals for each type.  See \cref{app:ce} for further discussion.}
\label{tab:ce_compare}
\setlength\tabcolsep{0pt} 
\centering
\begin{tabular*}{\columnwidth }{@{\extracolsep{\fill}}rccc}
\toprule
Algorithm & Resolve &  Budget & Arrivals\\
\midrule
\HopeGuardrail & \xmark & $B$ & $\Exp{N_\theta}$ \\
\CE & \checkmark & B & $N_{\leq t, \theta} + \Exp{N_{> t, \theta}}$ \\
\ResolveCE & \checkmark & $B_t^{alg}$ & $N_{t, \theta} + \Exp{N_{> t, \theta}}$ \\
\bottomrule
\end{tabular*}
\end{table}

\section{Certainty Equivalent Formulation}
\label{app:ceq}

In general sequential stochastic optimization problem, certainty equivalency formulations are widely used to design heuristic algorithms with good performance guarantees~\cite{bertsekas2012dynamic}.  These policies approximate the offline optimization problem by replacing stochastic unknown quantities (the arrivals $N_{t, \theta}$) with fixed determinsitic values (typically taken to be their means).
Here we describe the Certainty Equivalence (\CE) and Certainty Equivalence + Resolve (\ResolveCE), which respectively solve the Eisenberg-Gale program at every round $t$ using the current remaining budget and future sizes replaced with their expectation (in the case of the \ResolveCE), and the initial budget with past sizes as their realizations and future sizes replaced with their expectations (in the case of \CE).  See~\cref{tab:ce_compare} for a comparison between these algorithms and \HopeGuardrail. 

Interestingly, we see in \cref{sec:experiments} that these algorithms do not perform well for our fairness metrics of interest.  One potential reason is that these algorithms are designed to compete against the stochastic objective function (here chosen to be the Nash Social Welfare or its logarithm in the case of the Eisenberg-Gale program).  However, noticed in \cref{app:varian} approximate maximizers do not necessarily lead to fair outcomes. We also note that these algorithms require frequent solves of the Eisenberg-Gale program, and can potentially lead to larger inter-round envy since the allocations can vary more across rounds $t$ than our simple guardrail approach.

We start by considering the formulation of \cref{eq:eg} for values $n_\theta$ for the number of arrivals of type $\theta$, and budget $B$:
\begin{align}
\label{eq:eg_ce_appendix}
\max_{X \in \mathbb{R}_+^{\Theta \times K}} \, \sum_{\theta \in \Theta} n_\theta  \log\left( u(X_{\theta}, \theta) \right) 
\quad \quad \text{ s.t. } \, \sum_{\theta \in \Theta} n_\theta X_{\theta} \leq B 
\end{align}
Denote the optimal solution to this problem as $x((n_\theta)_{\theta \in \Theta}, B) \in \mathbb{R}^{\Theta \times K}$.

\subsection{Certainty Equivalence (\CE)}

We start off by describing \CE.  At every iteration, \CE solves the Eisenberg-Gale program using the initial budget, and arrivals as their realizations in the past with future sizes replaced with their expectation.  More concretely, the \CE algorithm works by allocating at round $t$:
\begin{align*}
    X_{t, \theta}^{alg} = \begin{cases}
                \frac{B_t^{alg}}{\sum_\theta N_{t, \theta}} & \text{ if } B_t^{alg} < \sum_{\theta} N_{t, \theta} x((N_{\leq t, \theta} + \Exp{N_{> t, \theta}})_{\theta \in \Theta}, B)_\theta \\
               x((N_{\leq t, \theta} + \Exp{N_{> t, \theta}})_{\theta \in \Theta}, B)_\theta & \text{ otherwise}
            \end{cases}
\end{align*}
In the first case statement we check whether the current solution estimate $x((N_{\leq t, \theta} + \Exp{N_{> t, \theta}})_{\theta \in \Theta}, B)$, i.e. the solution to the EG program with the initial budget, observed past arrivals, and mean future arrivals, is feasible.  If the current remaining budget is insufficient to allocate this amount, the algorithm divides all remaining resources evenly.  Otherwise, we allocate this solution earmarked by the EG program.

\subsection{Certainty Equivalence with Resolve (\ResolveCE)}

The \ResolveCE algorithm works similarly.  At every iteration, \ResolveCE solves the Eisenberg-Gale program using the current budget, and arrivals as the current arrival $N_{t, \theta}$ with future sizes replaced with their expectation.  More concretely, the \ResolveCE algorithm works by allocating at round $t$:
\begin{align*}
    X_{t,\theta}^{alg} = \begin{cases}
                \frac{B_t^{alg}}{\sum_\theta N_{t, \theta}} & \text{ if } B_t^{alg} < \sum_{\theta} N_{t, \theta} x((N_{t, \theta} + \Exp{N_{> t, \theta}})_{\theta \in \Theta}, B_t^{alg})_\theta \\
               x((N_{t, \theta} + \Exp{N_{> t, \theta}})_{\theta \in \Theta}, B_t^{alg})_\theta & \text{ otherwise}
            \end{cases}
\end{align*}

In the first case statement we again check whether the current solution estimate $x((N_{t, \theta} + \Exp{N_{> t, \theta}})_{\theta \in \Theta}, B_t^{alg})$, i.e. the solution to the EG program with the current budget, observed current arrivals, and mean future arrivals is feasible.  If the current remaining budget is insufficient to allocate this amount, the algorithm divides all remaining resources evenly.  Otherwise, we allocate this solution earmarked by the EG program.

    \section{Experiment Details}
\label{sec:experiment_details}

\begin{table}[!t]
\caption{Weights $w_k$ for the different products considered in the \textbf{Multi-FBST} experiments.  Here we use the weights taken from the historical prices used in the market mechanism to distribute food resources to food pantries across the United States~\citep{prendergast2017food}.}
\label{tab:weights}
\setlength\tabcolsep{0pt} 
\centering
\begin{tabular*}{\columnwidth }{@{\extracolsep{\fill}}rccccc}
\toprule
Resource & Cereal &  Pasta & Prepared Meals & Rice & Meat\\
\midrule
Weights (type $\theta = $ omnivore) & 3.9 & 3 & 2.8 & 2.7 & 1.9\\
Weights (type $\theta = $vegetarian) & 3.9 & 3 & .1 & 2.7 & .1\\
Weights (type $\theta = $prepared-only) & 3.9 & 3 & 2.8 & 2.7 & .1 \\
\bottomrule
\end{tabular*}
\end{table}

\begin{table}[!t]
\caption{Weights $w_k$ for the different products considered for the \textbf{Multi-Synthetic} experiments.}
\label{tab:weights_synthetic}
\setlength\tabcolsep{0pt} 
\centering
\begin{tabular*}{\columnwidth }{@{\extracolsep{\fill}}rccc}
\toprule
Resource & $k = 1$ &  $k = 2$ & $k = 3$\\
\midrule
Weights for type $\theta = 1$ & 1 & 2 & 3 \\
Weights for type $\theta = 2$ & 1 & 3 & 2 \\
Weights for type $\theta = 3$ & 4 & 1 & 5 \\
Weights for type $\theta = 4$ & 1 & 2 & .5 \\
Weights for type $\theta = 5$ & 3 & 7 & 5 \\
\bottomrule
\end{tabular*}
\end{table}

\begin{figure}
\centering     
\subfigure[\textbf{$\Denv$-$\Deff$}]{\label{fig:aaa}\includegraphics[width=.23\linewidth]{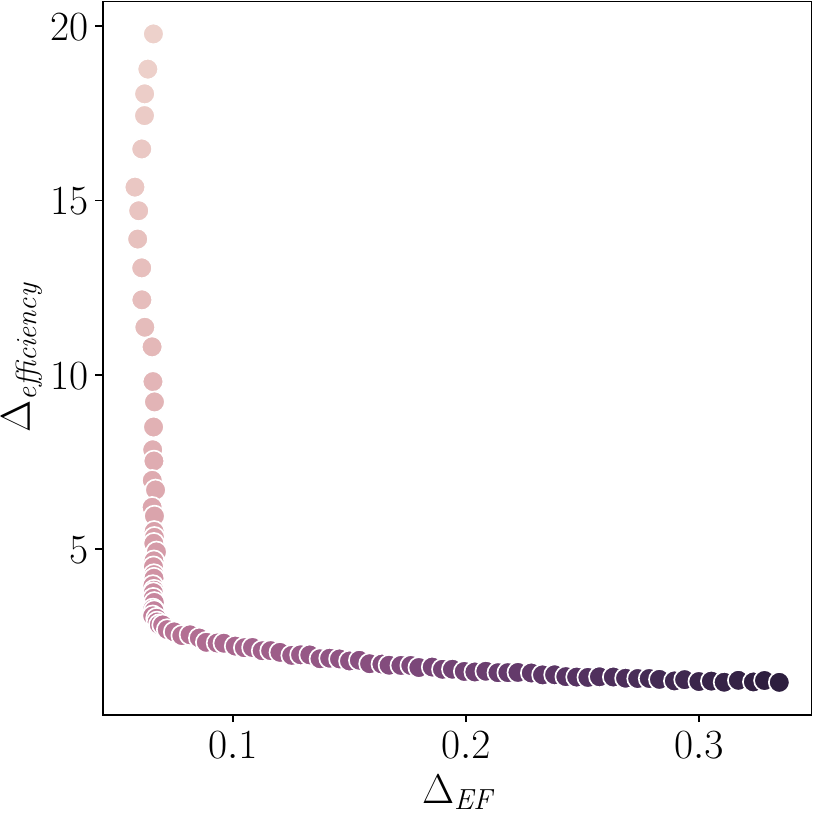}}
\subfigure[\textbf{$\Denvhind$-$\Deff$}]{\label{fig:bbb}\includegraphics[width=.23\linewidth]{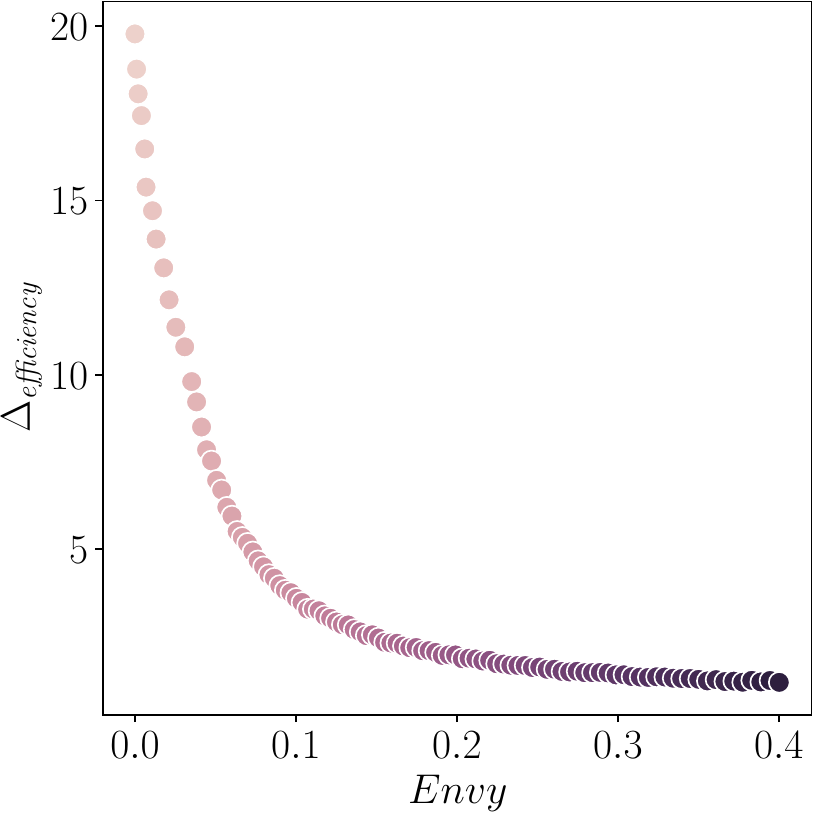}}
\subfigure[\textbf{$L_T$-$\Denv$}]{\label{fig:ccc}\includegraphics[width=.23\linewidth]{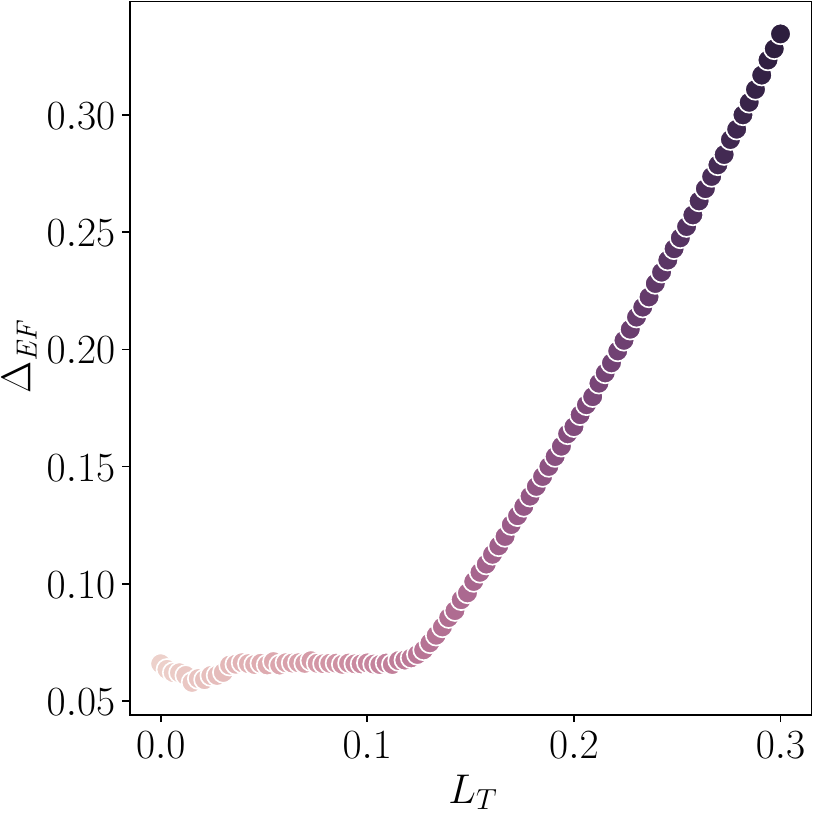}}
\subfigure[\textbf{$L_T$-$\Deff$}]{\label{fig:ddd}\includegraphics[width=.23\linewidth]{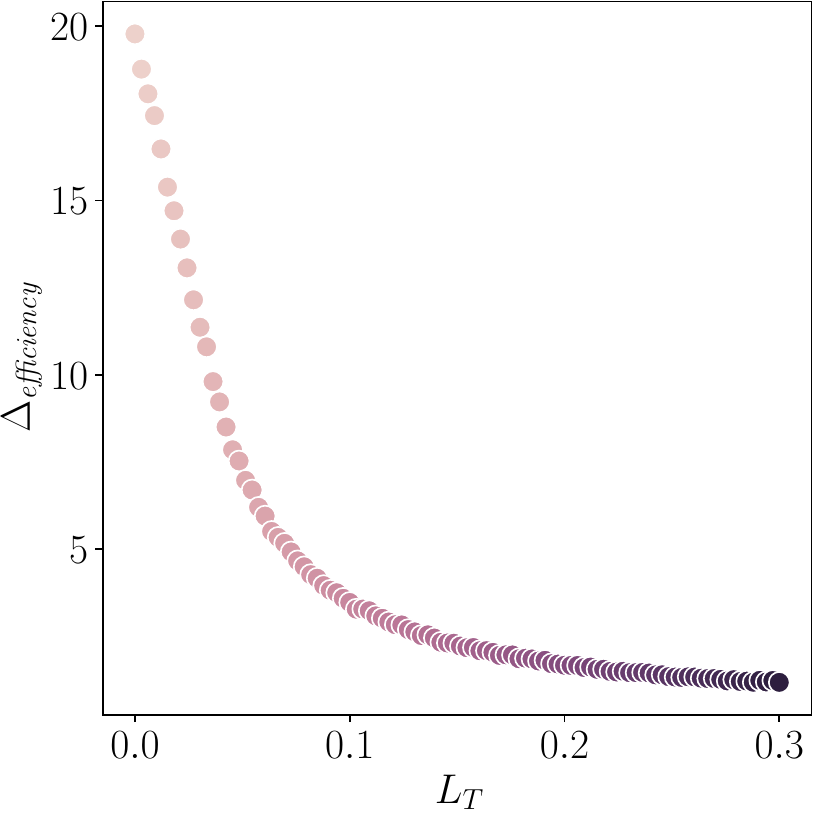}}
\caption{Empirical trade-off curves for the performance of \HopeGuardrail for various values of $L_T$ within $0$ and $0.3$.  Color of the dots corresponds to the value of $L_T$ with darker colors corresponding to larger values.  The number of rounds $T$ was taken to be $200$, with the distributions from the \textbf{Single-Synthetic} set-up.}
\label{fig:empirical_curve}
\end{figure}

\medskip \noindent \textbf{Experiment Setup}: Each experiment was run with $200$ iterations where the relevant plots are taking the mean of the related quantities.  In all experiments the budget $B = \sum_{t, \theta} \Exp{N_{t, \theta}}$ so that $\Bav$ scales as a constant as we vary the number of rounds $T$.  All randomness is dictated by a seed set at the start of each simulation for verifying results.

\medskip \noindent \textbf{Empirical Trade-off Curves}: In \cref{fig:empirical_curve} we show empirical trade-off curves for the performance of \HopeGuardrail for various values of $L_T$.  These plots serve to highlight an empirical version of \cref{fig:uncertainty_principle} highlighting the empirical uncertainty principles exhibited by the algorithm.

\medskip \noindent \textbf{Computing Infrastructure}: The experiments were conducted on a personal computer with an AMD Ryzen 5 3600 6-Core 3.60 GHz processor and 16.0GB of RAM. No GPUs were harmed in these experiments. 
    \section{Omitted Proofs}
\label{app:conc_proof}
We start by giving a full proof for the Lipschitz properties of the optimal solution to the Eisenberg Gale program.

\begin{lemma}[Sensitivity of solutions to the Eisenberg-Gale Program (\cref{lem:kkt_eg} of main paper)]
\label{lem:app_kkt_eg}
Let $x((N_\theta)_{\theta \in \Theta}), p((N_\theta)_{\theta \in \Theta})$ denote the optimal primal and dual solution to the Eisenberg-Gale program (\cref{eq:eg}) for a given vector of individuals of each type $(N_\theta)_{\theta \in \Theta}$.  Then we have that:
\begin{enumerate}
    \item \textit{Scaling}: If $\tilde{N}_\theta = (1 + \zeta) N_{\theta}$ for every $\theta \in \Theta$ and $\zeta \geq 0$ then we have that:
    \begin{align*}
    x((\tilde{N}_\theta)_{\theta \in \Theta}) & = \frac{x((N_\theta)_{\theta \in \Theta})}{1 + \zeta} \\
    p((\tilde{N}_\theta)_{\theta \in \Theta}) & = (1 + \zeta)p((N_\theta)_{\theta \in \Theta}).
    \end{align*}
    Moreover, we have that $$u(X((N_\theta)_{\theta \in \Theta})_\theta, \theta) - u(X((\tilde{N}_\theta)_{\theta \in \Theta})_\theta, \theta) = \left(1 - \frac{1}{1 + \zeta}\right) \max_k \frac{w_{\theta, k}}{p((N_\theta)_{\theta \in \Theta})_k}.$$
    \item \textit{Monotonicity}: If $N_\theta \leq \tilde{N}_{\theta}$ for every $\theta \in \Theta$ then we have 
    \begin{align*}
         p((\tilde{N}_{\theta})_{\theta \in \Theta}) & \geq p((N_{\theta})_{\theta \in \Theta}) \\
        u(X((\tilde{N}_\theta)_{\theta \in \Theta})_\theta, \theta) & \leq u(X((N_\theta)_{\theta \in \Theta})_\theta, \theta) \quad \forall \theta \in \Theta
    \end{align*}
\end{enumerate}
\end{lemma}

\begin{rproof}{\cref{lem:app_kkt_eg}}
For notational brevity we will use $x = x((N_{\theta})_{\theta \in \Theta}), p = p((N_{\theta})_{\theta \in \Theta})$ and $\tilde{x}, \, \tilde{p}$ for the solutions corresponding to $N_\theta$ and $\tilde{N}_\theta$ respectively.  We start off by taking the KKT conditions of the Eisenberg-Gale program.  Using variables $p_k$ for the constraint that $\sum_{\theta \in \Theta} N_{\theta} X_{\theta, k} \leq B_k$ we have the following conditions:
\begin{enumerate}
    \item \emph{Primal Feasibility}: $\sum_{\theta \in \Theta} N_{\theta} X_{\theta, k} \leq B_k$
    \item \emph{Dual Feasibility}: $p_k \geq 0$
    \item \emph{Complementary Slackness}: $p_k > 0$ implies that $\sum_{\theta \in \Theta} N_{\theta} X_{\theta, k} = B_k$
    \item \emph{Gradient Condition}: For every $\theta$ and $k$, $$\frac{w_{\theta, k}}{p_k} \leq \langle w_\theta, X_\theta \rangle$$ with equality whenever $X_{\theta, k} > 0$.
\end{enumerate}
\noindent \textbf{Scaling}: Suppose that $\tilde{N}_\theta = (1 + \zeta)N_\theta$ for every $\theta \in \Theta$.

Set $\tilde{p}_k = (1 + \zeta)p_k$ for each resource $k \in [K]$ and $\tilde{x}_{\theta, k} = x_{\theta, k} / (1 + \zeta)$.  Verifying the KKT conditions for the prices and solutions $\tilde{p}$ and $\tilde{x}$ verifies that these are in fact the optimal primal and dual variables assuming that $p$ and $x$ are.  To show the last property we note that
\begin{align*}
    u(x_\theta, \theta) - u(\tilde{x}_\theta, \theta) & = \max_k \frac{w_{\theta, k}}{p_k} - \max_k \frac{w_{\theta, k}}{\tilde{p}_k} \\
    & = \max_k \frac{w_{\theta, k}}{p_k} - \max_k \frac{w_{\theta, k}}{(1 + \zeta)p_k} \text{ via Scaling} \\
    & = \left(1 - \frac{1}{1 + \zeta}\right) \max_k \frac{w_{\theta, k}}{p_k}
\end{align*}

\noindent \textbf{Monotonicity}: Suppose that $N_\theta \leq \tilde{N}_\theta$ for every $\theta \in \Theta$.  

To show monotonicity with respect to the dual prices we use the t\^atonnement algorithm developed in \citep{nisan_roughgarden_tardos_vazirani_2007} to solve the Eisenberg-Gale program.  At a high level, the algorithm starts off with initial prices $p_k^0$ for $k \in [K]$ which satisfy a \emph{tight set} invariant (found by representing the allocations made via a network flow graph between resources and types).  The tight set invariant is defined as:
$$\forall S \subset [K]: \sum_{k \in S} p_k \leq \sum_{\theta \in \Gamma(S, p)} N_\theta$$
where $$\Gamma(S, p) = \left\{ \theta \in \Theta \mid \exists k \in S \text{ with } k \in \argmax_{k' \in [K]} \frac{w_{\theta, k'}}{p_{k'}} \right\}.$$
The algorithm then raises the prices $p_k^0$ until all of the types have no surplus money.

As such, consider initializing the algorithm to solve for the dual prices $\tilde{p}$ with the prices $p$.  To show monotonicity with respect for the dual prices it suffices to show that the prices $p$ satisfy the \emph{tight set} invariant when the number of individuals of each type is $\tilde{N}_\theta$.  However, this is trivially true due to the optimality of the prices $p$ (as they must satisfy the tight set invariant with respect to $N_\theta$).  Indeed, for any $S \subset [K]$ we must have that:
\begin{align*}
    \sum_{k \in S} p_k & \leq \sum_{\theta \in \Gamma(S, p)} N_\theta \text{ by optimality of the prices } p \\
    & \leq \sum_{\theta \in \Gamma(S, p)} \tilde{N}_\theta \text{ by assumption}.
\end{align*}
As a result we see that the prices $p_k$ satisfy the invariant and can serve as an initialization to the algorithm.  As the algorithm only increases the prices until reaching optimality, it gives us that $p_k \leq \tilde{p}_k$ for every $k \in [K]$.

With this result we are able to show the monotonicity guarantee with respect to the utilities as:

$$u(x_\theta, \theta) = \max_k \frac{w_{\theta, k}}{p_k} \geq \max_k \frac{w_{\theta, k}}{\tilde{p}_k} = u(\tilde{x}_\theta, \theta). \Halmos$$
\end{rproof}

Next we show the following technical lemma providing bounds on the dual prices and allocations made by the algorithm.  
\begin{lemma}
\label{lem:app_ef}
Suppose that $\underline{n}_\theta \leq \Exp{N_\theta}$ and $\overline{X} = x((\underline{n}_\theta)_{\theta \in \Theta}$.  Then we have that:
\begin{enumerate}
    \item $p((\underline{n}_\theta)_{\theta \in \Theta}) \geq \frac{\norm{w}_{min} \norm{\Bav}_{min}}{\norm{w}_\infty}$
    \item $\norm{\overline{X}}_{\infty} \geq \norm{\Bav}_{\infty}$
    \item For any allocation $X \in \mathbb{R}^{|\Theta| \times [K]}$ we have that $\norm{X}_\infty \leq \norm{B}_{\infty}$.
\end{enumerate}
\end{lemma}
\begin{rproof}{\cref{lem:app_ef}}
\noindent \textit{(1)}: For the first property we again use the t\^atonnement algorithm developed in \citep{nisan_roughgarden_tardos_vazirani_2007}.  Initially in the algorithm the prices are set as $p_k^0 = \frac{B_k}{\sum_{\theta} \underline{n}_\theta} \geq \frac{B_k}{\sum_\theta \Exp{N_\theta}}$.  However, if there is a resource which no type is interested in at the price $p_k^0$ then the prices is reduced to $p_k^1 = \max_{\theta \in \Theta} \frac{w_{\theta, k}}{\max_{k'} \frac{w_{\theta, k}}{p_{k'}^0}}.$

Thus we have that
\begin{align*}
    p_k & \geq \min(p_k^0, p_k^1)
     = \min\left(\Bav^{k}, \max_\theta \frac{w_{\theta, k}}{\max_{k'} \frac{w_{\theta, k'}}{\Bav^{k'}}} \right) \\
    & \geq \min\left( \Bav^{k}, \frac{\norm{w}_{min}}{\norm{w}_{max}} \norm{\Bav}_{min} \right)
     = \frac{\norm{w}_{min}}{\norm{w}_{max}} \norm{\Bav}_{min}.
\end{align*}

\noindent \textit{(2)}: This property follows as $\sum_{\theta} \underline{n}_\theta \overline{X}_\theta = B$ and Cauchy-Schwarz inequality.

\noindent \textit{(3)}: This property follows simply as the maximum allocation for any resource $k$ is bounded above by $B_k$. \Halmos
\end{rproof}

Lastly we include the following discussion for the proof on constructing the upper and lower threshold allocations and their resulting properties.

\begin{theorem}[Construction of Guardrail Allocations (\cref{thm:concentration} of main paper)]
\label{thm:concentration_app}
Let $X^{opt} = x((N_{\theta})_{\theta \in \Theta})$ denote the optimal solution to the Eisenberg-Gale program for a given vector of individuals of each type $(N_{\theta})_{\theta \in \Theta}$.  Further suppose that with probability at least $1 - \delta$ we have for all $\theta \in \Theta$: $|N_{\theta} - \Exp{N_{\theta}}| \leq \conf_\theta$.  Given any $L_T \geq 0$ and setting:
\begin{align*}
    \overline{n}_\theta & = \Exp{N_{\theta}}\left(1 + \max_\theta \frac{\conf_\theta}{\Exp{N_\theta}}\right)\\
    \underline{n}_\theta & = \Exp{N_{\theta}}\left(1 - c\right) \quad \text{ for } c = \frac{\norm{w}_{min} \norm{\Bav}_{min}}{\norm{w}^2_{\infty}} L_T \left(1 + \max_\theta \frac{\conf_\theta}{\Exp{N_\theta}}\right) - \max_\theta \frac{\conf_\theta}{\Exp{N_\theta}}
\end{align*}
then almost surely we have that: 
\begin{enumerate}
    \item $u(x((\underline{n}_{\theta})_{\theta \in \theta})_\theta, \theta) - u(x((\overline{n}_{\theta})_{\theta \in \theta})_\theta, \theta) \leq L_T$
    \item $\norm{x((\overline{n}_{\theta})_{\theta \in \Theta}) - x((\underline{n}_{\theta})_{\theta \in \theta})}_{\infty} \geq L_T \frac{\norm{\Bav}^2_{min} \norm{w}_{min}}{\norm{w}_\infty}$
    \item $\norm{x((\overline{n}_{\theta})_{\theta \in \Theta}) - x((\underline{n}_{\theta})_{\theta \in \theta})}_{\infty} \leq L_T \frac{\norm{B}_\infty \norm{\Bav}_{min} \norm{w}_{min}}{\norm{w}_\infty}$.
\end{enumerate}
If in addition $L_T \geq 2 \frac{\norm{w}^2_{\infty}}{\norm{w}_{min} \norm{\Bav}_{min}} \max_{\theta} \frac{\conf_\theta}{\Exp{N_\theta}}$ then with probability at least $1 - \delta$ we have:
\begin{enumerate}
    \item[\emph{4.}] $\underline{n}_\theta \leq N_\theta \leq \overline{n}_\theta$
    \item[\emph{5.}] $u(x((\overline{n}_{\theta})_{\theta \in \theta})_\theta, \theta) \leq u(X^{opt}_\theta, \theta) \leq u(x((\overline{n}_{\theta})_{\theta \in \theta})_\theta, \theta)$
\end{enumerate}
\end{theorem}

\begin{rproof}{\cref{thm:concentration_app}}
Define $\multconf = \max_{\theta} \conf_\theta / \Exp{N_\theta}$.  Then
$\overline{n}_\theta = \Exp{N_\theta}(1 + \multconf)$, $\underline{n}_\theta = \Exp{N_\theta}(1 - c)$, and so $\overline{n}_\theta = \frac{1 + \multconf}{1 - c} \underline{n}_\theta.$  Using \cref{lem:kkt_eg} and \cref{lem:app_ef} (from the Appendix) it follows that the difference in utilities for the guardrails and any $\theta$ is bounded by:
\begin{align*}
   u(x((\underline{n}_{\theta})_{\theta \in \theta})_\theta, \theta) - u(x((\overline{n}_{\theta})_{\theta \in \theta})_\theta, \theta) = & \left(1 - \frac{1 - c}{1 + \multconf}\right) \max_k \frac{w_{\theta, k}}{p((\underline{n}_\theta)_{\theta \in \Theta})} \\
   & \leq \frac{c + \multconf}{1 + \multconf} \frac{\norm{w}^2_\infty}{\norm{w}_{min} \norm{\Bav}_{min}}  = L_T.
\end{align*}
Moreover, again using \cref{lem:kkt_eg} we can bound the difference in allocations by
\begin{align*}
    \norm{x((\overline{n}_{\theta})_{\theta \in \Theta}) - x((\underline{n}_{\theta})_{\theta \in \theta})}_{\infty} & = \left(1 - \frac{1 - c}{1 + \multconf}\right) \norm{x((\underline{n}_{\theta})_{\theta \in \theta})}_\infty \geq \frac{c + \multconf}{1 + \multconf} \norm{\Bav}_{min} \\
    & = \frac{L_T \norm{\Bav}_{min} \norm{w}_{min}}{\norm{w}_\infty} \norm{\Bav}_{min} = L_T \frac{\norm{\Bav}^2_{min} \norm{w}_{min}}{\norm{w}_\infty}.
\end{align*}
The upper bound is similar where we use $\norm{x((\underline{n}_{\theta})_{\theta \in \theta})}_\infty \leq \norm{B}_\infty$.

Lastly we further suppose that $L_T \geq 2 \frac{\norm{w}^2_{\infty}}{\norm{w}_{min} \norm{\Bav}_{min}} \max_{\theta} \frac{\conf_\theta}{\Exp{N_\theta}}$.  First notice that via our concentration guarantees:
\begin{align*}
    N_{\theta} & = \Exp{N_\theta} + (N_\theta - \Exp{N_\theta}) = \Exp{N_\theta}\left(1 + \frac{N_\theta - \Exp{N_\theta}}{\Exp{N_\theta}}\right) \\
    & \leq \Exp{N_\theta}\left(1 + \frac{\conf_\theta}{\Exp{N_\theta}}\right) \leq \Exp{N_\theta}\left(1 + \max_{\theta} \frac{\conf_\theta}{\Exp{N_\theta}}\right)
\end{align*}
Similarly we can show that $N_\theta \geq \Exp{N_\theta}\left(1 - \max_{\theta} \frac{\conf_\theta}{\Exp{N_\theta}}\right).$  By construction then we have that $\overline{n}_\theta \geq N_\theta$.  Simple algebraic manipulations and the assumption on $L_T$ shows that $c \geq \max_\theta \frac{\conf_\theta}{\Exp{N_\theta}}$ and so $\underline{n}_\theta \leq N_\theta$ as well.  Moreover from the monotonicity property in \cref{lem:kkt_eg} we have for all $\theta \in \Theta$, $u(x((\overline{n}_{\theta})_{\theta \in \Theta})_\theta, \theta) \leq u(X^{opt}_{\theta}, \theta) \leq u(x((\underline{n}_{\theta})_{\theta \in \theta})_\theta, \theta)$.  \Halmos 
\end{rproof}

Next we give an example of the desired form of the concentration inequalities to obtain $\multconf$. For almost surely bounded demands, under the assumptions outlined in \cref{sec:preliminary}, a simple application of Hoeffding's inequality gives:
\begin{lemma}
\label{lem:hoeffding_app}
With probability at least $1 - \delta$, for every $\theta \in \Theta$ and $t \in [T]$ we have that $|N_{> t, \theta} - \Exp{N_{> t, \theta}}| \leq \conf_{t, \theta}$ where $$\conf_{t, \theta} = \sqrt{2 (T - t + 1) \rho_{max}^2 \log(T |\Theta| / \delta)}.$$
\end{lemma}
\begin{rproof}{\cref{lem:hoeffding_app}}
Let $t$ and $\theta$ be fixed.  By assumption we know that $N_{t, \theta} \in [\Exp{N_{t, \theta}} - \rho_{t, \theta}, \Exp{N_{t, \theta}} + \rho_{t, \theta}]$.  From a simple application of Hoeffding's inequality:
$$\Pr(|N_{> t, \theta} - \Exp{N_{> t, \theta}}| \geq \epsilon) \leq 2 \exp\left( - \frac{2 \epsilon^2}{\sum_{t' > t} 4 \rho_{t', \theta}^2}\right)$$
Setting the right hand side equal to $\delta$, and relabeling $\delta$ to $\delta / T |\Theta|$ via a union bound shows the result.
\end{rproof}

\end{document}